%% file: ismd08-all-hepph.tex
\pdfoutput=1

\documentclass[twoside]{report}
\usepackage{longtable,tabularx,array,url,ifthen,multicol}
\usepackage{fancyhdr,desyhead}
\usepackage{graphpap}
\usepackage[fit]{truncate}

\newcommand{\BLKP}{% add blank page when on even page after PDF file
  \ifthenelse{\isodd{\value{page}}}{\relax}{\mbox{}\thispagestyle{empty}\newpage}}
\newcommand{\CLDP}{\newpage % add blank page when on even page after PDF file
  \ifthenelse{\isodd{\value{page}}}{\relax}{\mbox{}\thispagestyle{empty}\newpage}}
\usepackage{geometry}
\usepackage{fancyhdr}
\newcommand{\ARTauthor}{~}
\newcommand{\ARTtitle}{~}
\usepackage{pdfpages}
\newenvironment{papers}{\clearpage}{\clearpage}
\newcommand*\coltoctitle[1]{\def\CTIT{#1}}
\newcommand*\coltocauthor[1]{\def\CAUT{#1}}
\makeatletter
\renewcommand{\l@section}{\@dottedtocline{1}{2em}{0em}}
\renewcommand{\@dotsep}{1000}
\makeatother
\newcommand{\Includeart}[4][]{%
\def\AAA{#2}\def\TTT{#3}%
\renewcommand{\ARTauthor}{~}
\renewcommand{\ARTtitle}{~}
   \includepdf[%frame,
               pages=1,
               noautoscale,
               pagecommand={\pagestyle{fancy}},
               offset=0mm 0mm,
               addtotoc={1, subsubsection, 3, ~,  S#4},
               trim=19mm 21mm 19mm 27mm, clip]
               {#4.pdf}%
\addtocontents{toc}{\protect\contentsline{chapter}{\textbf{\TTT}}{\textbf{\pageref{S#4}}}}
\addtocontents{toc}{\protect\par}
\addtocontents{toc}{\protect\contentsline{section}{\AAA}{~}}
\ifthenelse{\equal{#1}{OnePage}}{%One page only, exit
                                }{%Handle pages 2 to end
  \renewcommand{\ARTauthor}{\truncate{.9\linewidth}{\AAA}}
  \renewcommand{\ARTtitle}{\truncate{.9\linewidth}{\TTT}}
    \includepdf[%frame,
                pages=2-,
                noautoscale,
                pagecommand={\pagestyle{fancy}},
                offset=0mm 0mm,
                trim=19mm 21mm 19mm 27mm, clip]
                {#4.pdf}}
}
\begin{document}
\pagestyle{empty}
\setlength{\fboxsep}{0pt}
\setlength{\fboxrule}{0.04pt}
%%%% here inser cover page
\setlength{\unitlength}{1mm}
%\begin{picture}(0.001,0.001)
%%\put(-20,-240){\includegraphics[bb=60 60 538 736,width=16.3cm]{title-deckblatt.ps}}
%\put(-20,-230){\includegraphics[width=16.3cm]{title-deckblatt.jpg}}
%\put(-20,-230){\includegraphics[width=16.3cm]{ismd-deckblatt-new.pdf}}
%\end{picture}
%\newpage
%%%% end cover page
\pagestyle{plain}
\pagenumbering{roman}
\setcounter{page}{1}
\CLDP
\parindent 0mm
\hspace*{12 cm} {\small DESY--PROC--2009--01}\\
\vspace*{2cm}
{\sffamily\LARGE
Proceedings of the\\
\Huge
38th International Symposium on \\
Multiparticle Dynamics \\
ISMD08
\vspace{7mm}

\LARGE
September 15-20, 2008\\[1ex]
DESY, Hamburg, Germany}
\vfill

{\Large
Editors:  
Jochen Bartels,
Kerstin Borras,
G\"{o}sta Gustafson,
Hannes Jung,
Krzysztof Kutak,
Serguei Levonian,
Joachim Mnich

\vspace{1cm}
Verlag Deutsches Elektronen-Synchrotron}

\newpage

{\bfseries\Large
Impressum
\vspace{10mm}

Proceedings of the 38th International Symposium on Multiparticle Dynamics
(ISMD08)\\[1ex]
September 15-20, 2008, Hamburg, Germany}
\vspace{10mm}

Conference homepage\\
\url{http://ismd08.desy.de}

\vspace{5mm}

Slides at\\
\url{hhttps://indico.desy.de/conferenceOtherViews.py?confId=754}

\vspace{5mm}
Online proceedings at\\
\url{http://ismd08.desy.de/proceedings.html}

\vfill

The copyright is governed by the Creative Commons agreement, which allows
for free use and distribution of the articels for non-commercial activity,
as long as the title, the authors' names and the place of the original are
referenced.

\vspace{10mm}

Editors:\\
Jochen Bartels (University Hamburg) ,
Kerstin Borras (DESY),
G\"{o}sta Gustafson (University Hamburg, University Lund),
Hannes Jung (DESY, University Antwerp),
Krzysztof Kutak (DESY),
Serguei Levonian (DESY),
Joachim Mnich (DESY) \\
%Cover Photo: \copyright %Rainer Mankel, DESY
%Photo of Participants: Marta Mayer (DESY, Hamburg) \\
January 2009\\
DESY-PROC-2009-01  \\
ISBN 978-3-935702-31-7\\
ISSN 1435-8077

\vspace{5mm}

Published by\\
Verlag Deutsches Elektronen-Synchrotron\\
Notkestra{\ss}e 85\\
22607 Hamburg\\
Germany
\newpage

\begin{flushleft} 
\mbox{}\\[1cm]
{\bfseries\large Organizing Committee:}\\[3mm]
Jochen Bartels (University Hamburg),
Kerstin Borras (DESY), 
G\"{o}sta Gustafson (cochair) (University Hamburg, University Lund),
Hannes Jung (cochair) (DESY, University Antwerp), 
Krzysztof Kutak (DESY),
Serguei Levonian (DESY),
Joachim Mnich (DESY)  
\\[6mm]
{\bfseries\large Conveners:}\\[3mm]
{\bfseries Dilute systems:} \\
A. Cooper-Sarkar (Oxford), A. Kulesza (DESY), K. Hatakeyama (Rockefeller)

{\bfseries High density systems in ep, pp and heavy ions:}\\
D. d'Enterria (CERN), T. Cs\"org\H{o} (Budapest), E. Iancu (Saclay)

{\bfseries Interpolation region:}\\
M. Grothe (Wisconsin), M Albrow (FNAL), K. Werner (Nantes)

{\bfseries Strategies and analysis methods:}\\
E. DeWolf (Antwerp), A. Geiser (DESY), M. Sj\"odahl (Manchester)

{\bfseries New physics:}\\
G. Weiglein (Durham), A. DeRoeck (CERN, University Antwerp)
\\[6mm]

{\bfseries\large Advisory Committee:}\\[3mm]
A. Bia{\l}as (Cracow), 
T.  Cs\"org\H{o} (Budapest), 
I. Dremin (Moscow), 
K. Fia{\l}kowski (Cracow), 
B. Gary (Riverside), 
R. Godbole (Bangalore), 
T. Hallman (Upton), 
Y. Hama (Sao Paulo), 
W. Kittel (Nijmegen), 
T. Kodama (Rio de Janeiro), 
V. Kuvshinov (Minsk), 
L. Liu (Wuhan), 
L. McLerran (Brookhaven), 
N. Schmitz (Munich), 
A. Valkarova (Prague), 
E. De Wolf (Antwerp),
N. Xu (Berkeley) \\[6mm]
%{\bfseries\large Permanent Committee of the ISMD Symposia:}\\[3mm]
% \\[6mm]
{\bfseries\large Supported by:}\\[3mm]
Deutsches Elektronen-Synchroton DESY\\
Deutsche Forschungsgemeinschaft \\
SFB 676 (Particles, Strings and the Early Universe)
\end{flushleft}

\newpage

\begin{center}
\mbox{}\\[5mm]
{\bfseries\Large Preface }\\[1cm]
\end{center}

The {\it XXXVIII International Symposium on Multiparticle Dynamics}, 
ISMD~2008, was held on 15 - 20 September at DESY, Hamburg. This series of 
symposia attracts participants from different areas, with a common interest in 
reactions where a large number of particles are produced. Traditionally 
this has mainly included three communities, hadronic collisions and DIS, 
high energy heavy ion collisions, and cosmic ray physics. With the increasing 
accelerator energies, in particular awaiting the startup of the LHC, these 
communities have been brought even closer. Effects of gluon saturation is 
not only of interest in heavy ion collisions; saturation and multiple 
subcollisions are possibly seen at HERA, are essential features at the Tevatron, 
and will be a major effect in minimum bias and underlying events at the LHC. 
With the LHC the energy in collider experiments will also become comparable to that 
in high energy cosmic rays. As the cosmic ray detectors are mostly sensitive 
to the forward region, the increased forward coverage in the LHC 
experiments brings these communities closer together.

To further encourage contacts and exchanges of information across the community 
bounderies, the organizers chose to test a new way to schedule the sessions. 
Thus five cross-diciplinary sessions were organized in which a predefined  set
of questions ought to be addressed in the sessions.

\begin{itemize}
\item {\bf Dilute systems}\\
\emph{Conveners}: A. Cooper-Sarkar (Oxford), A. Kulesza (DESY), K. Hatakeyama (Rockefeller).\\
\emph{Topics}: 
High $Q^2$, structure functions, jets, central production.\\
\emph{Questions}:
How well do we know PDFs from HERA and the TeVatron and how well can we predict cross sections at the LHC?
           Are NLO PDFs all we need?
           Which PDFs should be used in Monte Carlo event generators?
           How small in $x$ can we trust the linear evolution equations (BFKL/DGLAP)?

\item {\bf High density systems in ep, pp and heavy ions}\\
\emph{Conveners}: D. d'Enterria (CERN), T. Cs\"org\H{o} (Budapest), E. Iancu (Saclay).\\
\emph{Topics}: 
Saturation, hydrodynamics, CGC, perfect fluids, AdS/CFT.\\
\emph{Questions}:
           What are the expected effects of multiple interactions and saturation at the LHC?
           What lesson from heavy ion collisions can shed light on saturation in $pp$ and $ep$ and vice versa?
           What can heavy ion physics learn from $pp$ and $ep$?
           Can ideas from hydrodynamics, classical fields and quenching be used in $pp$?
           What from hard QCD calculations can be used in heavy ion physics?

\item {\bf Interpolation region}\\
\emph{Conveners}: M. Grothe (Wisconsin), M. Albrow (FNAL), K. Werner (Nantes).\\
\emph{Topics}: Forward production at highest energies, diffraction, glasma.\\
\emph{Questions}:
           What can cosmic rays say about forward physics at the LHC and vice versa?
           Transition from dense to dilute systems, as discussed in HI \& Cosmic Rays-relevance for pp:
           forward production, diffractive event (shadow of a dense system) and PDFs?

\item {\bf Strategies and analysis methods}\\
\emph{Conveners}: E. DeWolf (Antwerp), A. Geiser (DESY), M. Sj\"odahl (Manchester).\\
\emph{Topics}: Correlations, heavy quark production, MC techniques.\\
\emph{Questions}:
           How can forward detectors improve our understanding of QCD effects and identify signals for new physics?
           Are existing tools sufficient and where should they be improved?
           How can correlations be used to determine the size of the interaction and phase transitions?

\item {\bf New physics}\\
\emph{Conveners}: G. Weiglein (Durham), A. DeRoeck (CERN).\\
\emph{Topics}: Possible signals for Higgs, Susy, etc.\\
\emph{Questions}:
           Scenarios for new physics: how probable are the different scenarios?
           How to discriminate new physics from complicated background like multiparton interactions?
           How to discriminate different scenarios?
           How well can the  LHCexperiments cope with the different scenarios?
\end{itemize}

The new scheme would not have worked so well without the excellent work by 
the conveners, who invited good speakers and prepared a very interesting 
program. This apparently looked attractive, and the symposium had a total
of 127 registered participants.
An unavoidable consequence of this success  was, that 
it became necessary to include two evening sessions, one of them even with 
paralell talks. 

The program also included a discussion session about the applicability and 
limitations of collinear factorization, and of linear parton evolution. 
As a preparation before the real symposium, we had two introductory talks
by H. Meyer and L. McLerran. 
 
Another new feature was that all plenary talks were video recorded, 
and together with the transparencies are available from:\\
\verb+https://indico.desy.de/conferenceOtherViews.py?confId=754+

Unfortunately, not all presentations during the workshop appear as a writeup in
these proceedings.  Ch. Anastasiou, H. B\"usching and  V.V. Khoze were not able 
to deliver a written version of their contribution.  The online
version of the proceedings can be found at:\\
\verb$http://ismd08.desy.de$.

We wish to thank all the participants of ISMD08 for making this symposium so interesting and lively. 
We thank especially the conveners for their enormous work in the preparation of this symposium.
We are in particular indebted to the summary speakers,
P. van Mechelen and Y. Kovshegov, who in an excellent way fulfilled their 
difficult tasks. 

Last but not least we wish to thank 
A. Grabowksy, S. Platz and L. Schmidt for their continuous help and support during
all the meeting week. We thank B. Liebaug for the design of the poster.  
We are grateful to R. Eisberg, O. Knak and S. K\"onig for recording the talks and all technical help. We thank M.
Mayer, K. Sachs and M. Stein for their help in  printing the proceedings.
We are grateful to the DESY directorate for financial support
of this workshop  and for the hospitality which they extended to all
participants of the workshop, and to the DFG and the SFB for financial support.

\begin{flushleft} 
{The Organizing Committee:}\\
Jochen Bartels,
Kerstin Borras,
G\"{o}sta Gustafson(cochair),
Hannes Jung(cochair),
Krzysztof Kutak,
Serguei Levonian,
Joachim Mnich
\end{flushleft}

%\begin{picture}(0.001,0.001)
%\put(0,-100){\includegraphics[width=15cm]{../../pictures/IMG_01191.jpg}}
%\end{picture}

\CLDP
\tableofcontents % main ToC
\CLDP

\pagestyle{fancy}
\setcounter{page}{1}
\pagenumbering{arabic}

\include{include-intro}

\include{include-dilute}

\include{include-dense}

\include{include-interpolation}

\include{include-methods}

\include{include-new-physics}

\include{include-recent-developments}

\include{include-discussion}

\include{include-summary}

\pagestyle{desyplain}
\CLDP 

%\begin{papers} % start of individual articles/papers
%\coltoctitle{Authors}
%\Includeart{}{\CTIT}{include-authors} 
%\end{papers} % start of individual articles/papers
%\CLDP 

%\centerline{\bfseries\Large Participants}
\begin{papers} % start of individual articles/papers
\coltoctitle{Participants}
\Includeart{}{\CTIT}{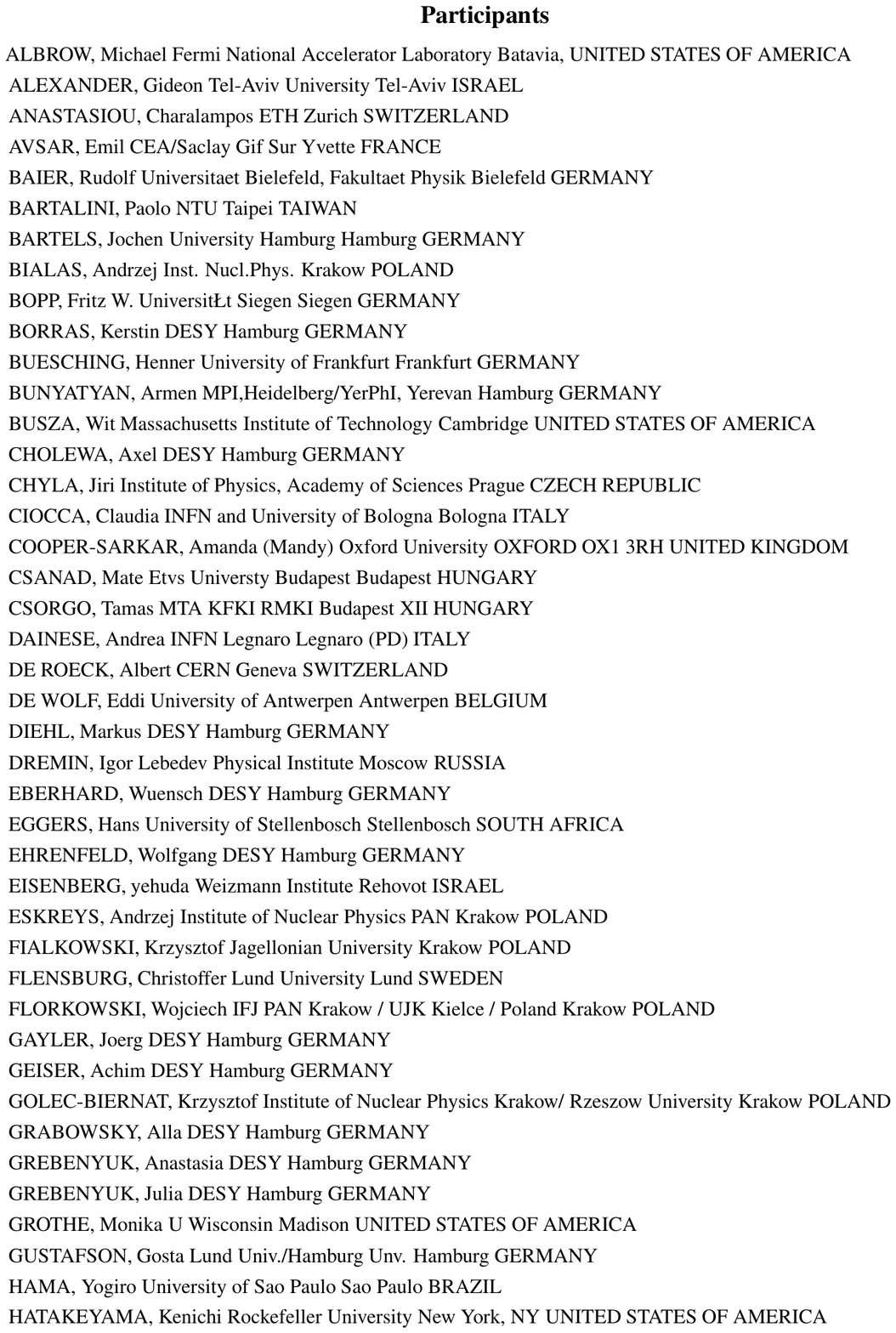} 
\end{papers} % start of individual articles/papers
\end{document}

%% file: include-intro.tex
\chapter[\large Introductory Lecture]{ Introductory Lecture}

\CLDP
\begin{papers} % start of individual articles/papers

\coltoctitle{A Brief Introduction to the Color Glass Condensate and the Glasma}
\coltocauthor{L. McLerran}
\Includeart{\CAUT}{\CTIT}{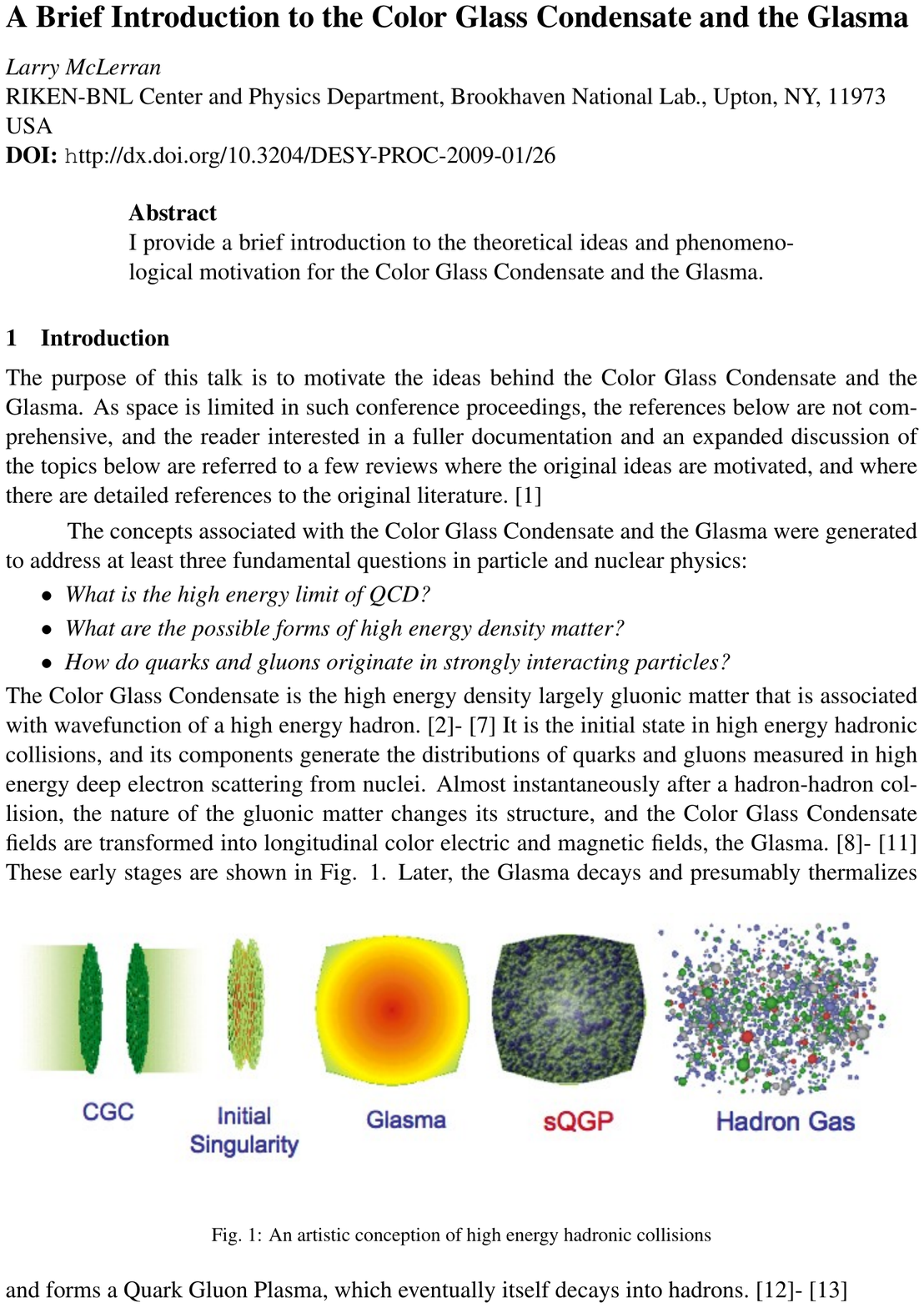}

\end{papers} % end of individual articles/papers

%% file: include-dilute.tex
\chapter[\large  WG: Dilute systems]{ Working Group \\ Dilute Systems}
{\Large\bfseries Convenors:}
\vspace{10mm}

{\Large\itshape 
Mandy Cooper-Sarkar (Oxford)\\
Anna Kulesza (DESY) \\
Ken Hatakeyama (Rockefeller)}

\CLDP
\begin{papers} % start of individual articles/papers

\coltoctitle{Cross section measurements in DIS}
\coltocauthor{K. Papageorgiou}
\Includeart{\CAUT}{\CTIT}{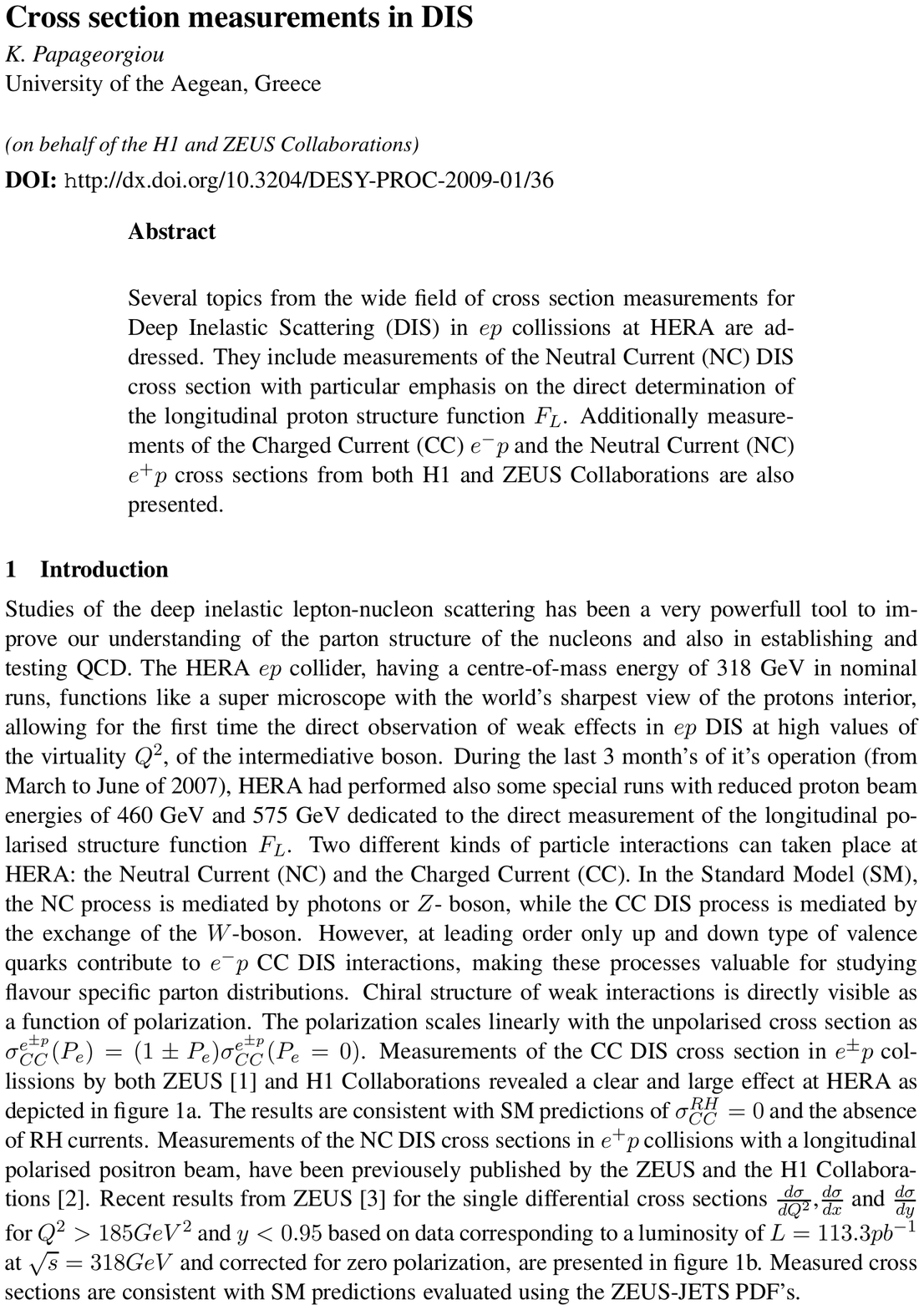} 

\coltoctitle{Jet Production at HERA}
\coltocauthor{A. Savin}
\Includeart{\CAUT}{\CTIT}{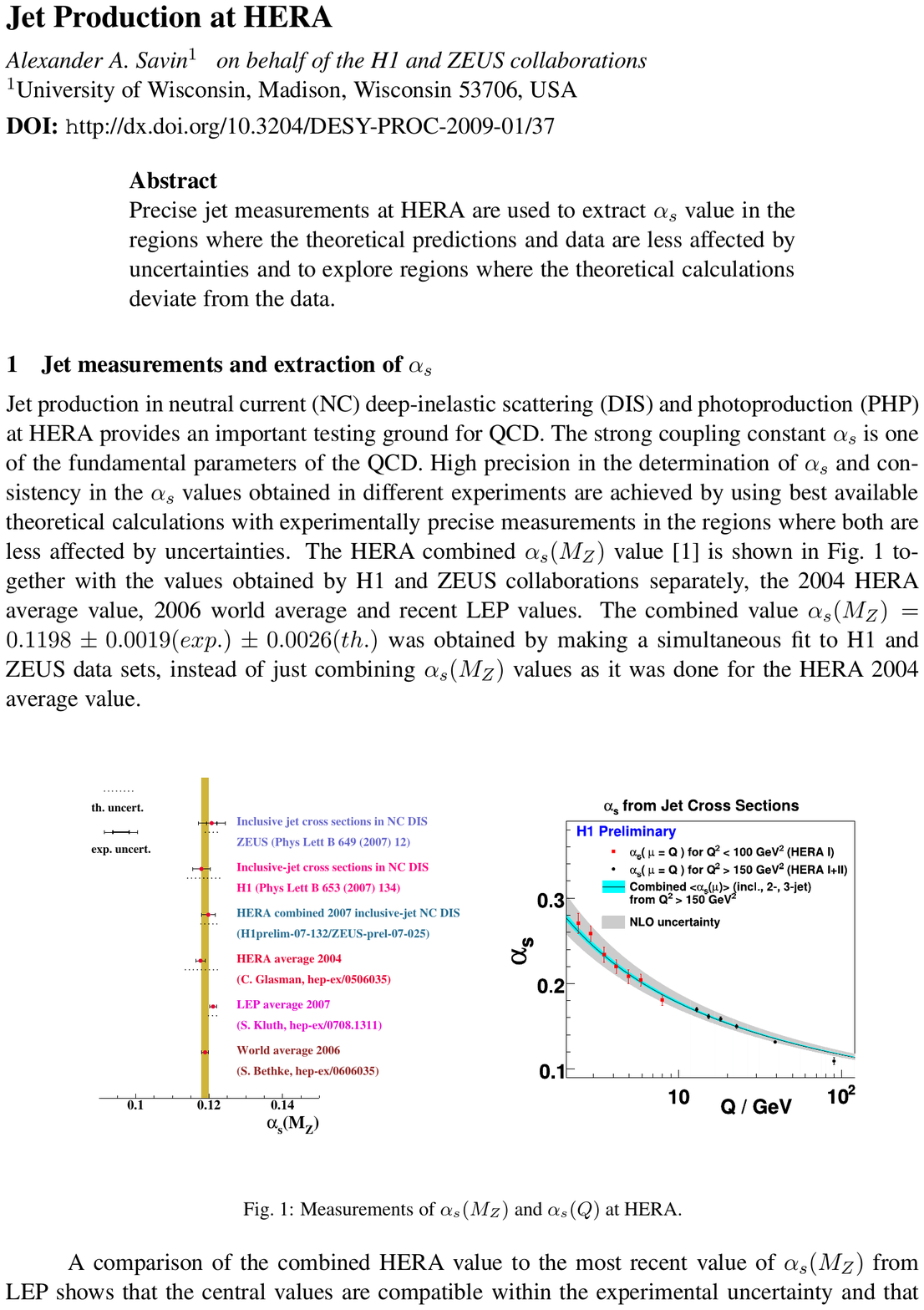} 

\coltoctitle{Extraction of the proton parton density functions using
a NLO-QCD fit of the combined H1 and ZEUS inclusive
DIS cross sections}
\coltocauthor{G. Li}
\Includeart{\CAUT}{\CTIT}{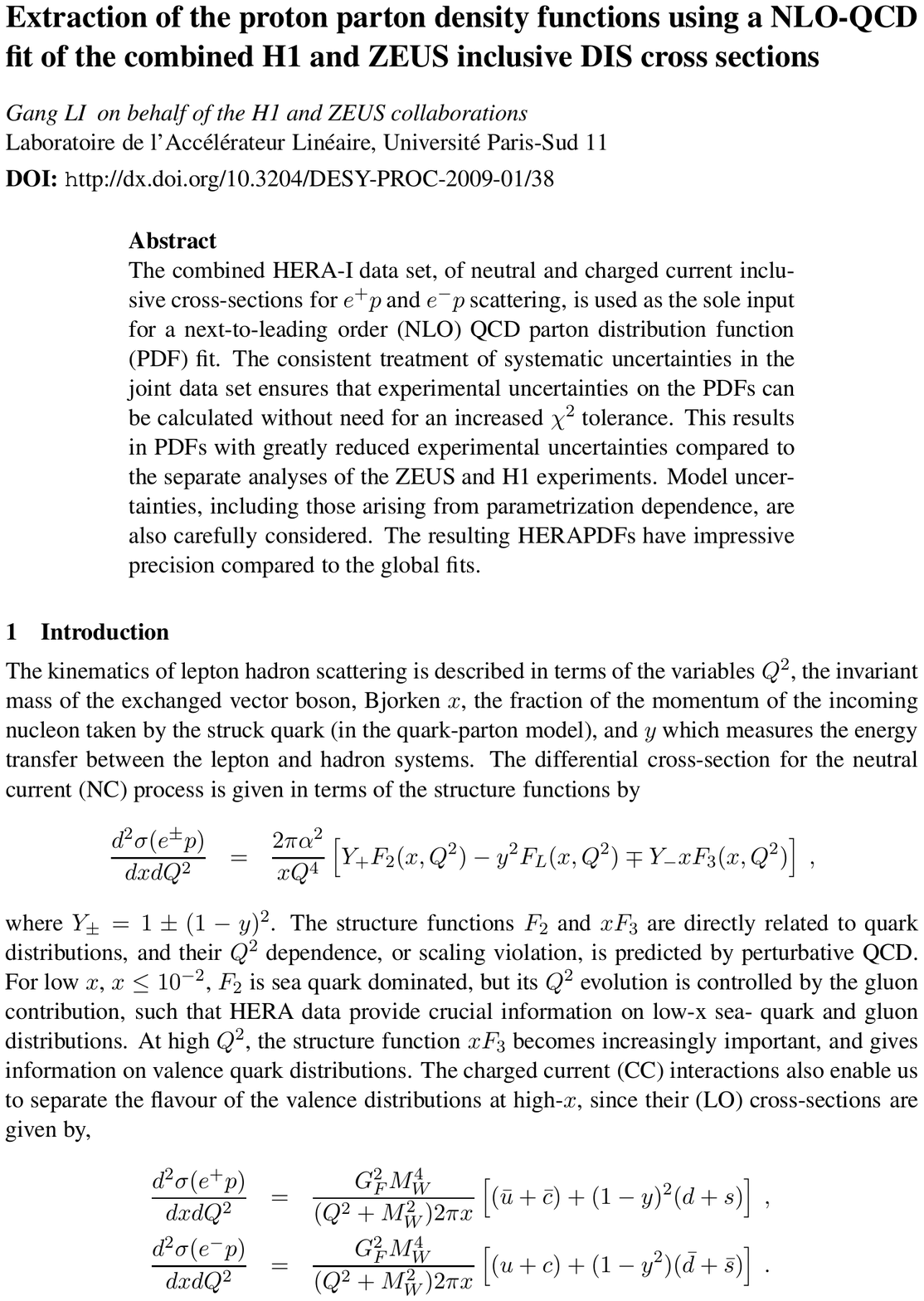} 

\coltoctitle{PDF Constraints From Tevatron Data}
\coltocauthor{M. Lancaster}
\Includeart{\CAUT}{\CTIT}{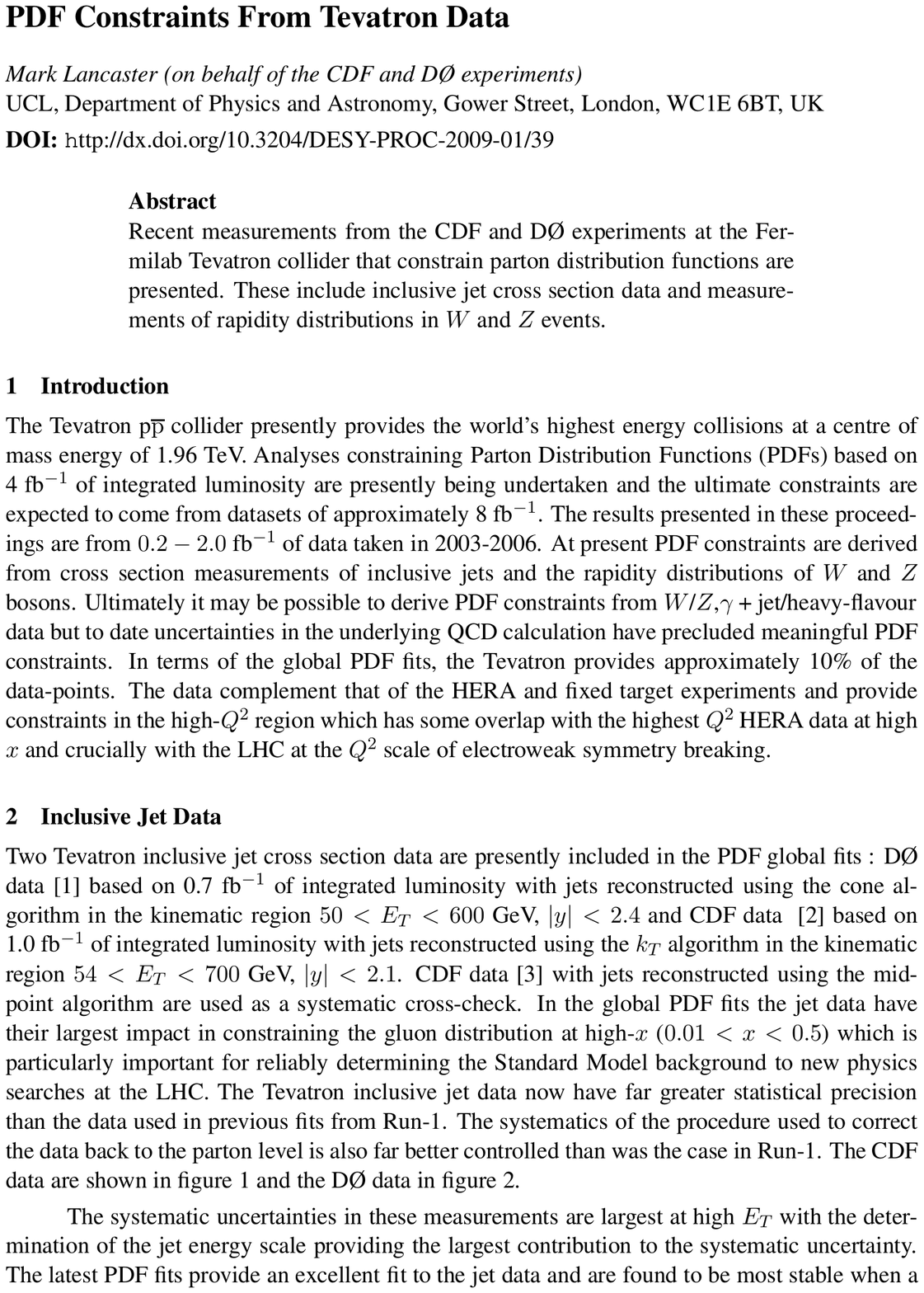} 

\coltoctitle{Small $x$ Resummation - An Overview}
\coltocauthor{Ch. White}
\Includeart{\CAUT}{\CTIT}{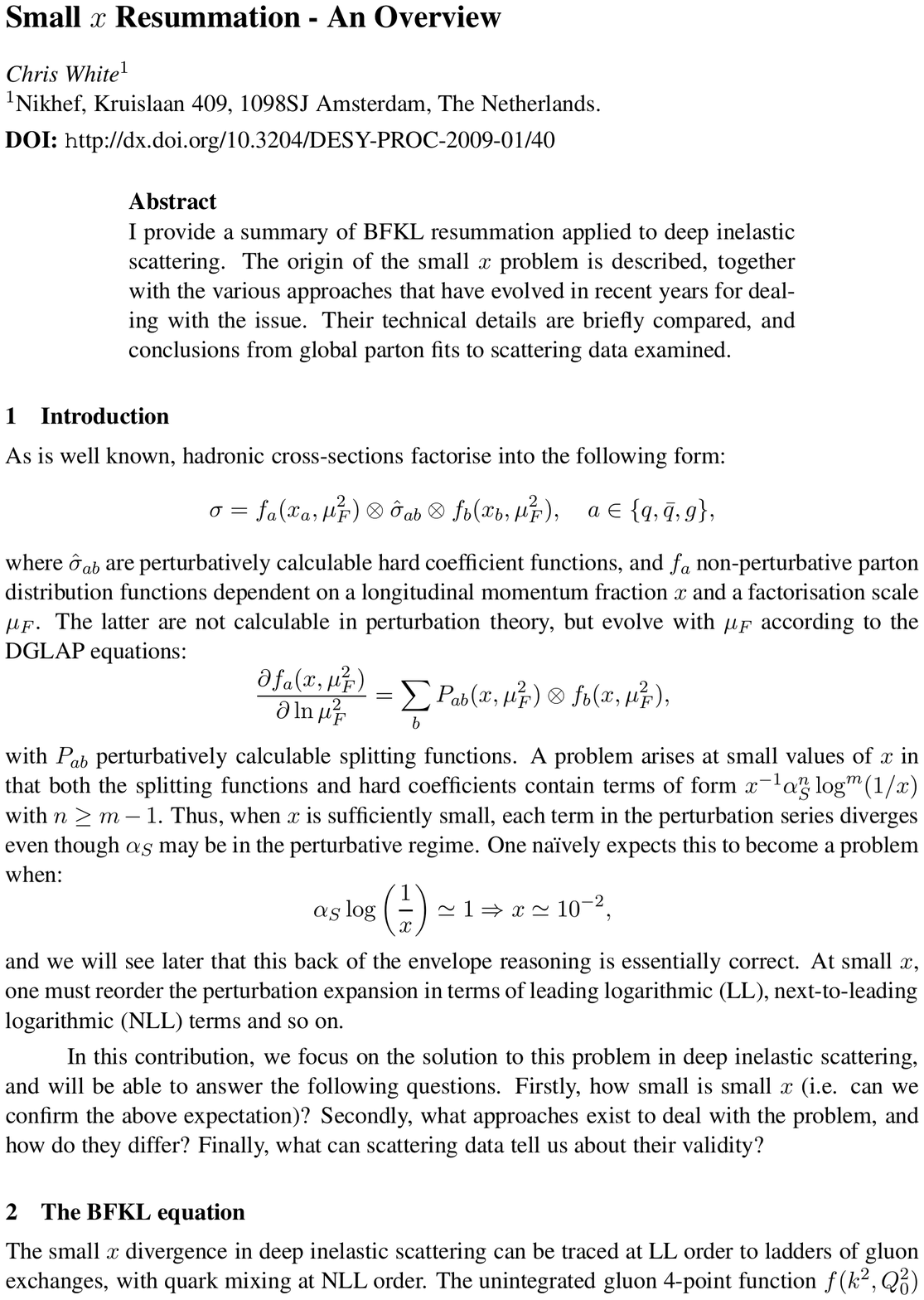} 

\coltoctitle{Progress in Parton Distribution Functions 
and Implications for LHC}
\coltocauthor{J. Stirling}
\Includeart{\CAUT}{\CTIT}{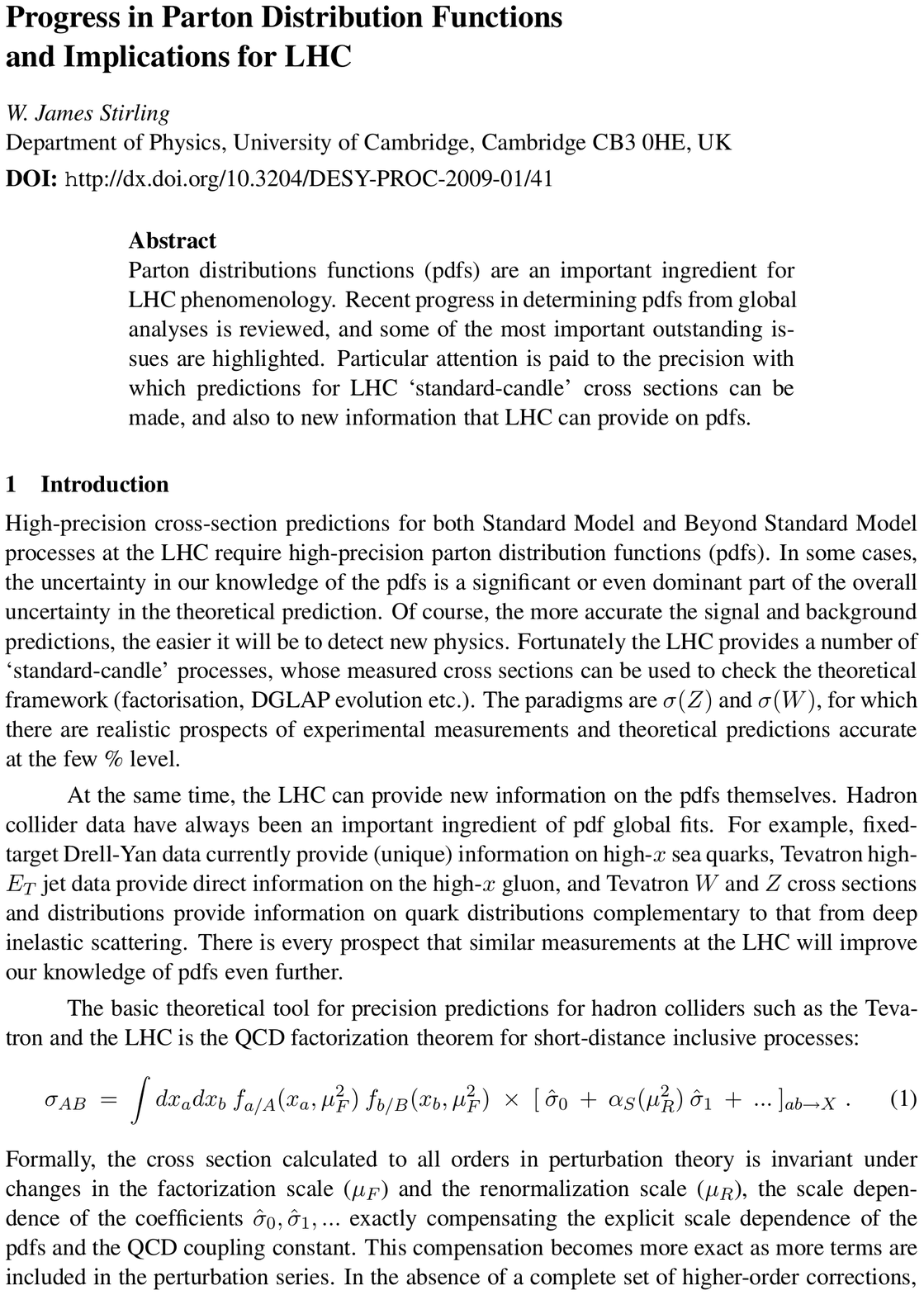} 

\coltoctitle{Theoretical predictions for the LHC}
\coltocauthor{S. Moch}
\Includeart{\CAUT}{\CTIT}{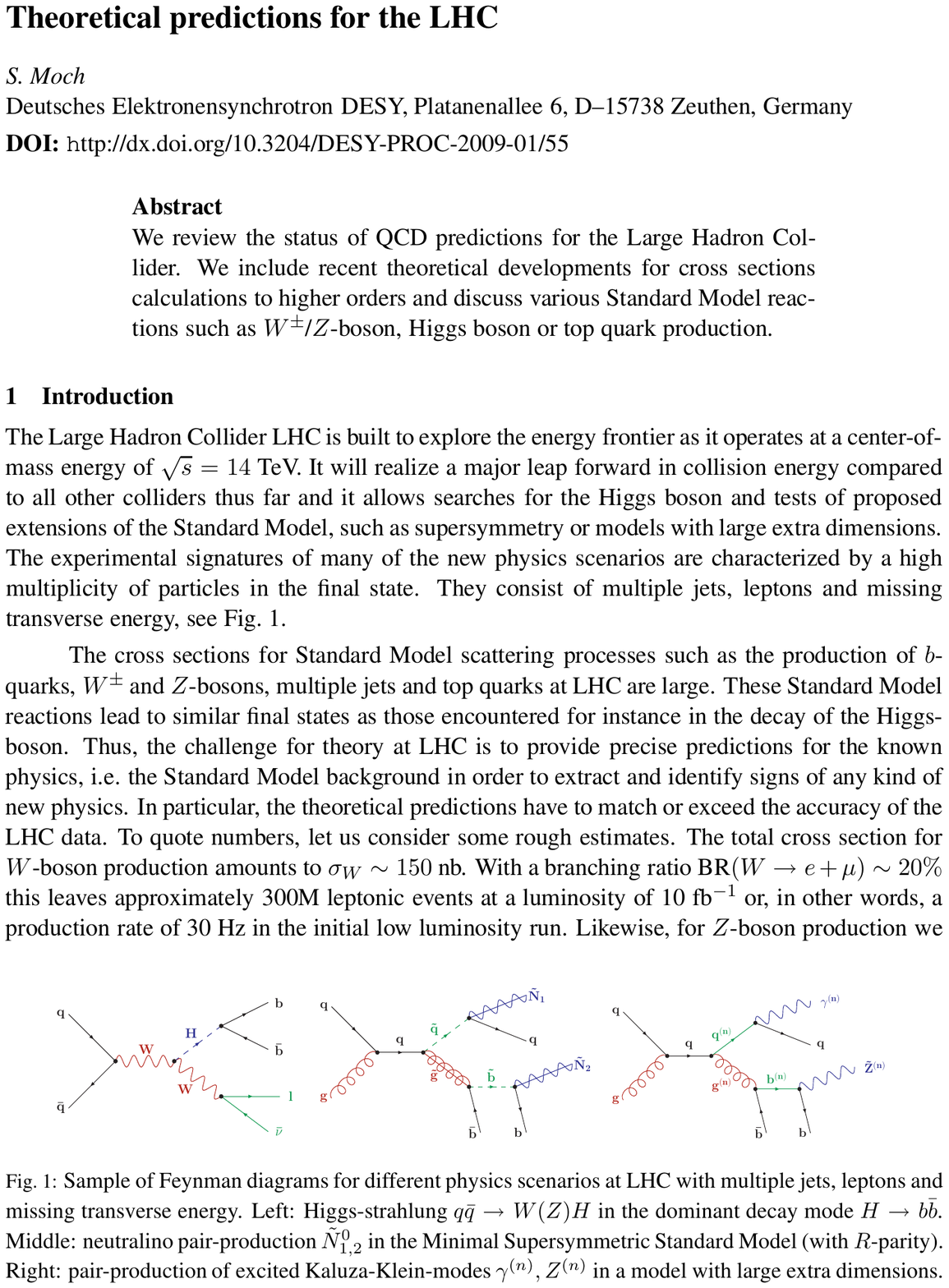} 

\coltoctitle{First physics prospects with  the ATLAS detector at LHC}
\coltocauthor{J. Katzy}
\Includeart{\CAUT}{\CTIT}{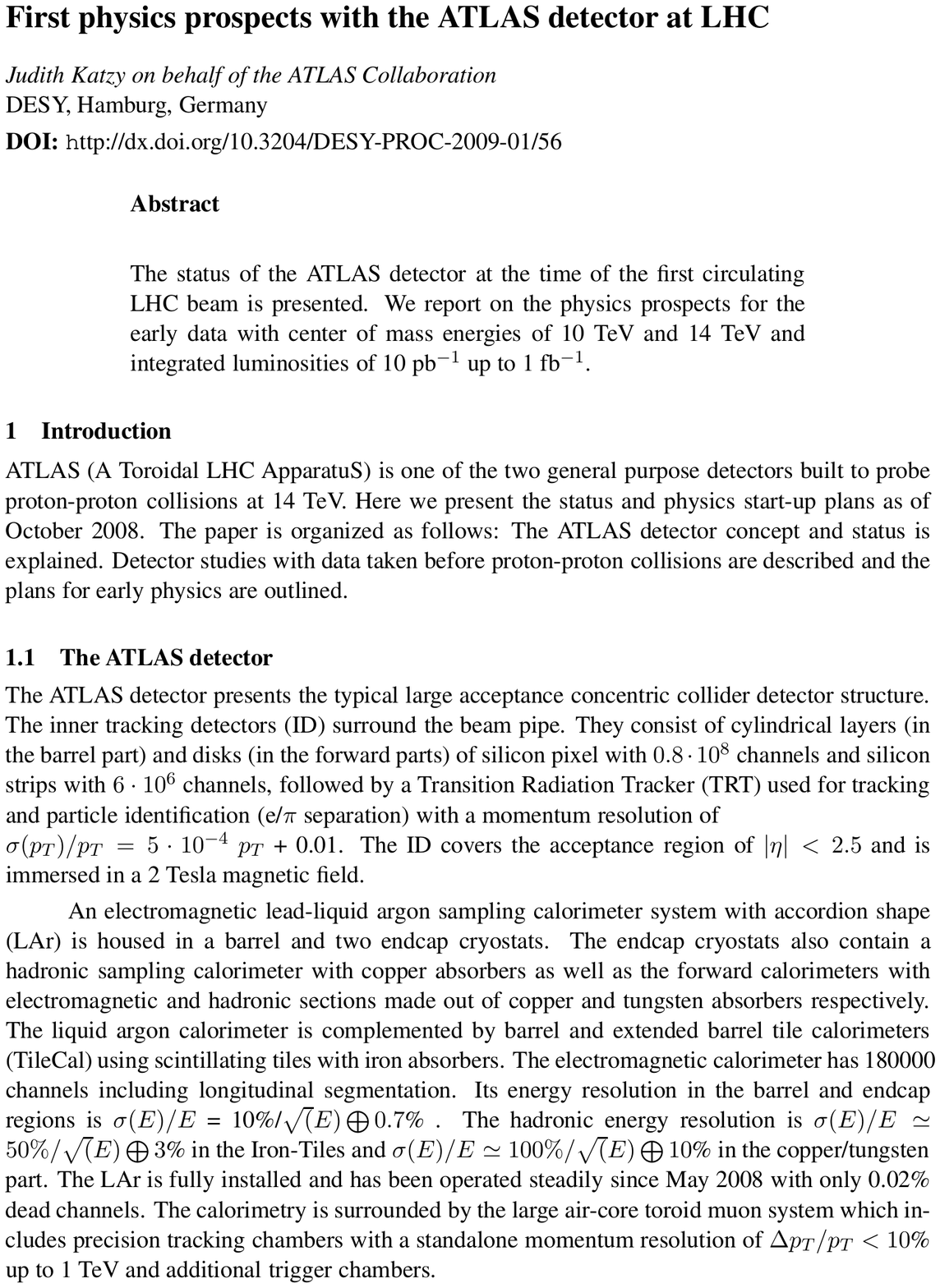} 

\coltoctitle{The Heavy Flavour Content of the Proton}
\coltocauthor{P. Thompson}
\Includeart{\CAUT}{\CTIT}{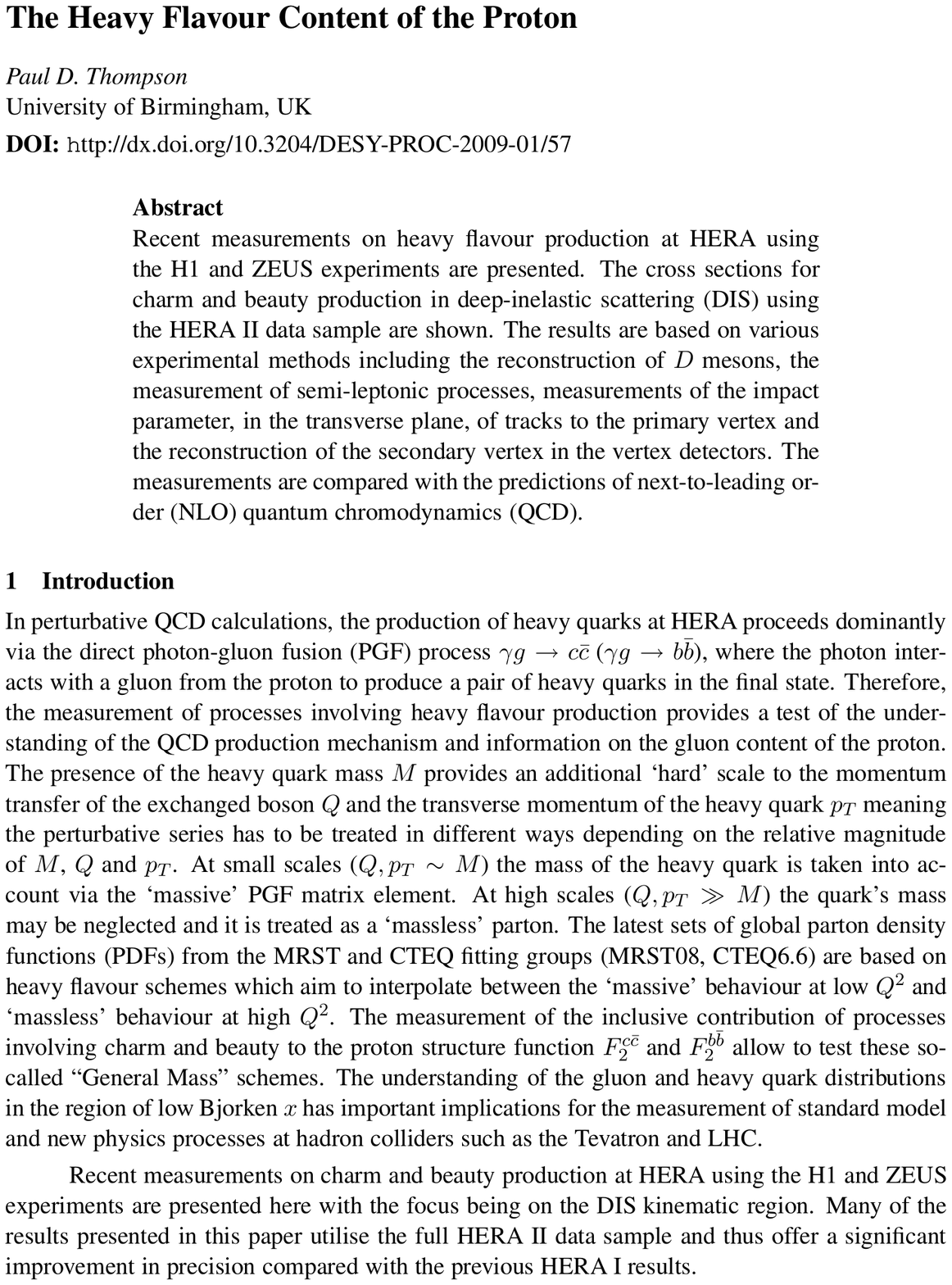} 

\coltoctitle{Update on Neural Network Parton Distributions:  NNPDF1.1}
\coltocauthor{J. Rojo}
\Includeart{\CAUT}{\CTIT}{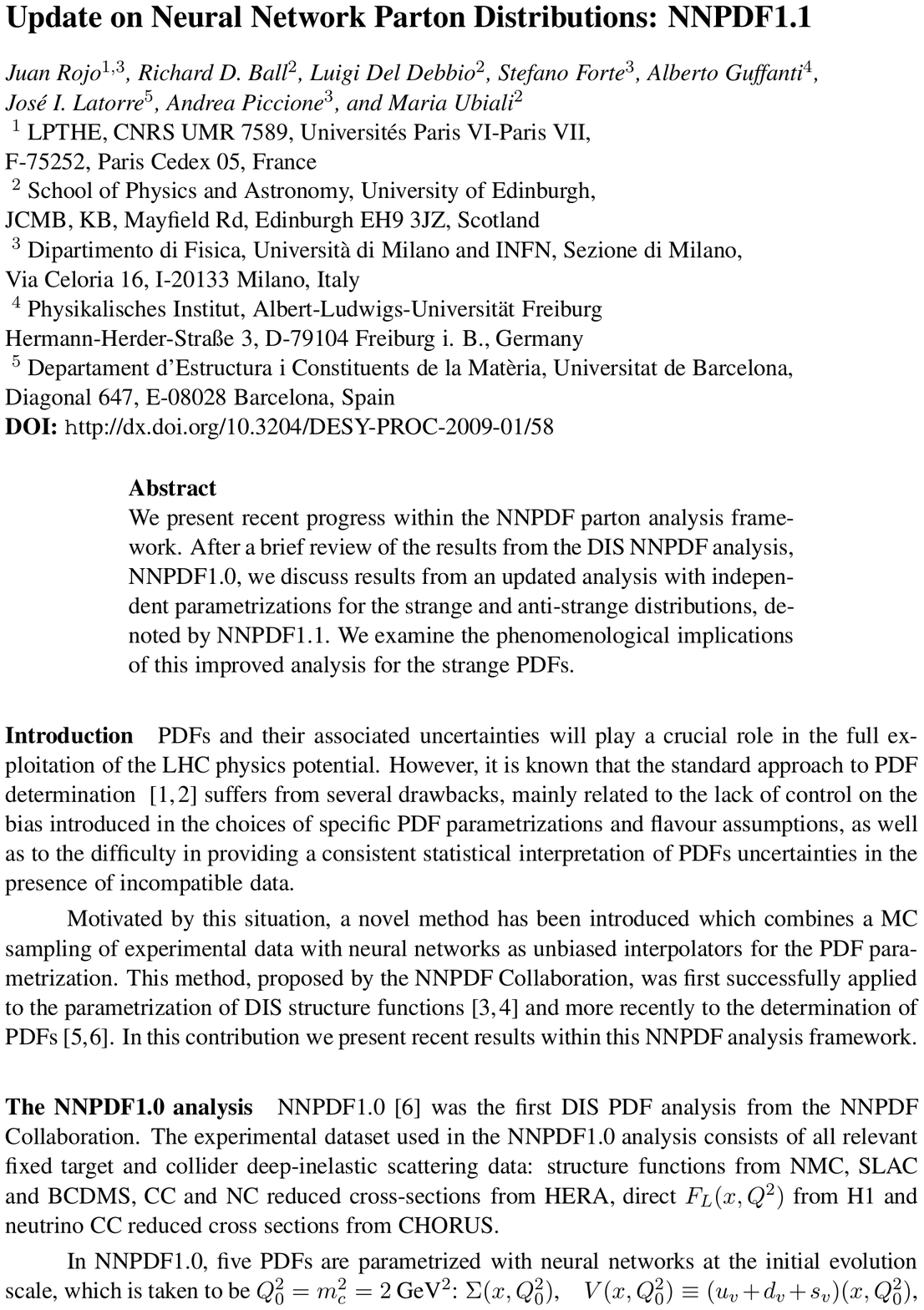}

% \coltoctitle{Second Talk}
% \coltocauthor{David Summerset}
% \Includeart{\CAUT}{\CTIT}{summerset_david}

\end{papers} % end of individual articles/papers

%% file: include-dense.tex
\chapter[\large  WG: Dense systems]{ Working Group \\ Dense Systems}
{\Large\bfseries Convenors:}
\vspace{10mm}

{\Large\itshape 
 D. d'Enterria (CERN) \\
  T. Cs\"org\H{o}  (Budapest)  \\
  E.Iancu (Saclay)}

\CLDP
\begin{papers} % start of individual articles/papers

\coltoctitle{High-energy heavy-ion collisions: from CGC to Glasma}
\coltocauthor{ K. Itakura}
\Includeart{\CAUT}{\CTIT}{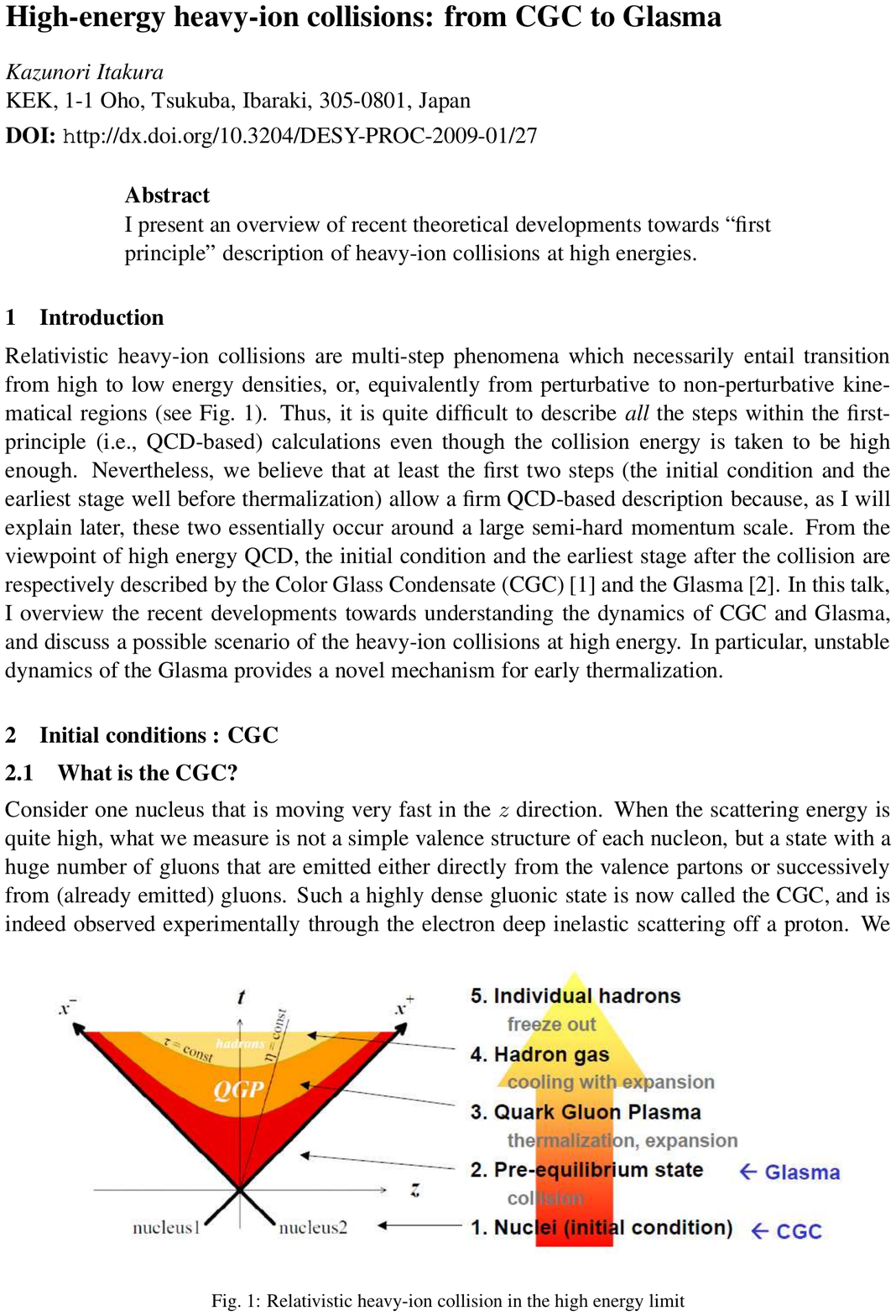} 

\coltoctitle{Particle production and saturation at RHIC and LHC}
\coltocauthor{ C. Marquet}
\Includeart{\CAUT}{\CTIT}{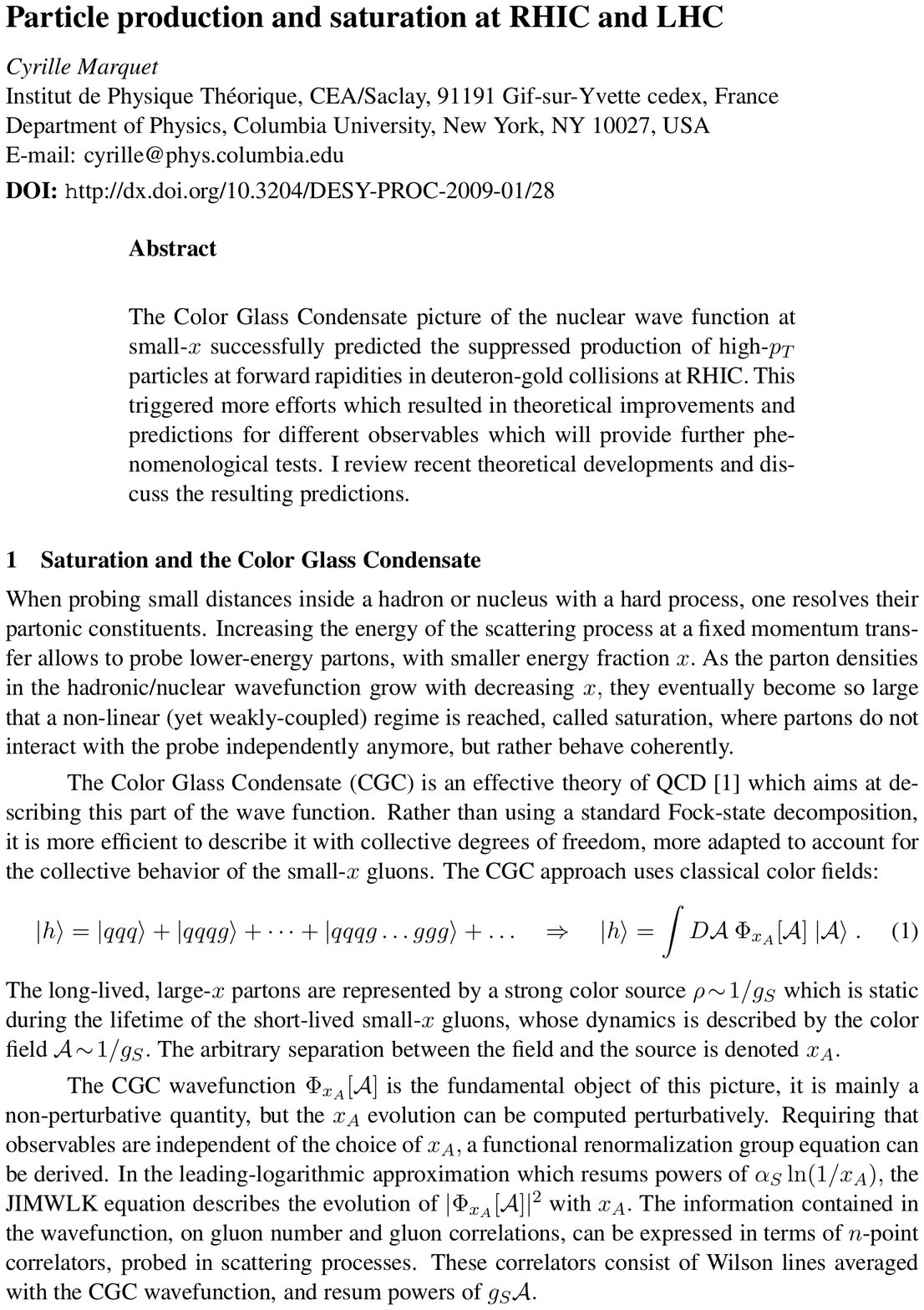} 

\coltoctitle{Introducing Saturation Effects into Event Generators}
\coltocauthor{E. Avsar}
\Includeart{\CAUT}{\CTIT}{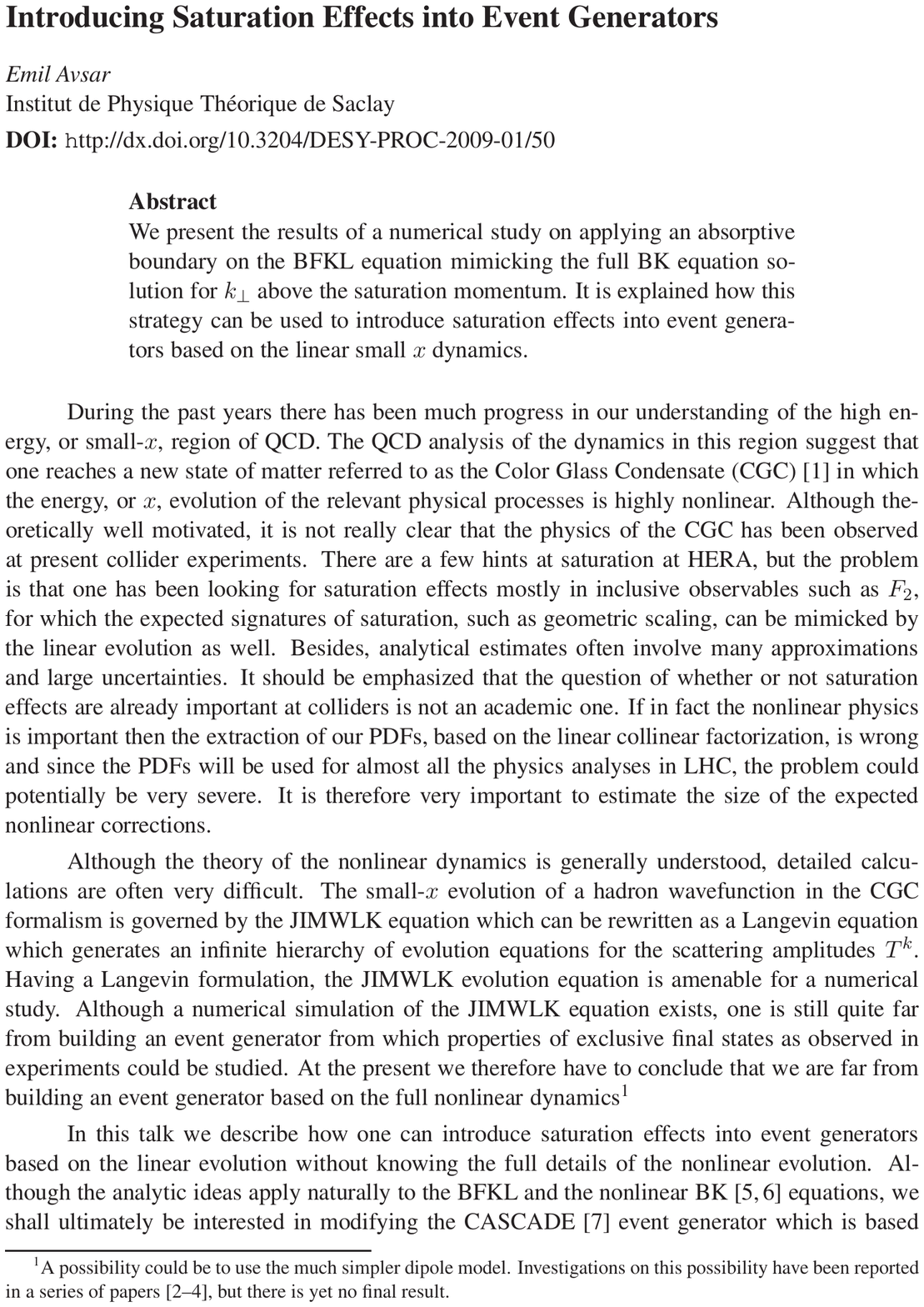} 

\coltoctitle{The Ridge, the Glasma and Flow}
\coltocauthor{L. McLerran}
\Includeart{\CAUT}{\CTIT}{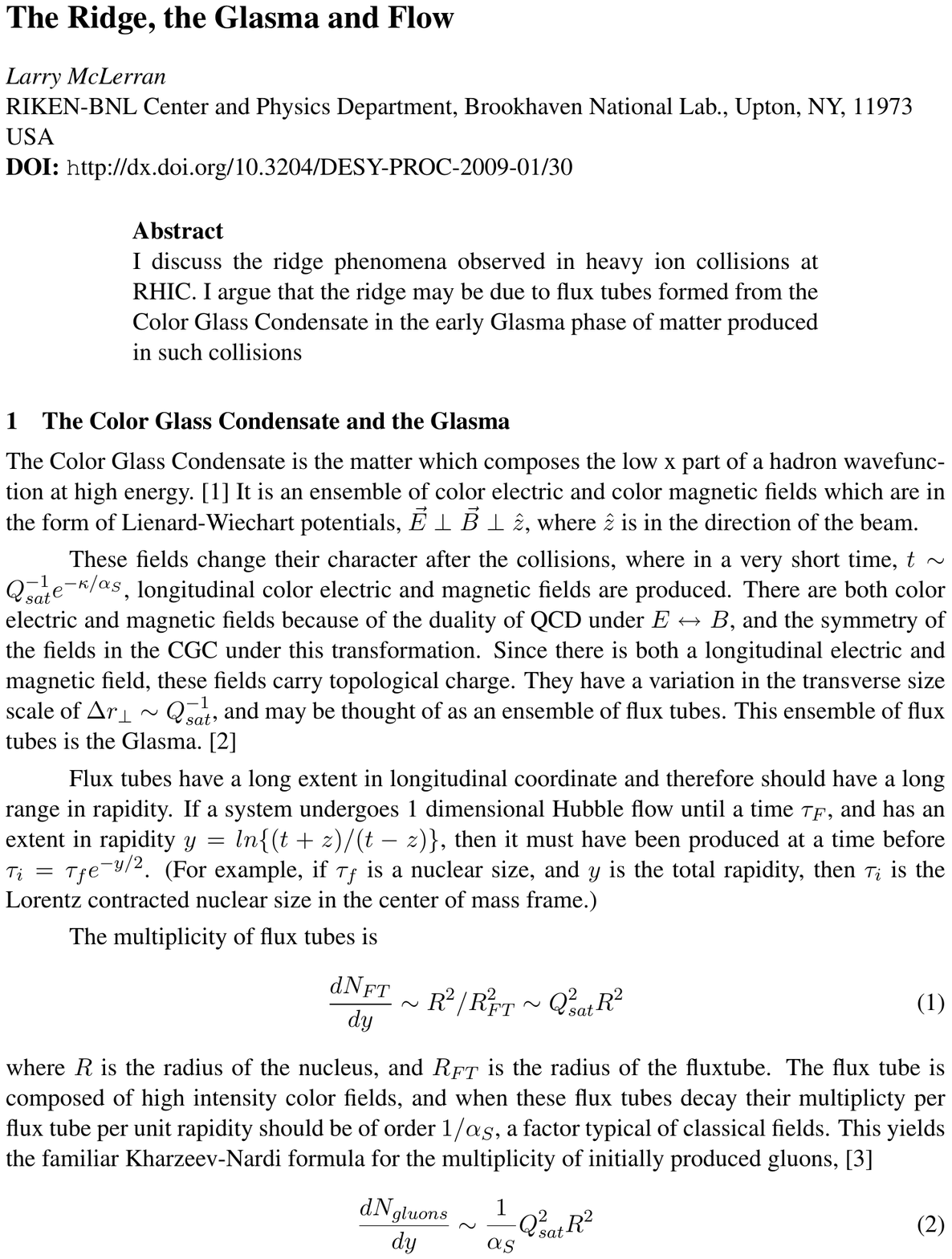} 

%\coltoctitle{}
%\coltocauthor{H. B|"usching}
%\Includeart{\CAUT}{\CTIT}{../itakura/} 

\coltoctitle{Partons and jets at strong coupling from AdS/CFT}
\coltocauthor{E. Iancu}
\Includeart{\CAUT}{\CTIT}{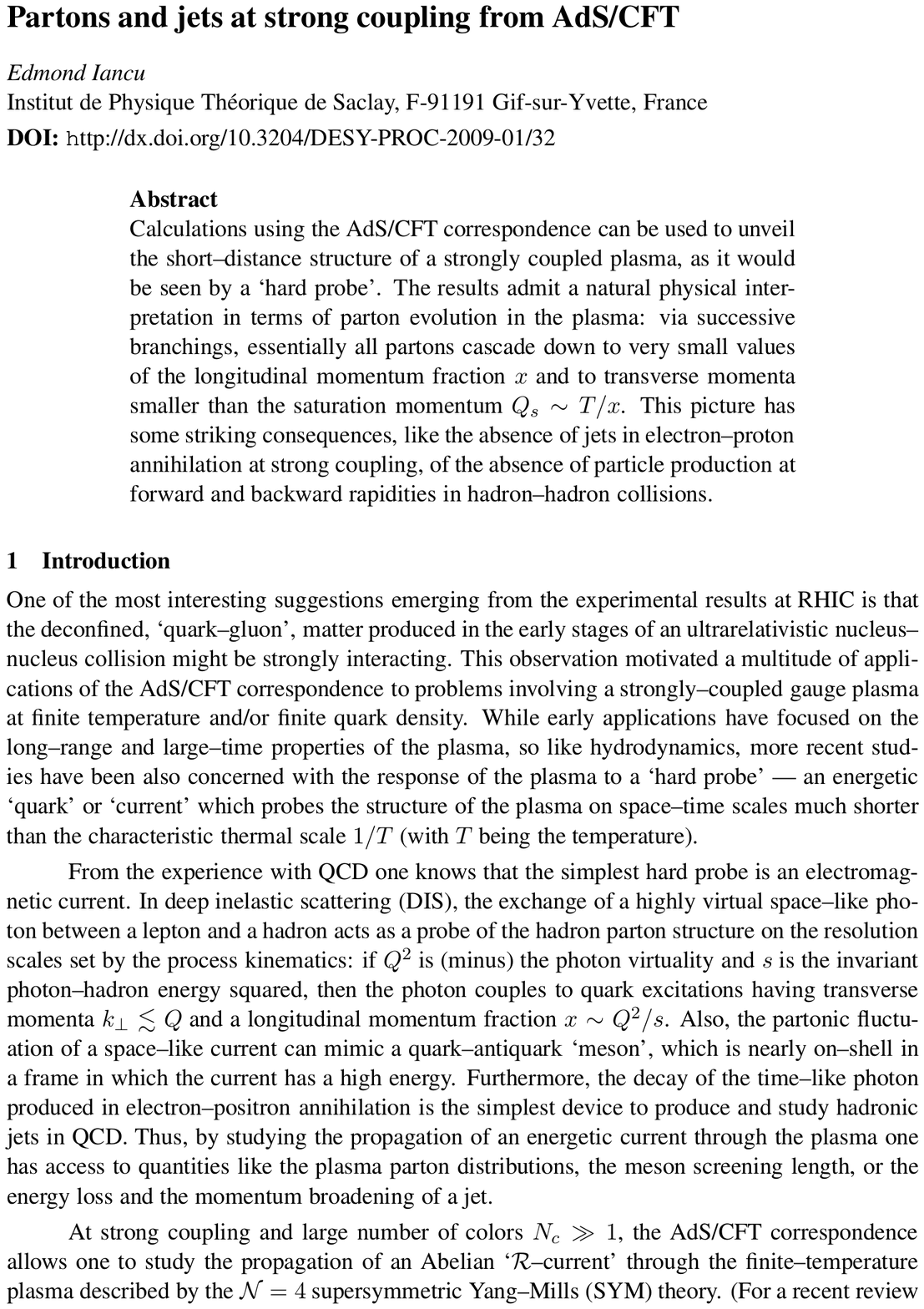} 

\coltoctitle{QCD EoS, initial conditions and final state from relativistic hydrodynamics in heavy-ion collisions}
\coltocauthor{M. Nagy}
\Includeart{\CAUT}{\CTIT}{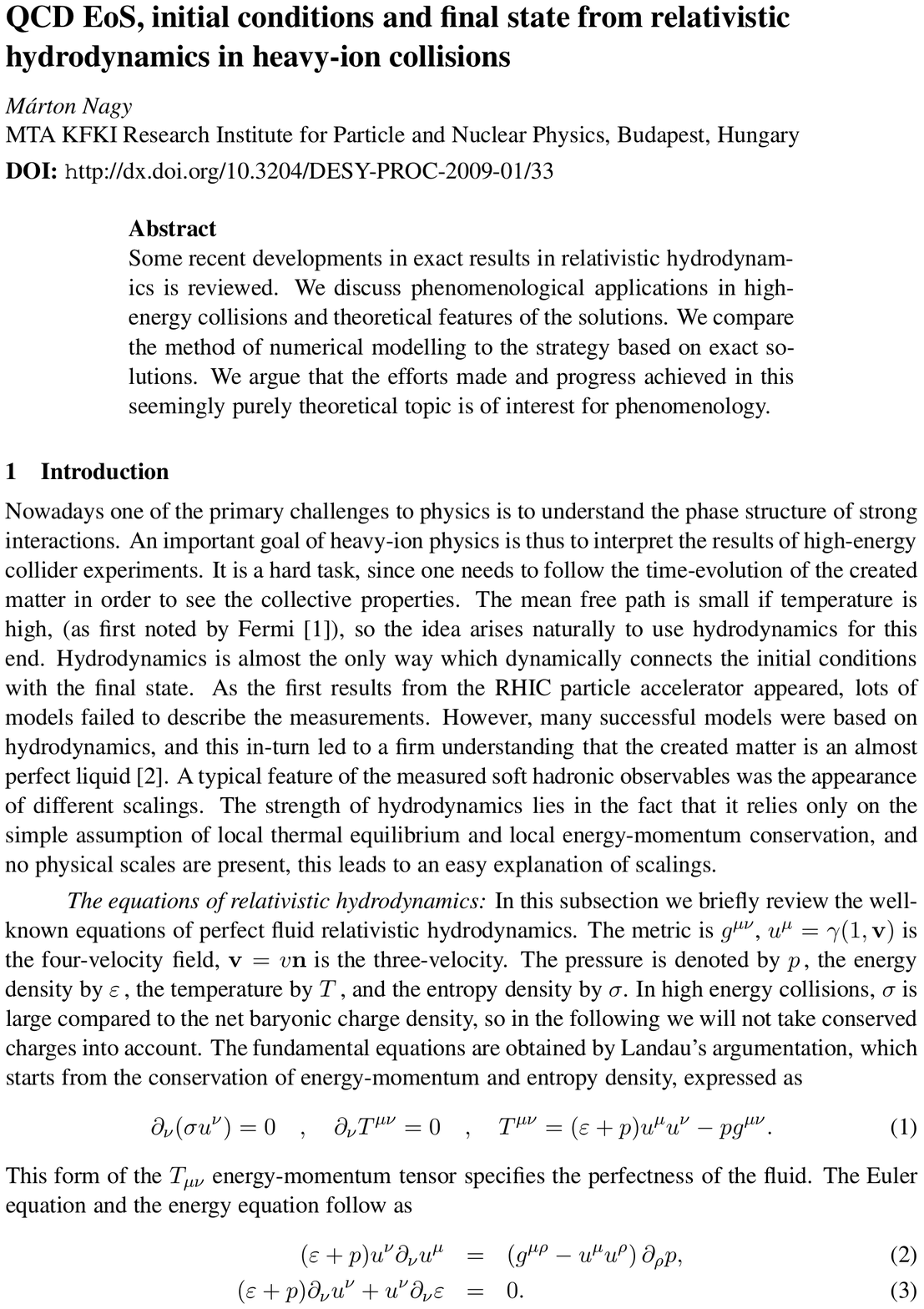} 

\coltoctitle{Hydrodynamics \& perfect fluids: uniform description of soft
 observables 
in Au+Au collisions at RHIC}
\coltocauthor{W. Florkowski, 
W Broniowski, M. Chojnacki, A. Kisiel}
\Includeart{\CAUT}{\CTIT}{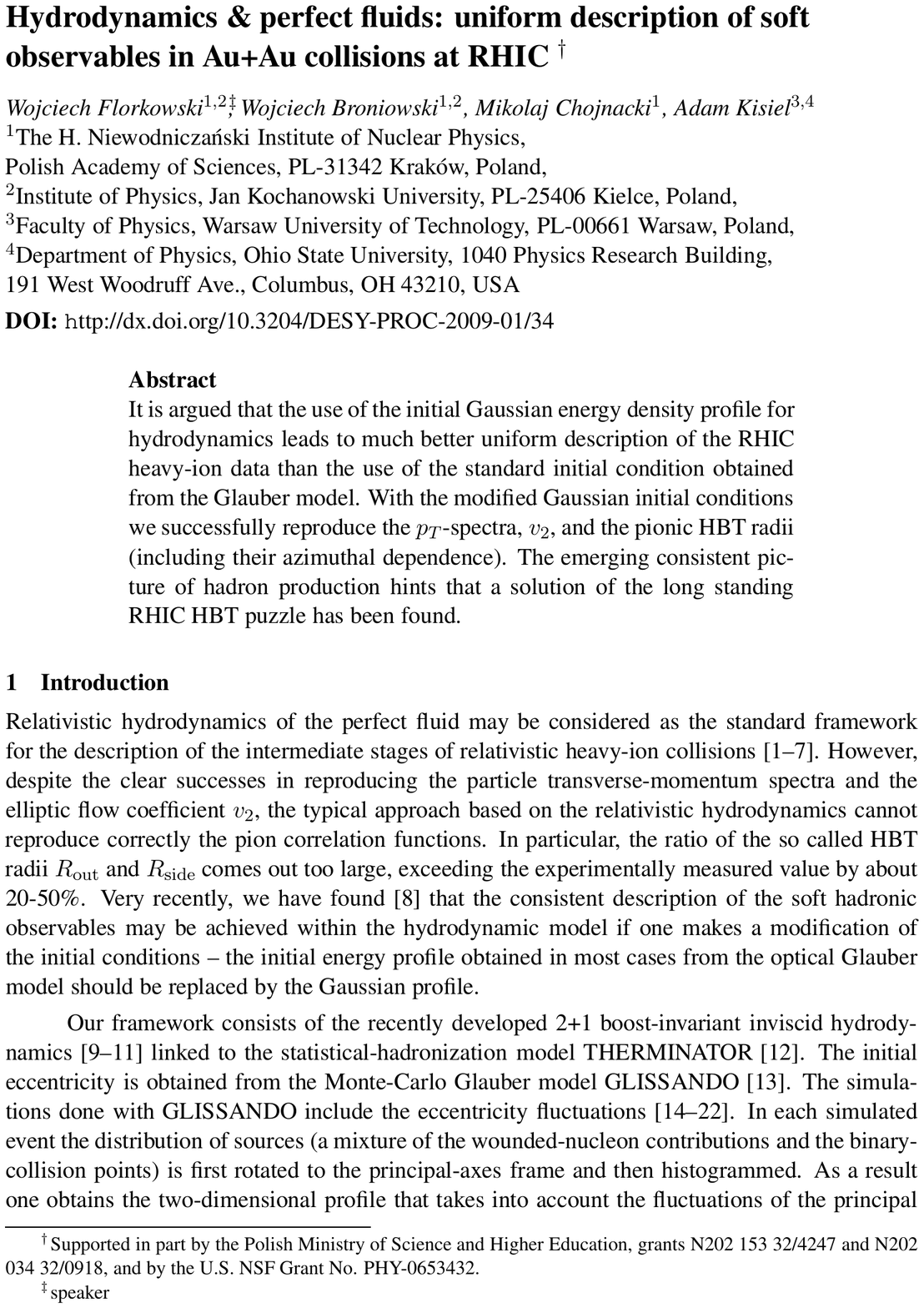} 

\coltoctitle{The Remarkable Simplicity and Universality of Multiparticle 
Production Data}
\coltocauthor{W. Busza}
\Includeart{\CAUT}{\CTIT}{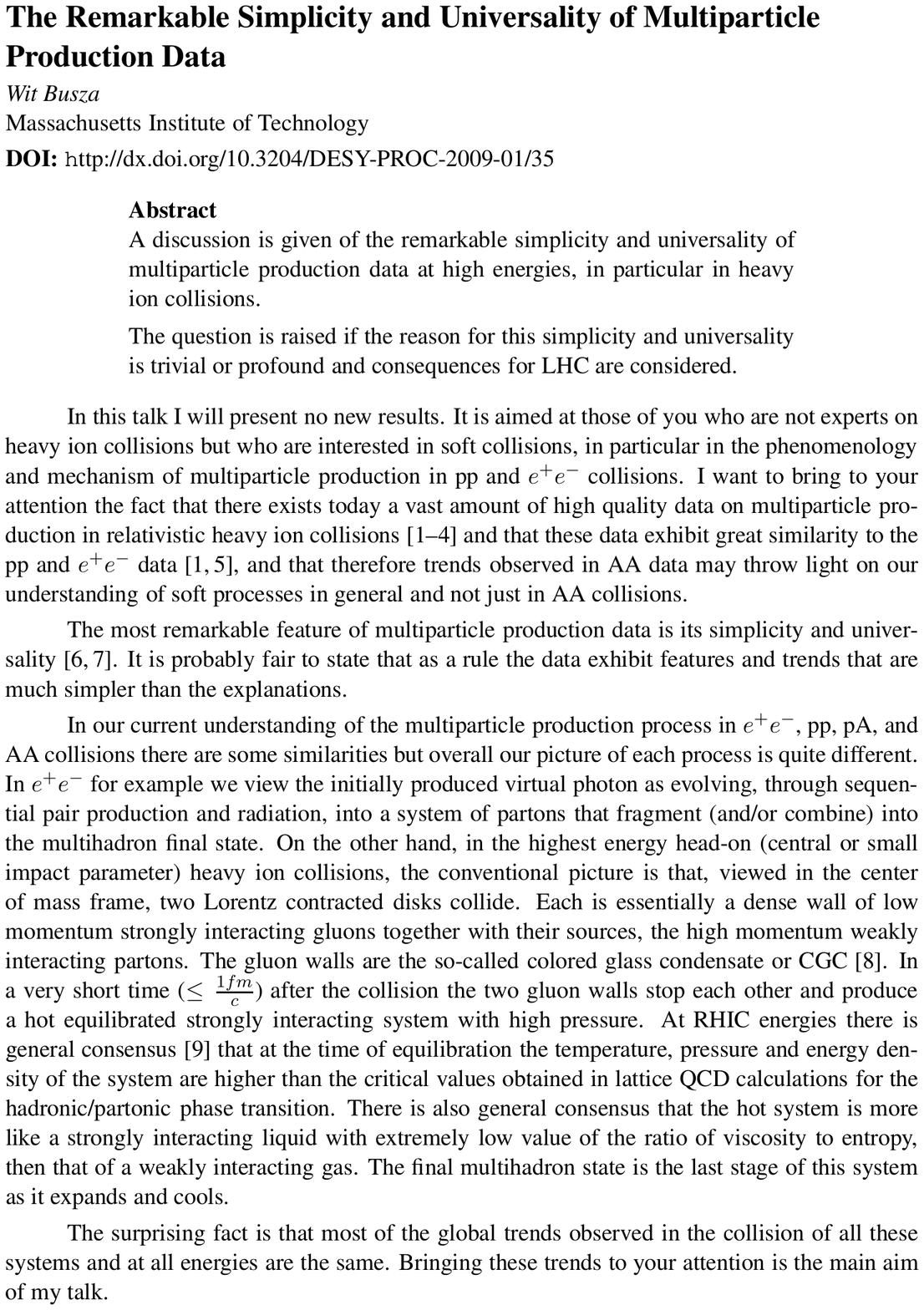} 

\coltoctitle{Heavy ions and parton saturation from RHIC to LHC}
\coltocauthor{A.~Dainese}
\Includeart{\CAUT}{\CTIT}{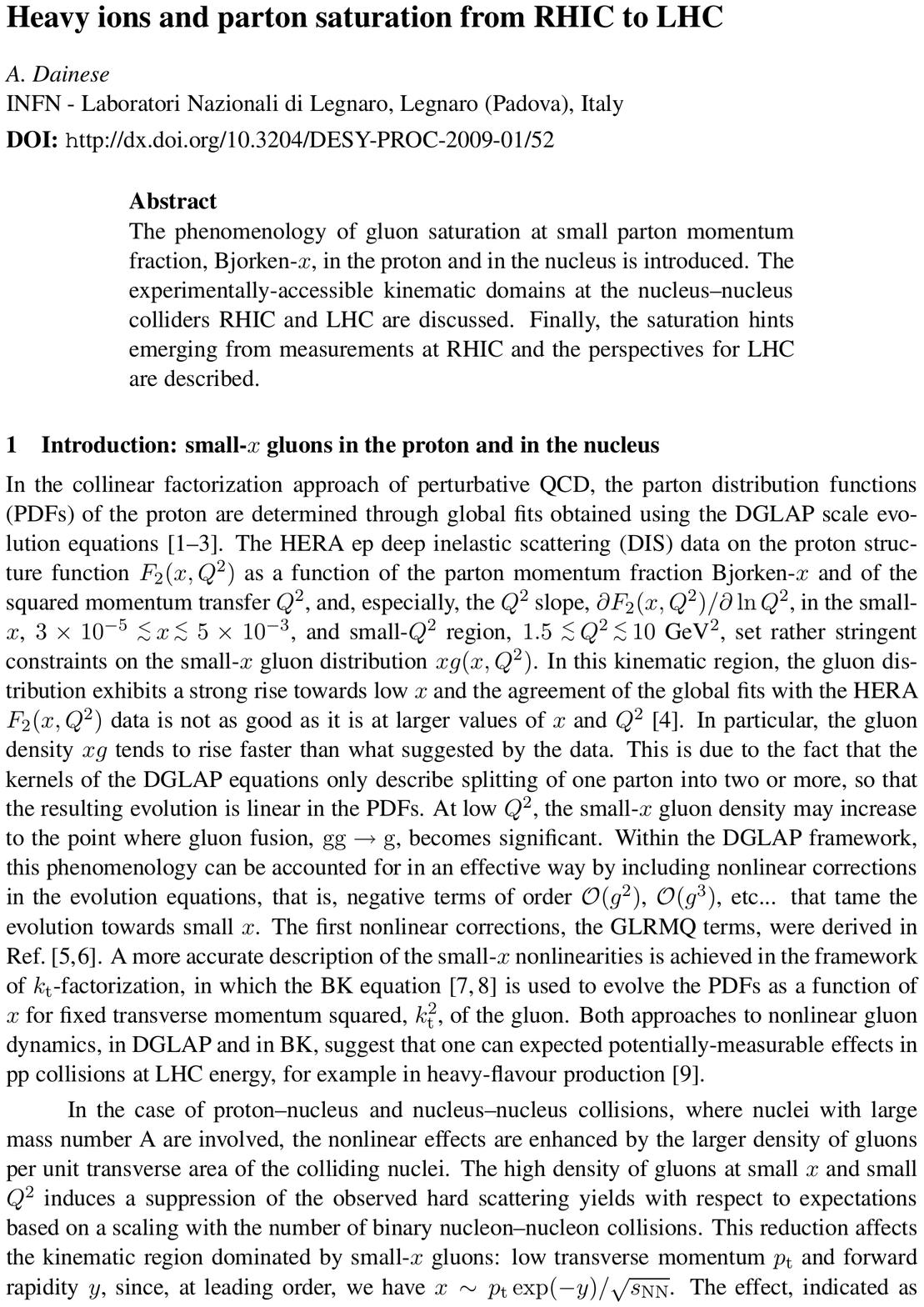} 

\coltoctitle{Probing the Properties of the Matter Created at RHIC}
\coltocauthor{H. Caines}
\Includeart{\CAUT}{\CTIT}{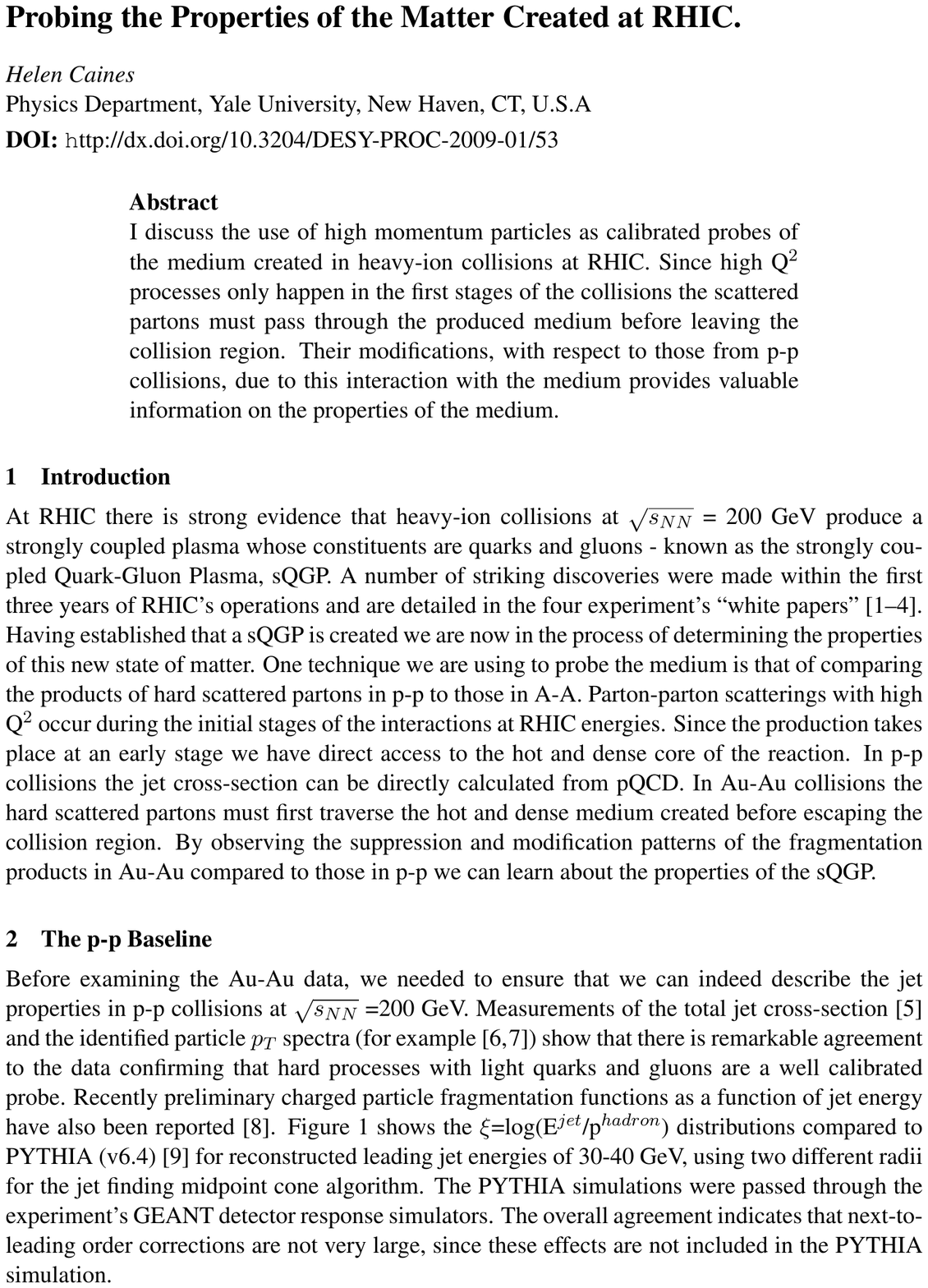} 

\coltoctitle{Relativistic viscous hydrodynamics and  
AdS/CFT correspondence at finite temperature }
\coltocauthor{R. Baier}
\Includeart{\CAUT}{\CTIT}{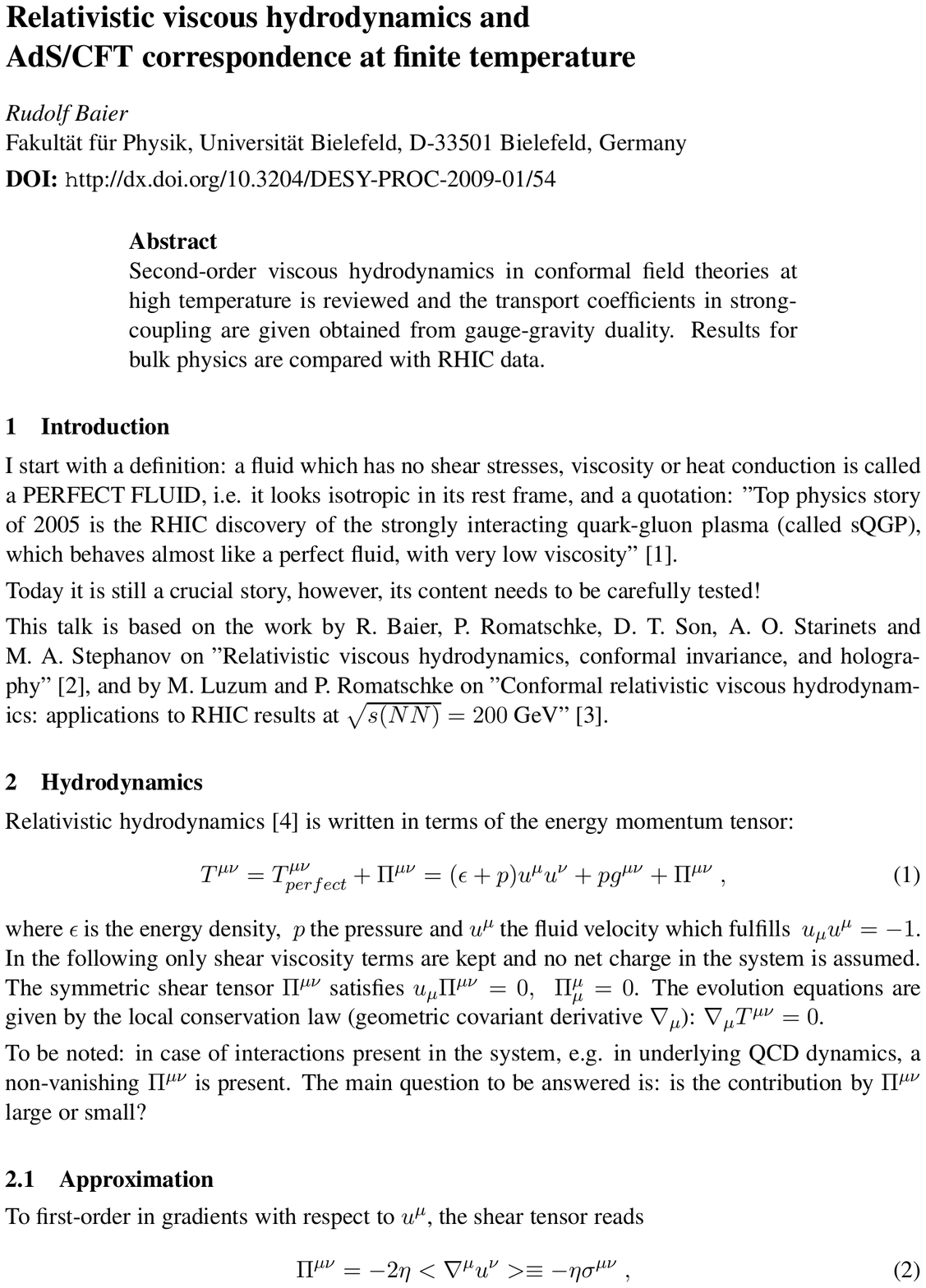} 

\coltoctitle{AdS/CFT Correspondence in Heavy Ion Collisions}
\coltocauthor{J. L.\ Albacete, Y. V.\  Kovchegov, A. Taliotis}
\Includeart{\CAUT}{\CTIT}{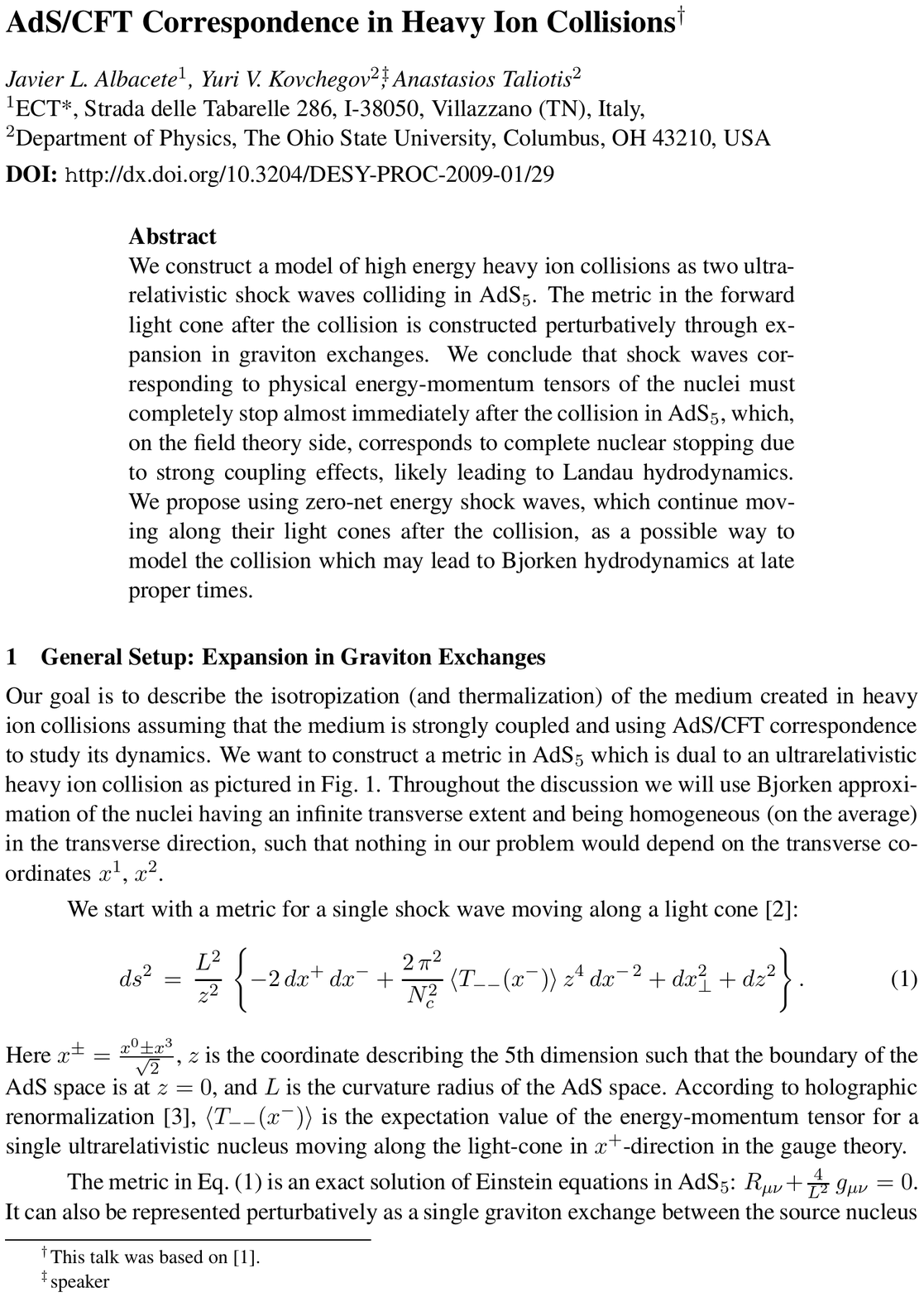} 

\coltoctitle{Saturation and Confinement: Analyticity, Unitarity and AdS/CFT Correspondence}
\coltocauthor{R. Brower, M. Djuric, C-I Tan}
\Includeart{\CAUT}{\CTIT}{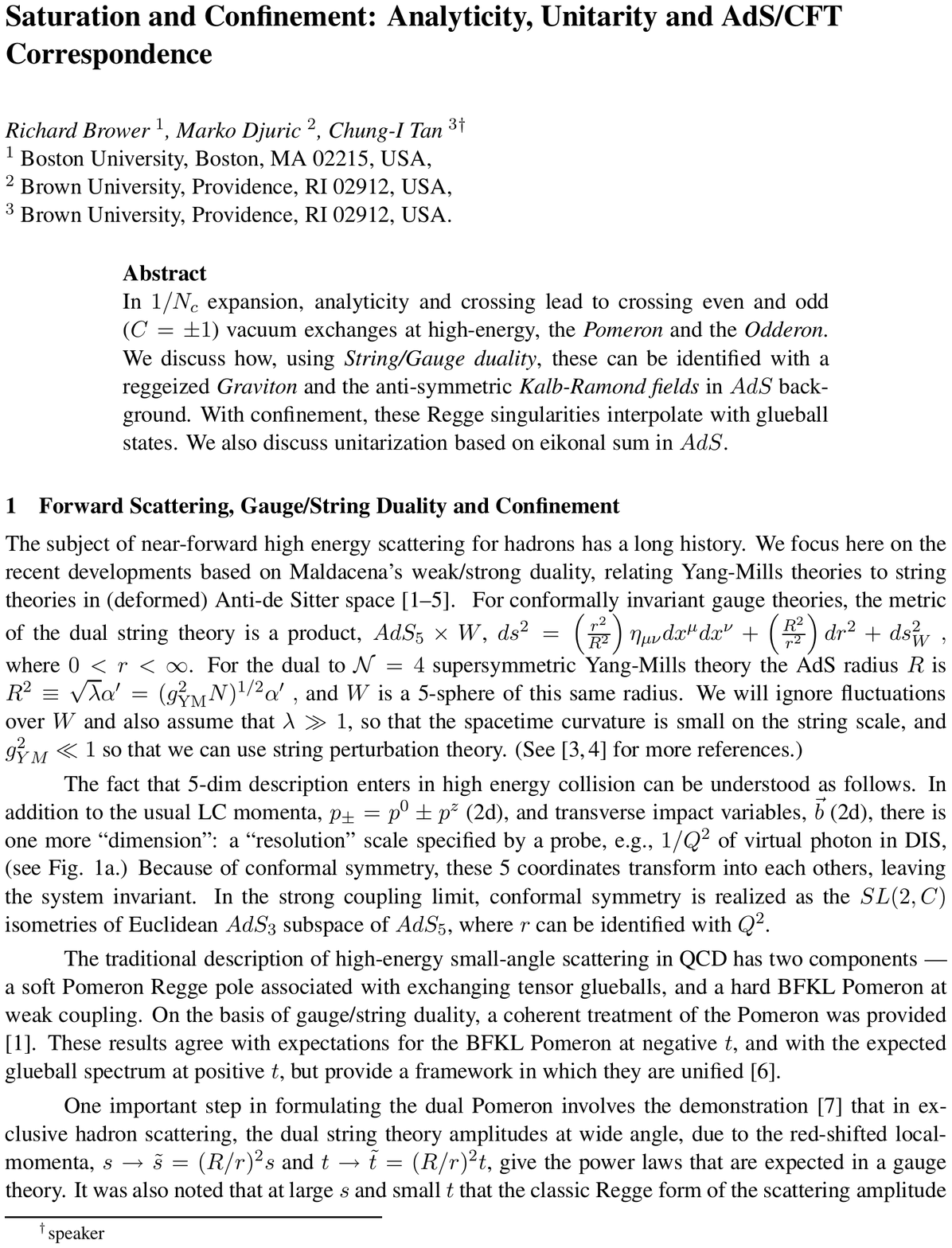} 

\coltoctitle{Light, Strange and Charm Hadron Measurements in  \boldmath{$ep$}}
\coltocauthor{A. Kropivnitskaya}
\Includeart{\CAUT}{\CTIT}{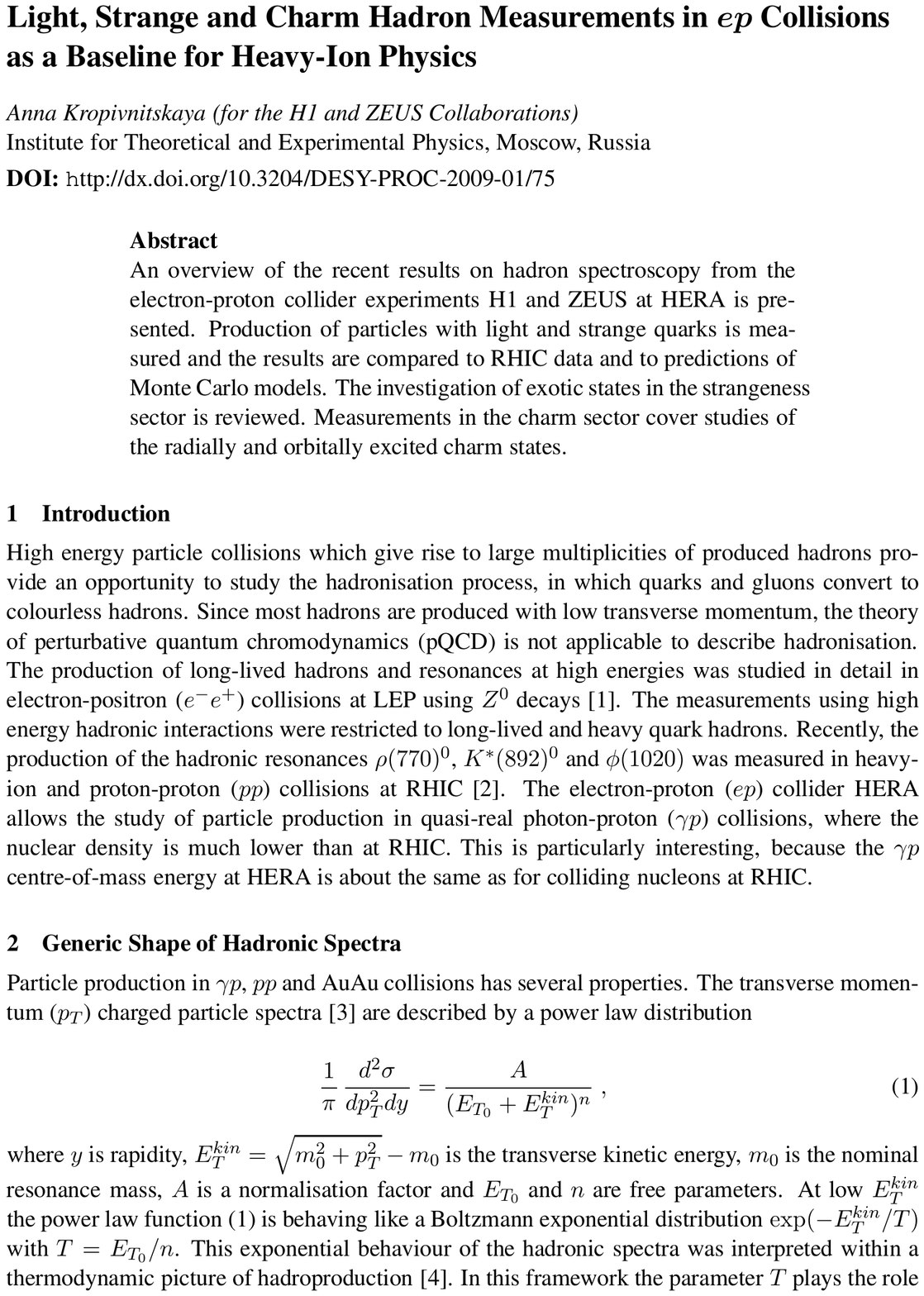}

% \coltoctitle{Second Talk}
% \coltocauthor{David Summerset}
% \Includeart{\CAUT}{\CTIT}{summerset_david}

\end{papers} % end of individual articles/papers

%% file: include-interpolation.tex
\chapter[\large  WG: Interpolation Region]{ Working Group \\ Interpolation Region}
{\Large\bfseries Convenors:}
\vspace{10mm}

{\Large\itshape
M. Grothe (University Wisconsin, Madison) \\
M. Albrow (FNAL) \\
K. Werner (Nantes)}

\CLDP
\begin{papers} % start of individual articles/papers

\coltoctitle{Diffractive Production of Jets and
  Vector Bosons at the Tevatron}
\coltocauthor{K.  Hatakeyama}
\Includeart{\CAUT}{\CTIT}{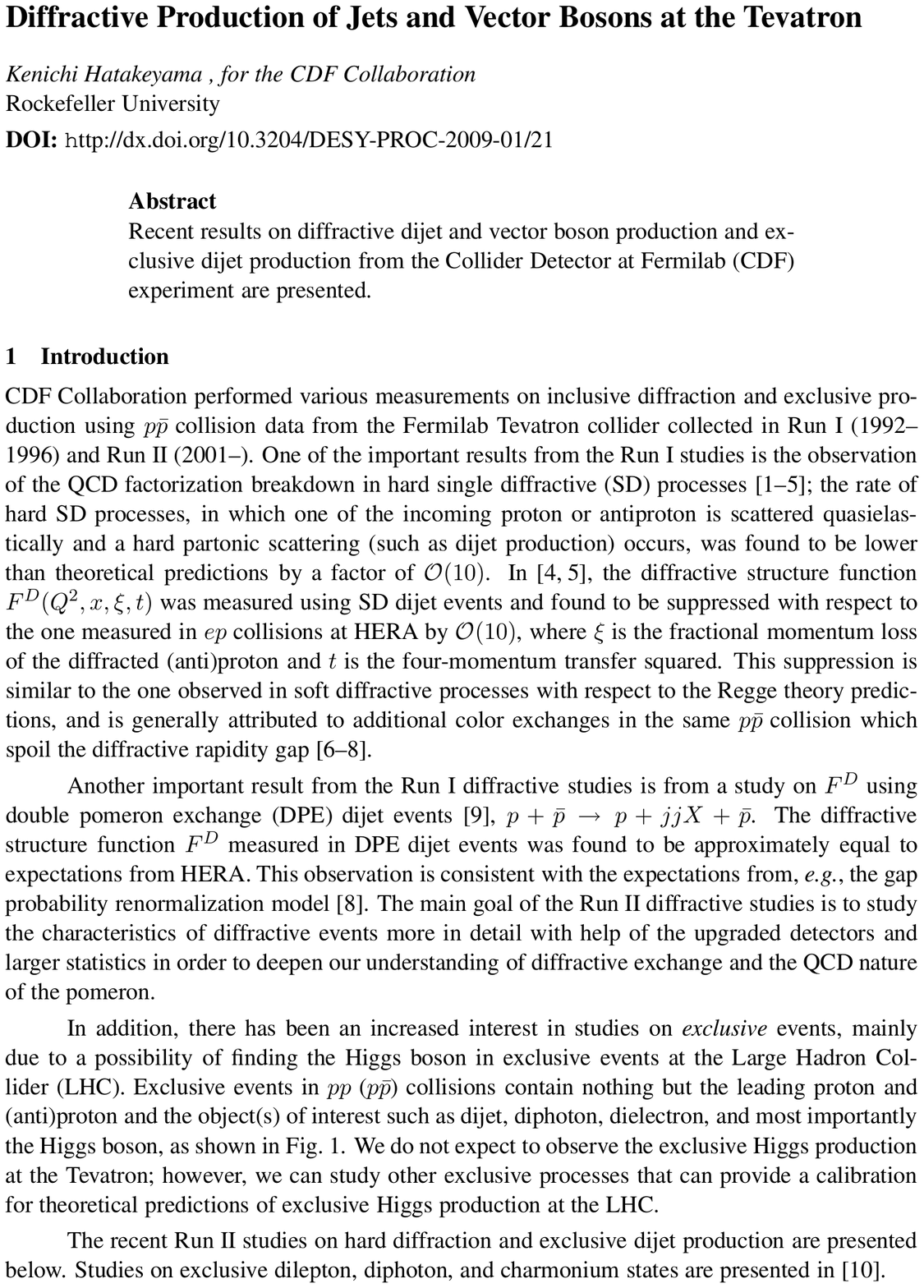} 

\coltoctitle{Central Exclusive Production at the Tevatron}
\coltocauthor{M. G. Albrow}
\Includeart{\CAUT}{\CTIT}{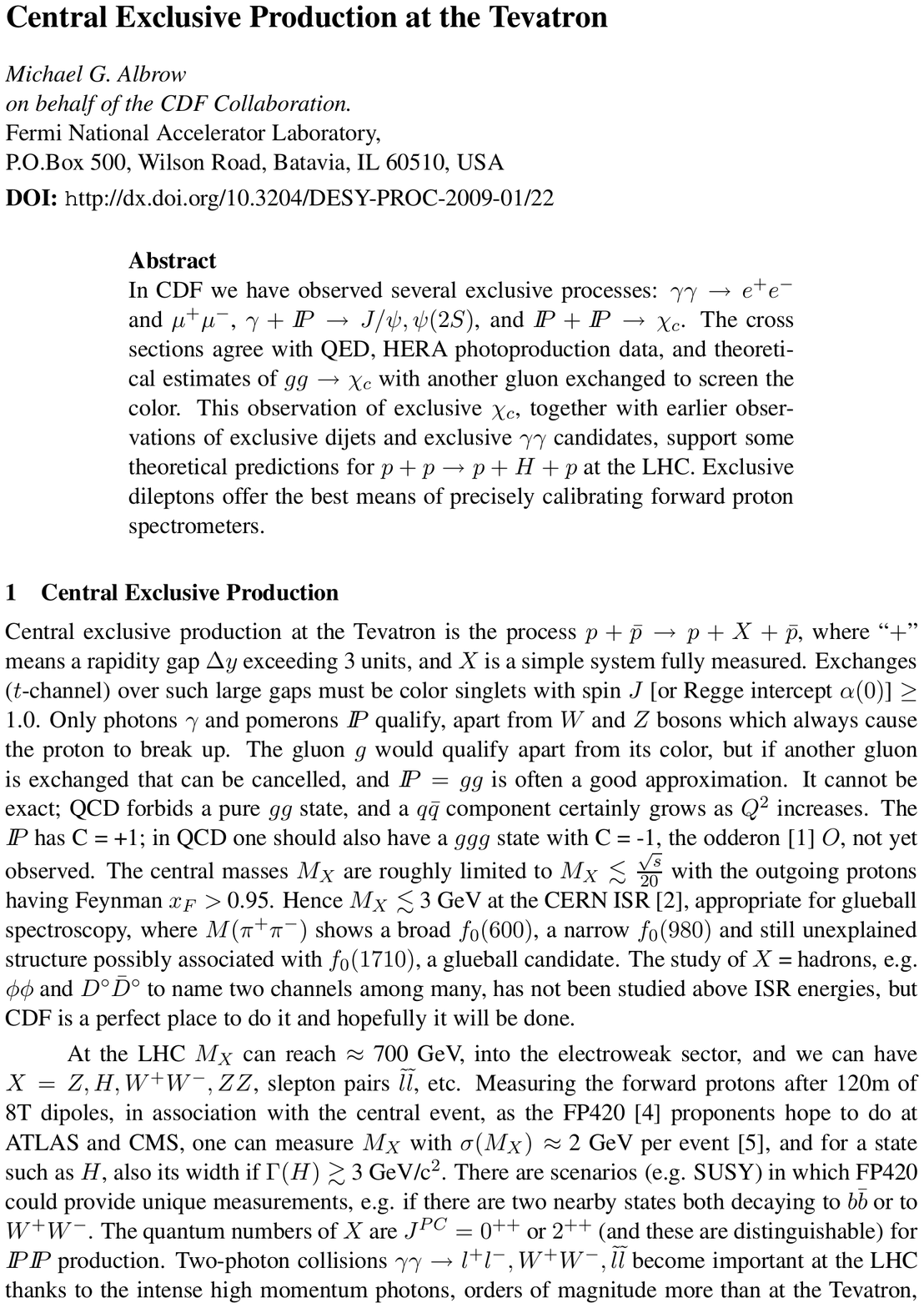} 

\coltoctitle{Overview on rapidity gap survival predictions for LHC}
\coltocauthor{A.B. Kaidalov}
\Includeart{\CAUT}{\CTIT}{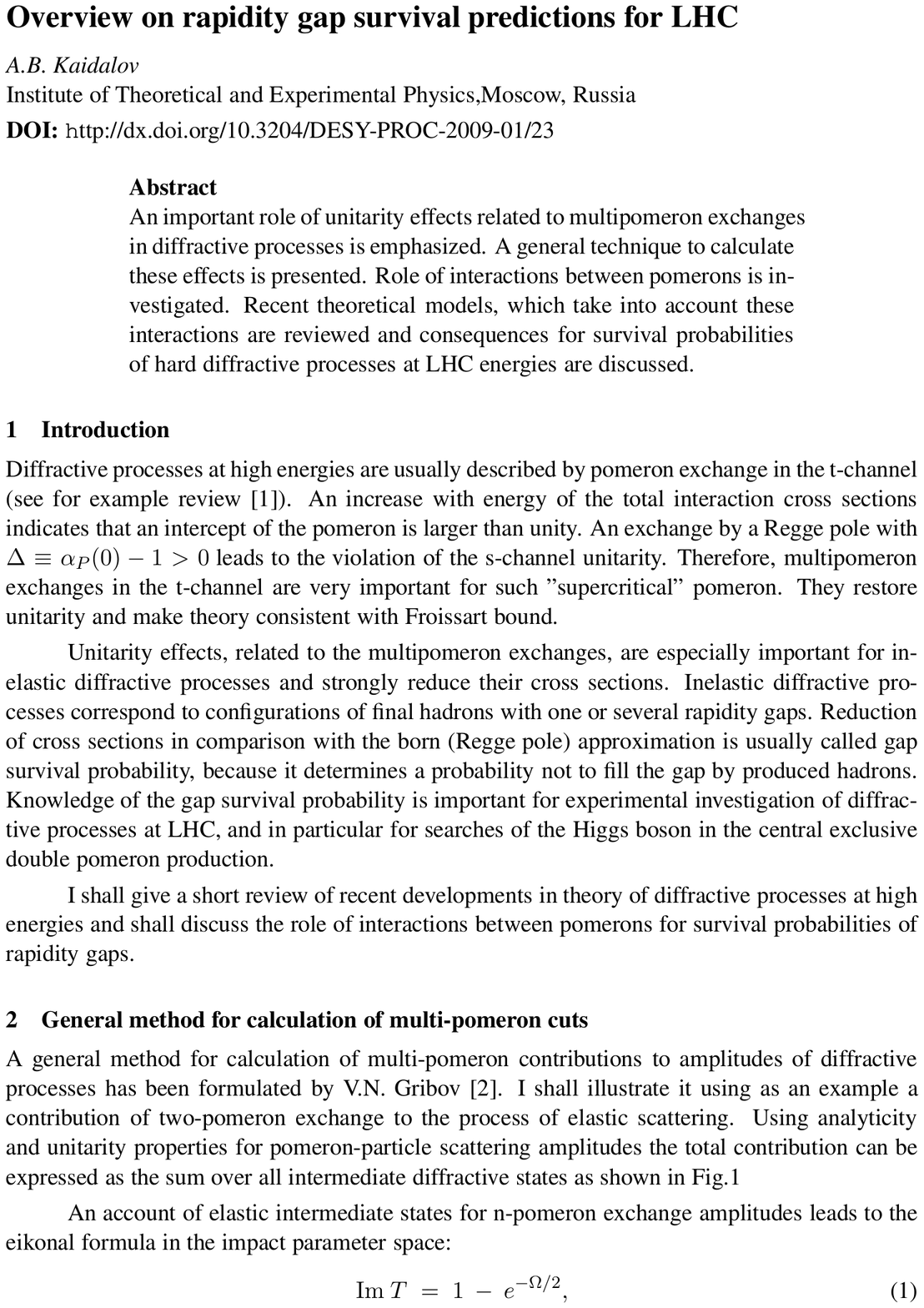} 

\coltoctitle{Access to small x Parton Density Functions at the LHC}
\coltocauthor{T. Shears}
\Includeart{\CAUT}{\CTIT}{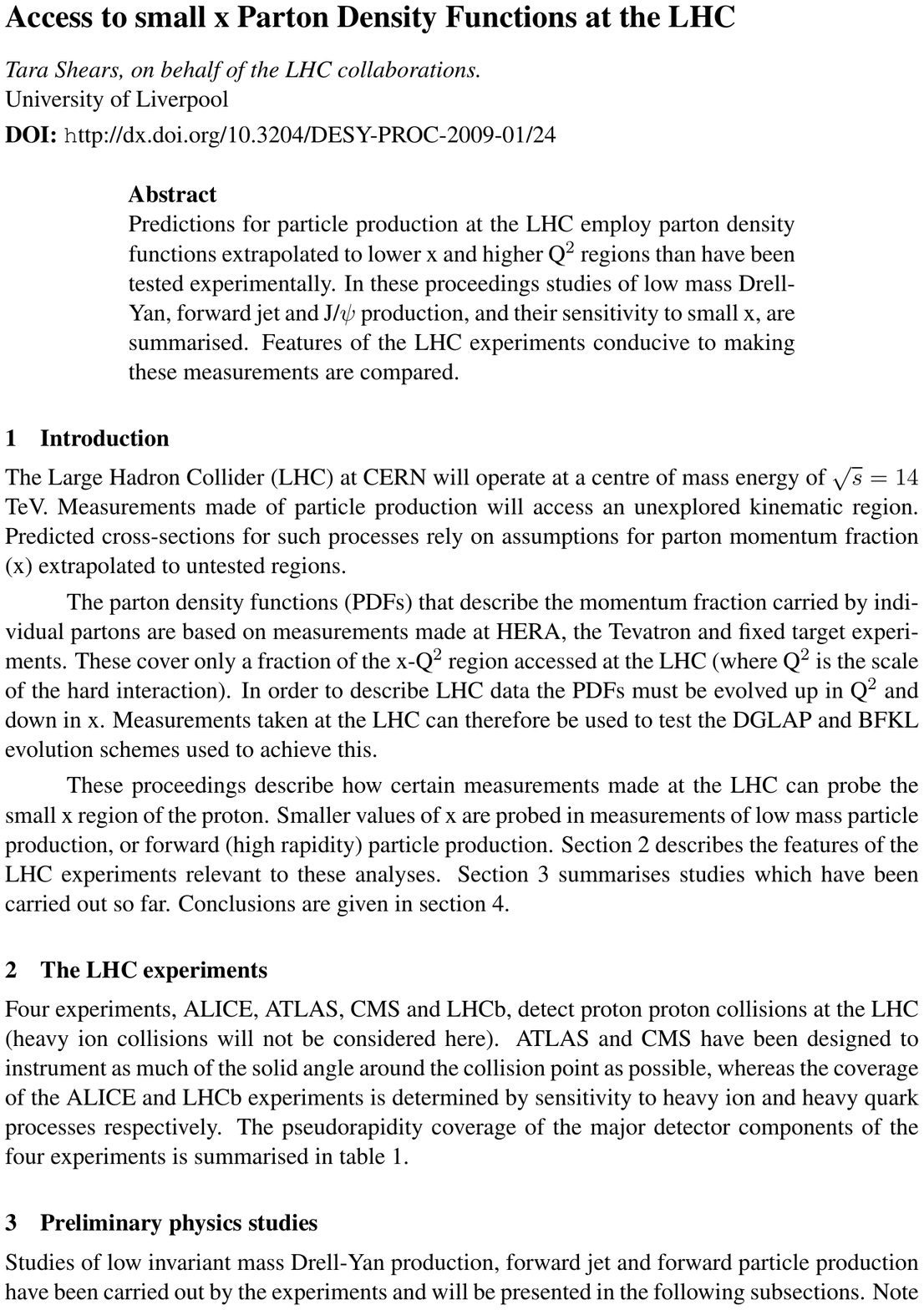} 

\coltoctitle{Theoretical concepts of parton saturation - from HERA to LHC}
\coltocauthor{K. Golec-Biernat}
\Includeart{\CAUT}{\CTIT}{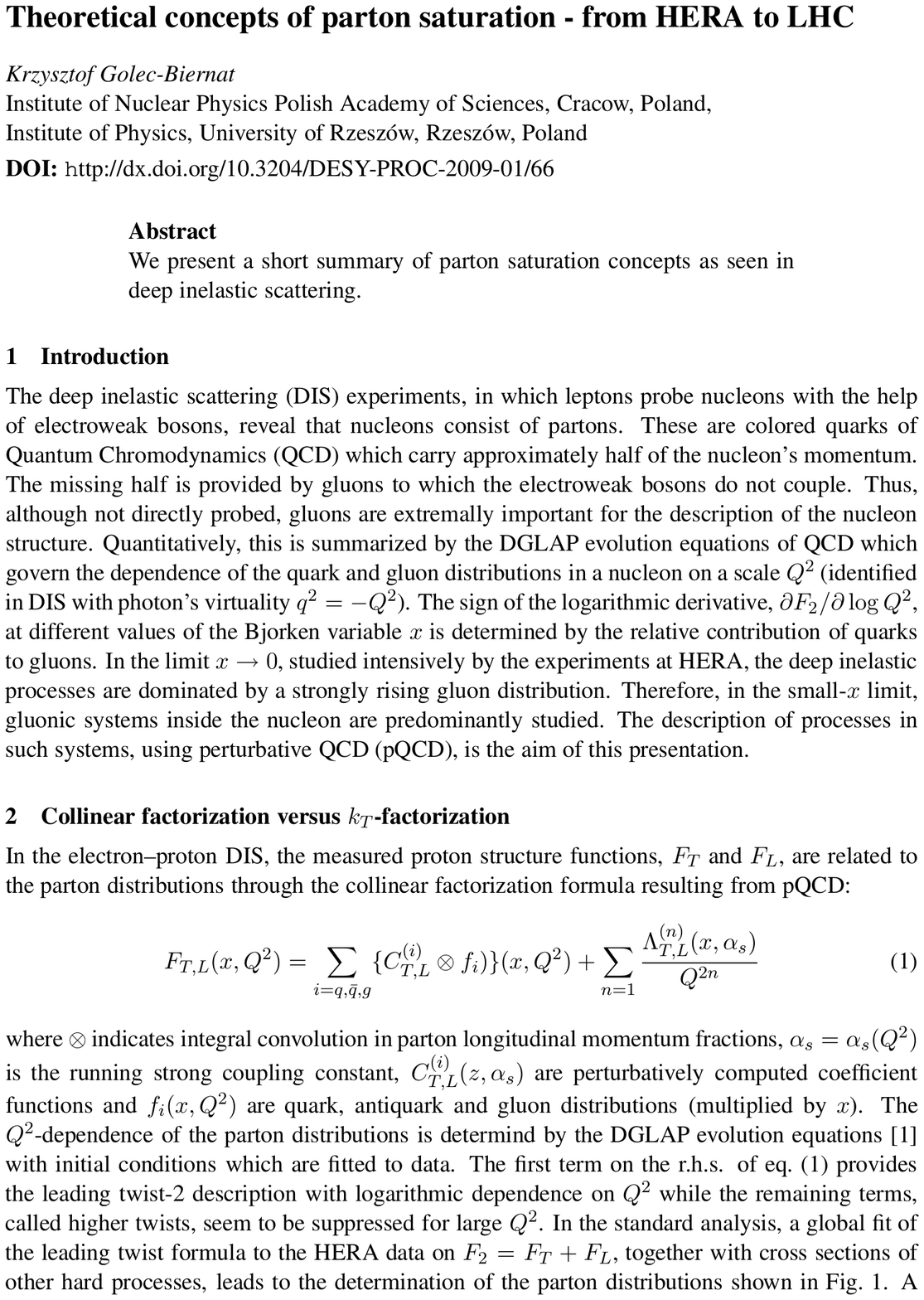} 

\coltoctitle{What HERA can tells us about saturation}
\coltocauthor{R. Yoshida}
\Includeart{\CAUT}{\CTIT}{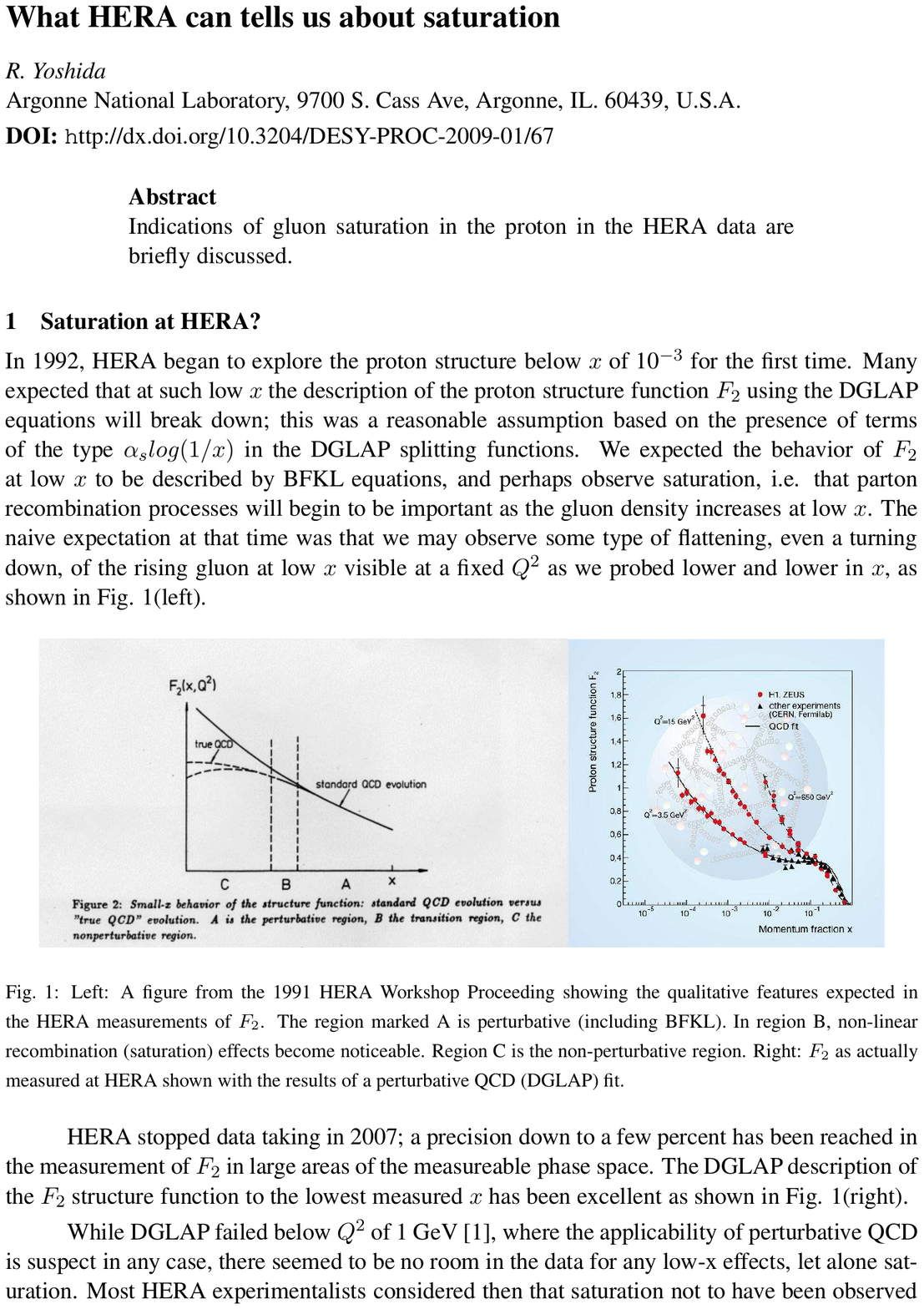} 

\coltoctitle{Inclusive diffraction and factorisation at HERA}
\coltocauthor{M. Wing}
\Includeart{\CAUT}{\CTIT}{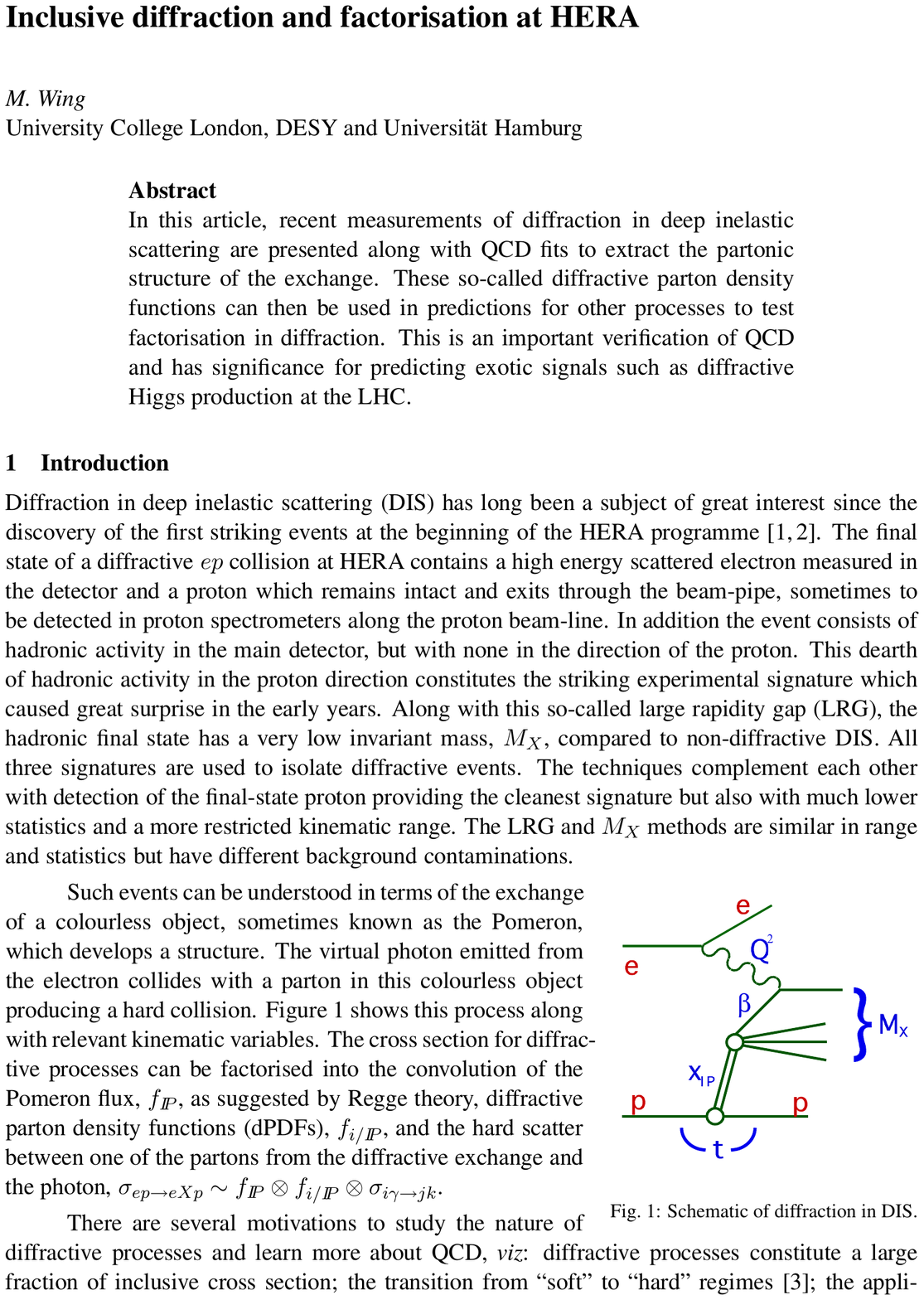} 

\coltoctitle{Exclusive Diffraction and Leading baryons at HERA}
\coltocauthor{D. Wegener}
\Includeart{\CAUT}{\CTIT}{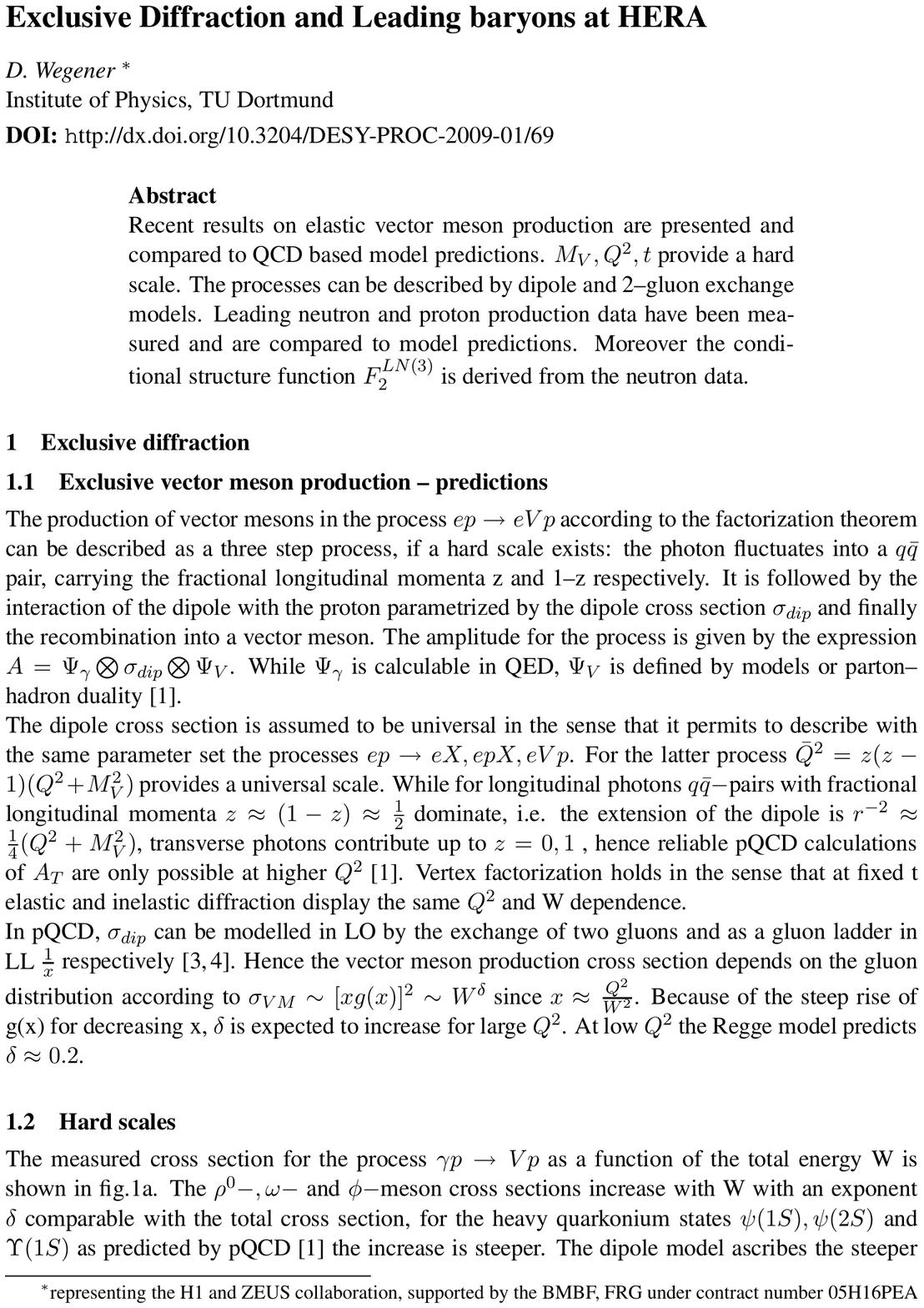} 

\coltoctitle{Quasielastic Scattering in the Dipole Model}
\coltocauthor{Ch. Flensburg}
\Includeart{\CAUT}{\CTIT}{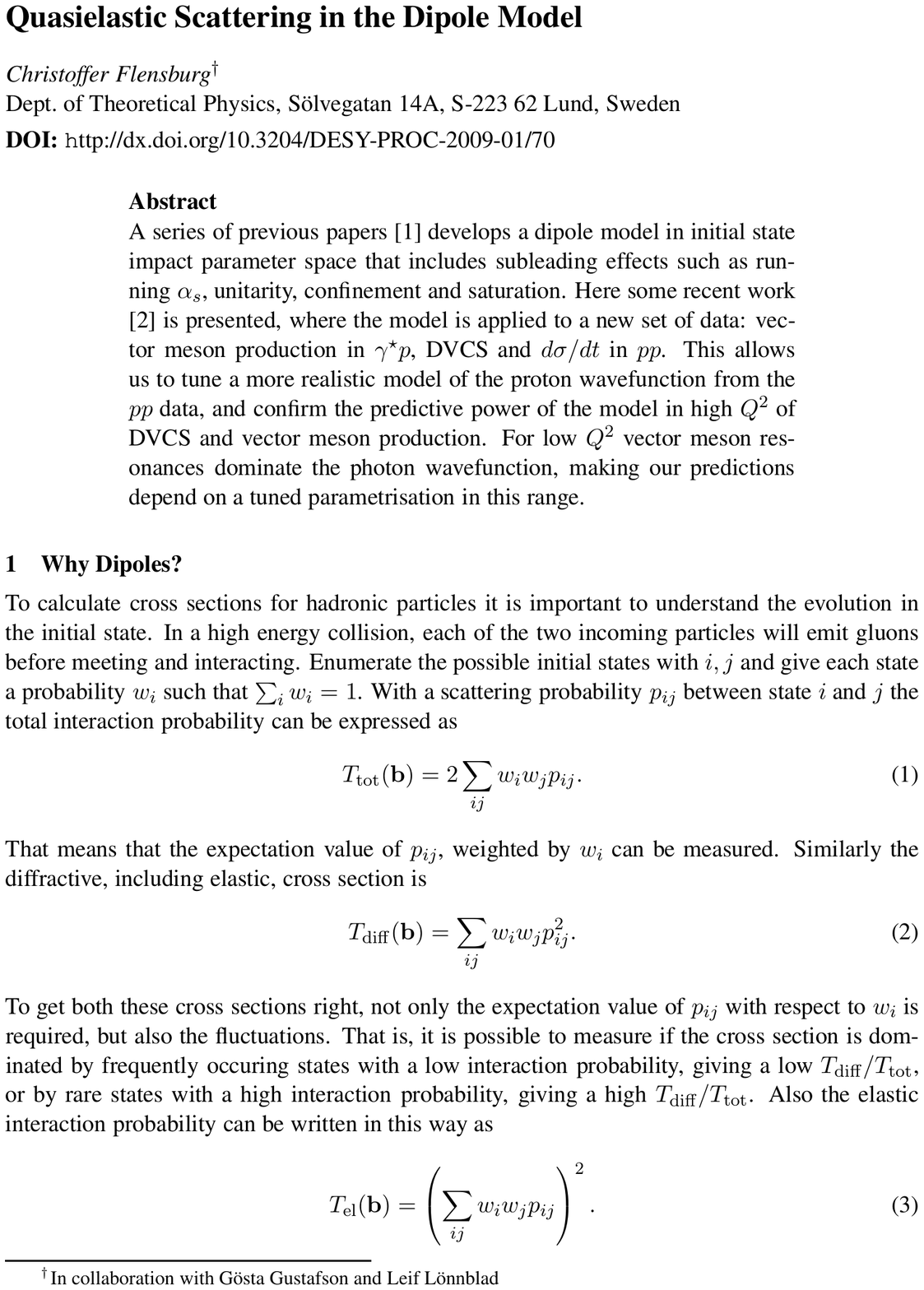} 

\coltoctitle{Hadronic interaction models in the light of the color glass 
condensate}
\coltocauthor{S.~Ostapchenko}
\Includeart{\CAUT}{\CTIT}{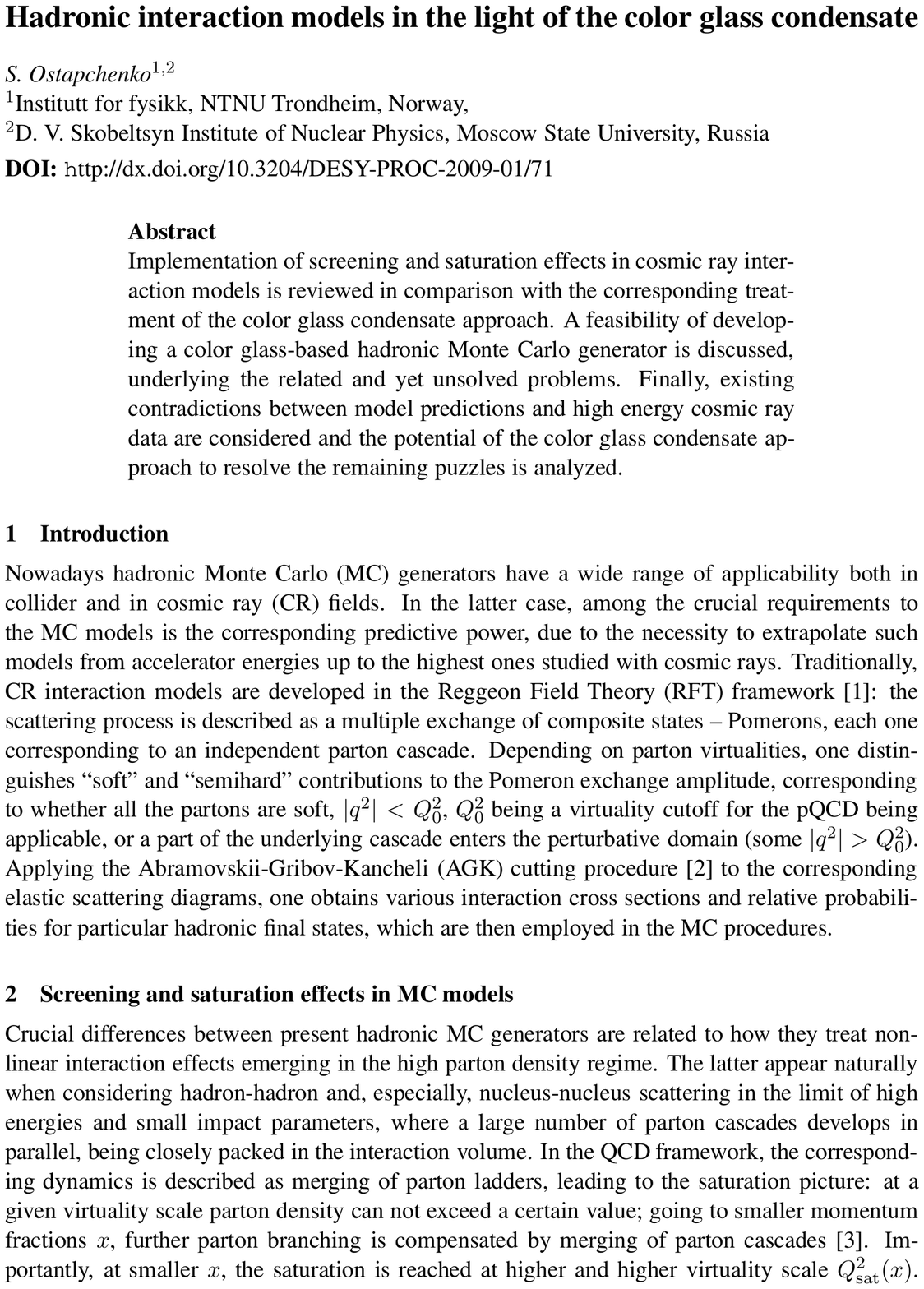} 

\coltoctitle{Test of interaction models via accelerator data}
\coltocauthor{T. Pierog}
\Includeart{\CAUT}{\CTIT}{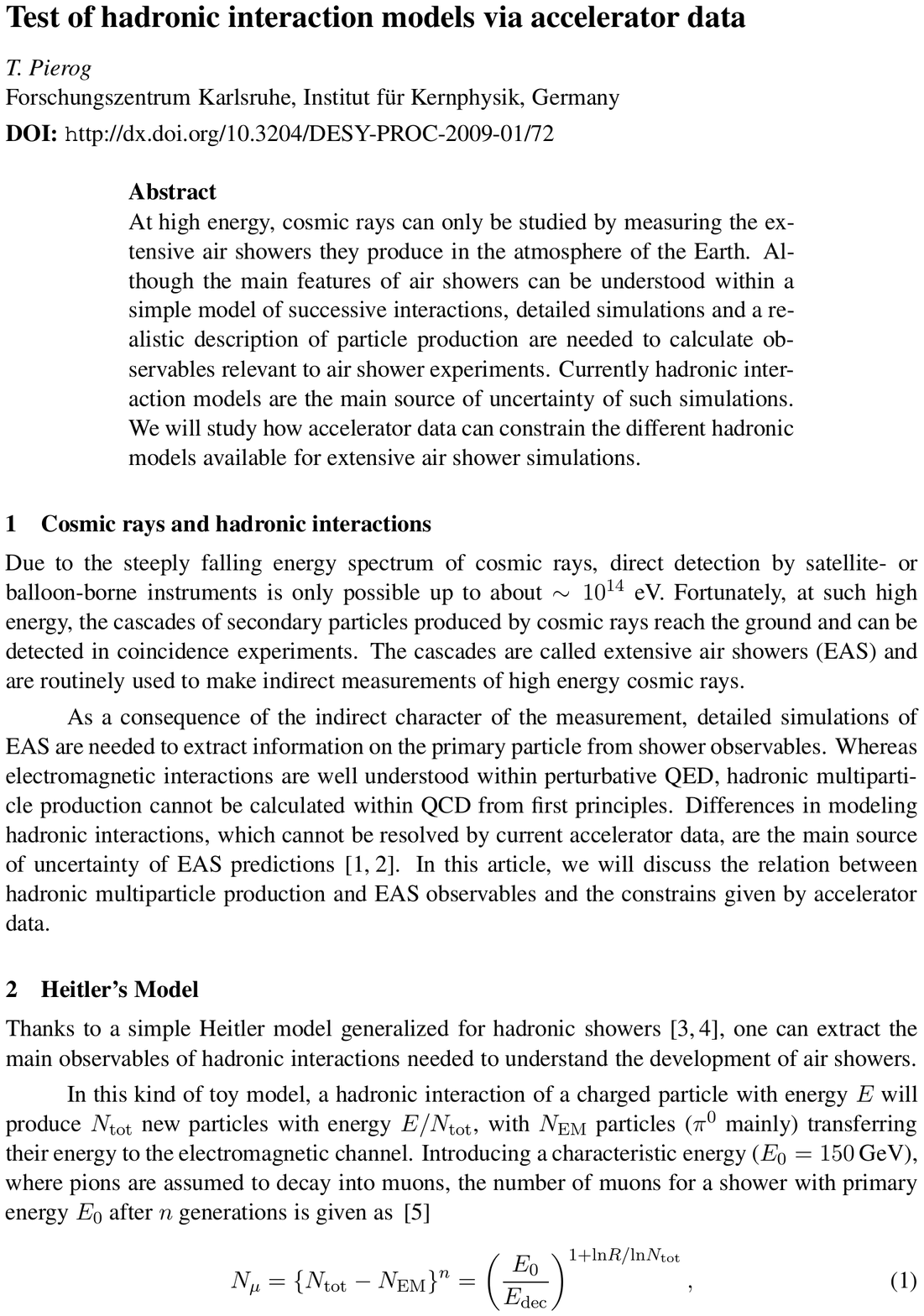} 

\coltoctitle{Results of the Pierre Auger Observatory
 - aspects  related to hadronic interaction models}
\coltocauthor{P. Travnicek}
\Includeart{\CAUT}{\CTIT}{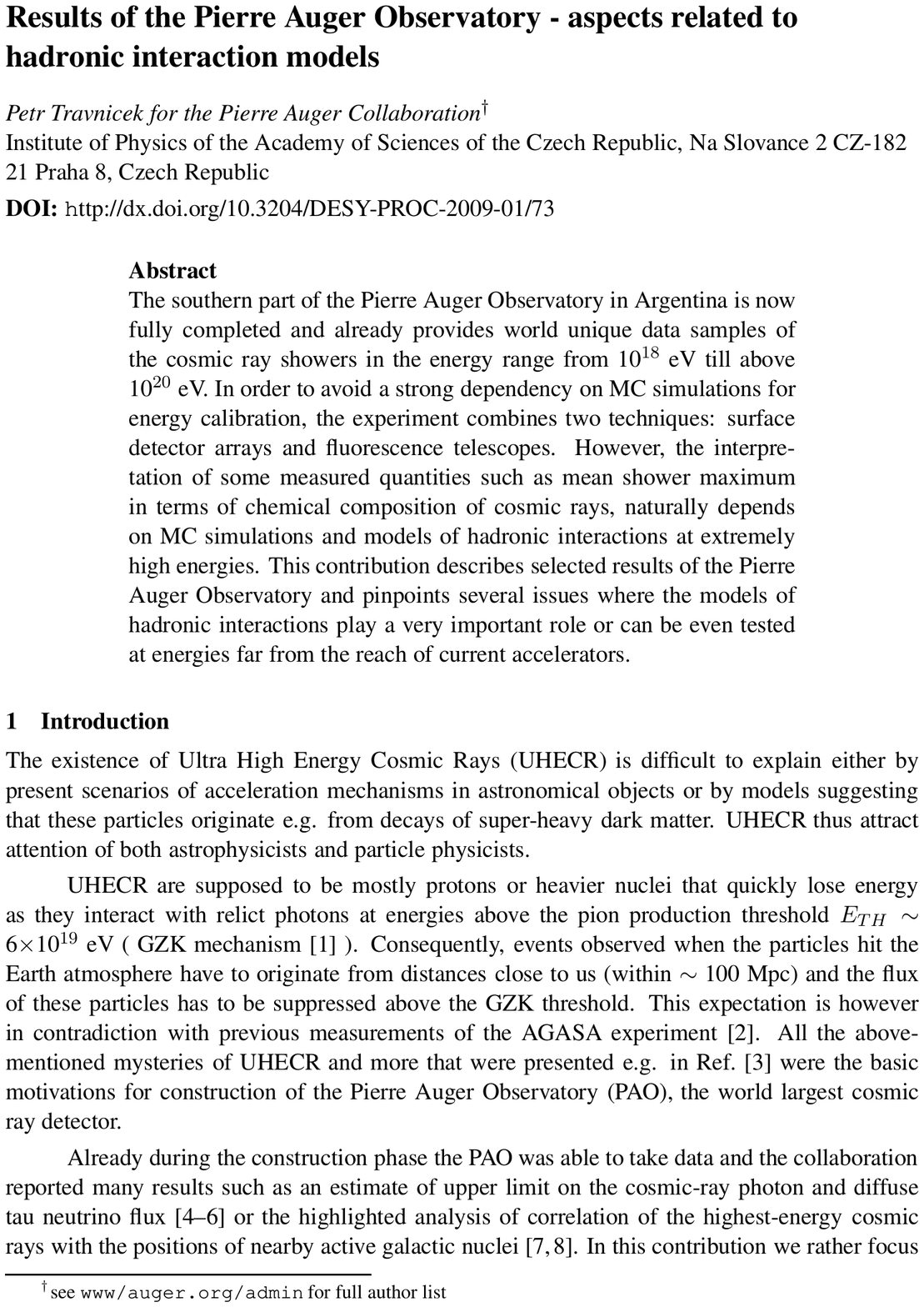} 

\coltoctitle{$D\overline{D}$ momentum correlations versus relative
azimuth as a sensitive probe for thermalization}
\coltocauthor{G.~Tsiledakis}
\Includeart{\CAUT}{\CTIT}{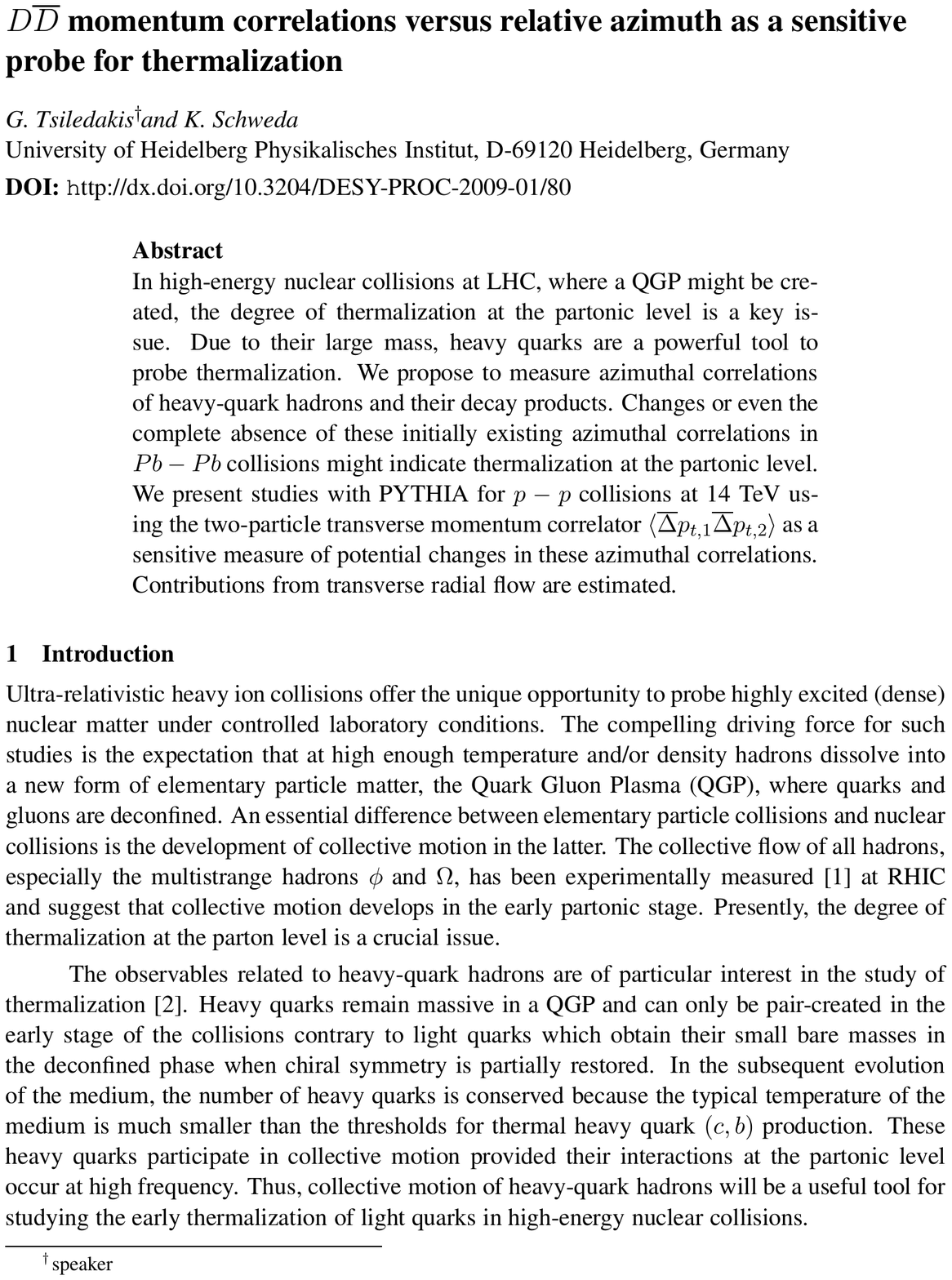} 

\coltoctitle{Hard Diffraction at the LHC }
\coltocauthor{A. De Roeck}
\Includeart{\CAUT}{\CTIT}{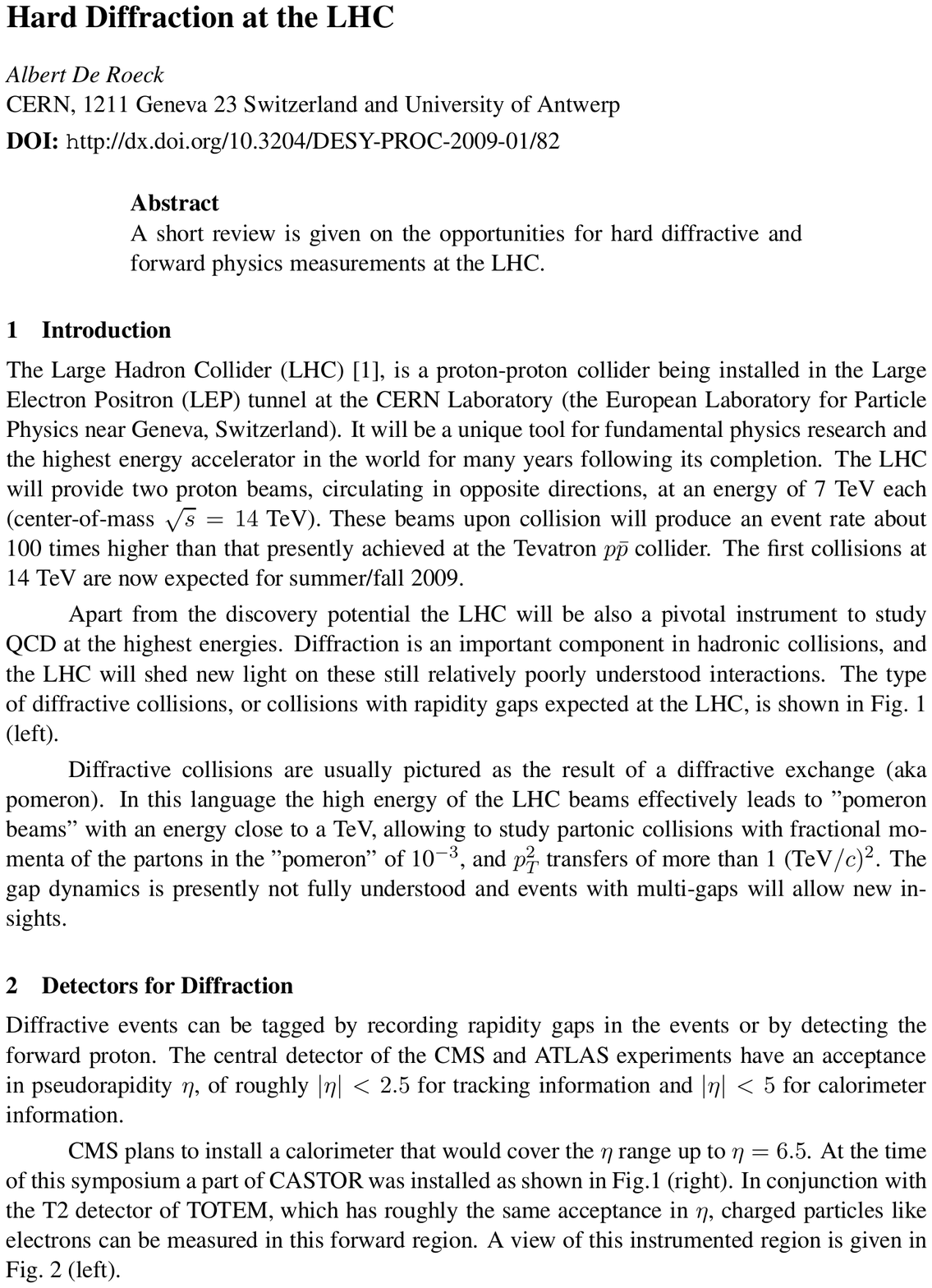} 

\coltoctitle{Exclusive photoproduction of dileptons at high energies}
\coltocauthor{M.V.T. Machado}
\Includeart{\CAUT}{\CTIT}{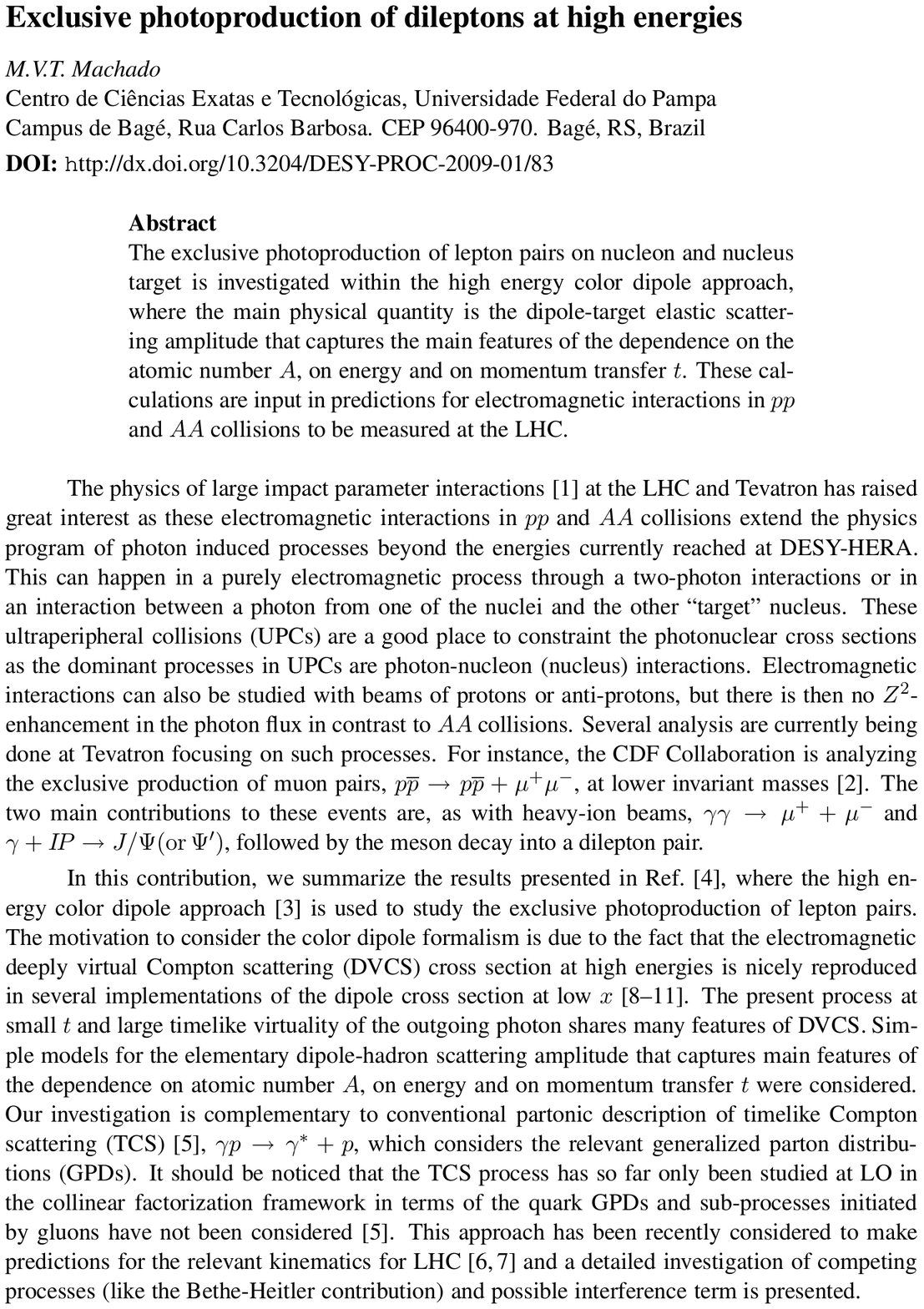}

% \coltoctitle{Second Talk}
% \coltocauthor{David Summerset}
% \Includeart{\CAUT}{\CTIT}{summerset_david}

\end{papers} % end of individual articles/papers

%% file: include-methods.tex
\chapter[\large WG: Strategies and Analysis Methods]{ Working Group \\ Strategies and Analysis Methods}
{\Large\bfseries Convenors:}
\vspace{10mm}

{\Large\itshape 
E.DeWolf (Antwerp)\\
A.Geiser (DESY)\\
M.Sjodahl (Manchester)

\CLDP
\begin{papers} % start of individual articles/papers

\coltoctitle{Colour Reconnections and Top Physics}
\coltocauthor{D. Wicke}
\Includeart{\CAUT}{\CTIT}{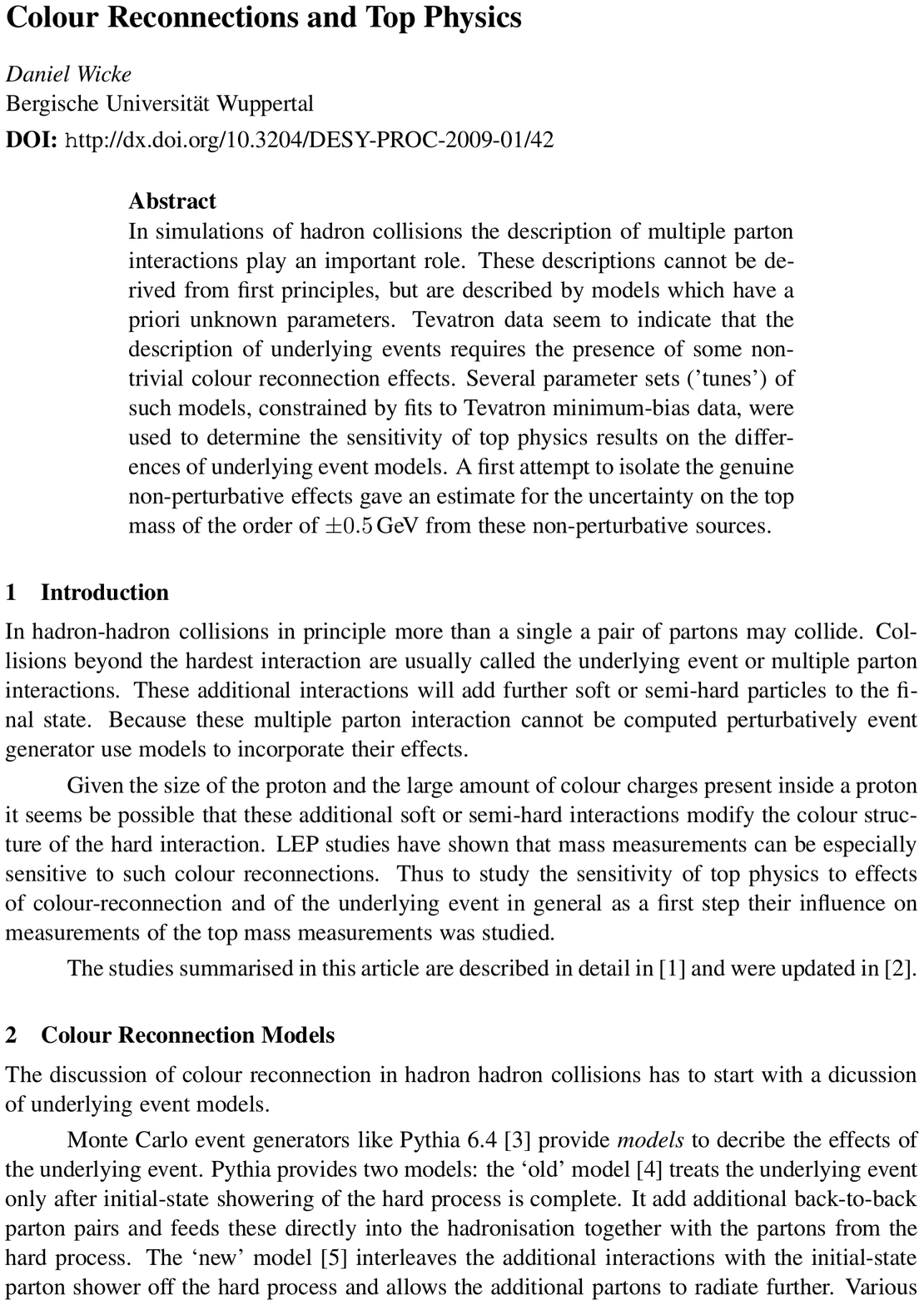} 

\coltoctitle{Heavy Quark Production at HERA as a Probe of Hard QCD}
\coltocauthor{R. Shehzadi}
\Includeart{\CAUT}{\CTIT}{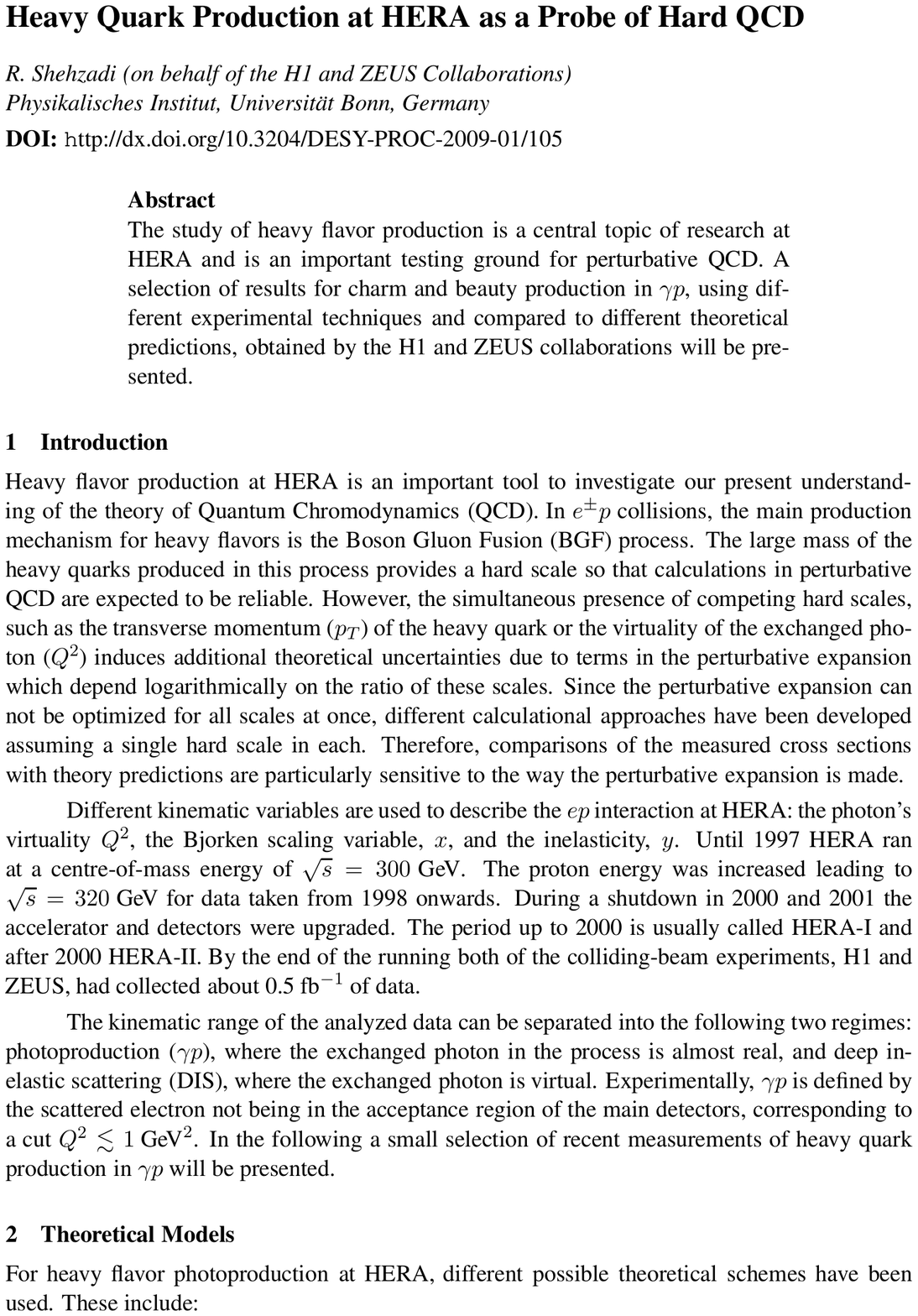} 

\coltoctitle{What do we learn from forward detectors at LHC ?}
\coltocauthor{A. Bunyatyan}
\Includeart{\CAUT}{\CTIT}{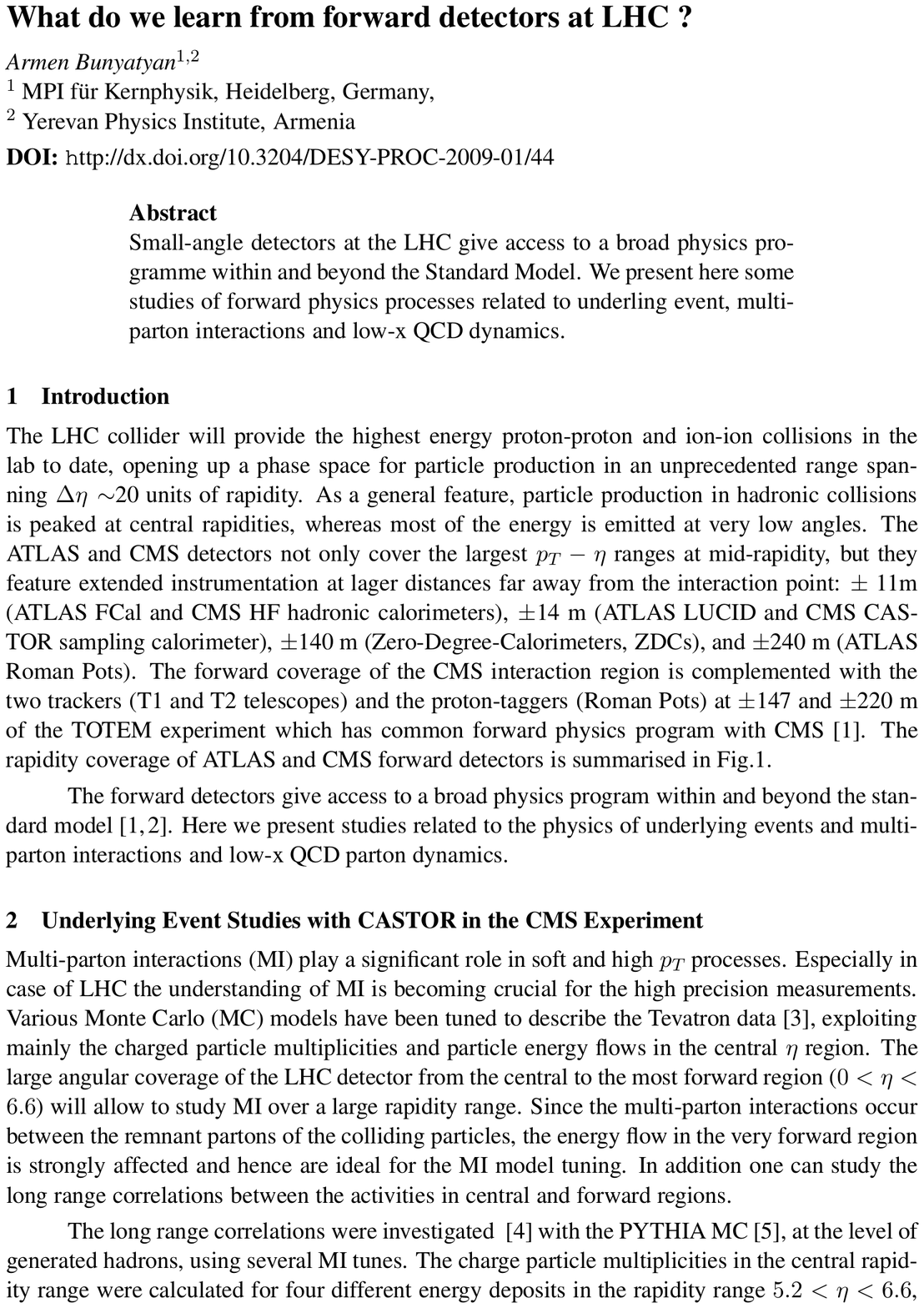} 

\coltoctitle{Recent L3 Results (and Questions) on BEC at LEP}
\coltocauthor{W. J. Metzger}
\Includeart{\CAUT}{\CTIT}{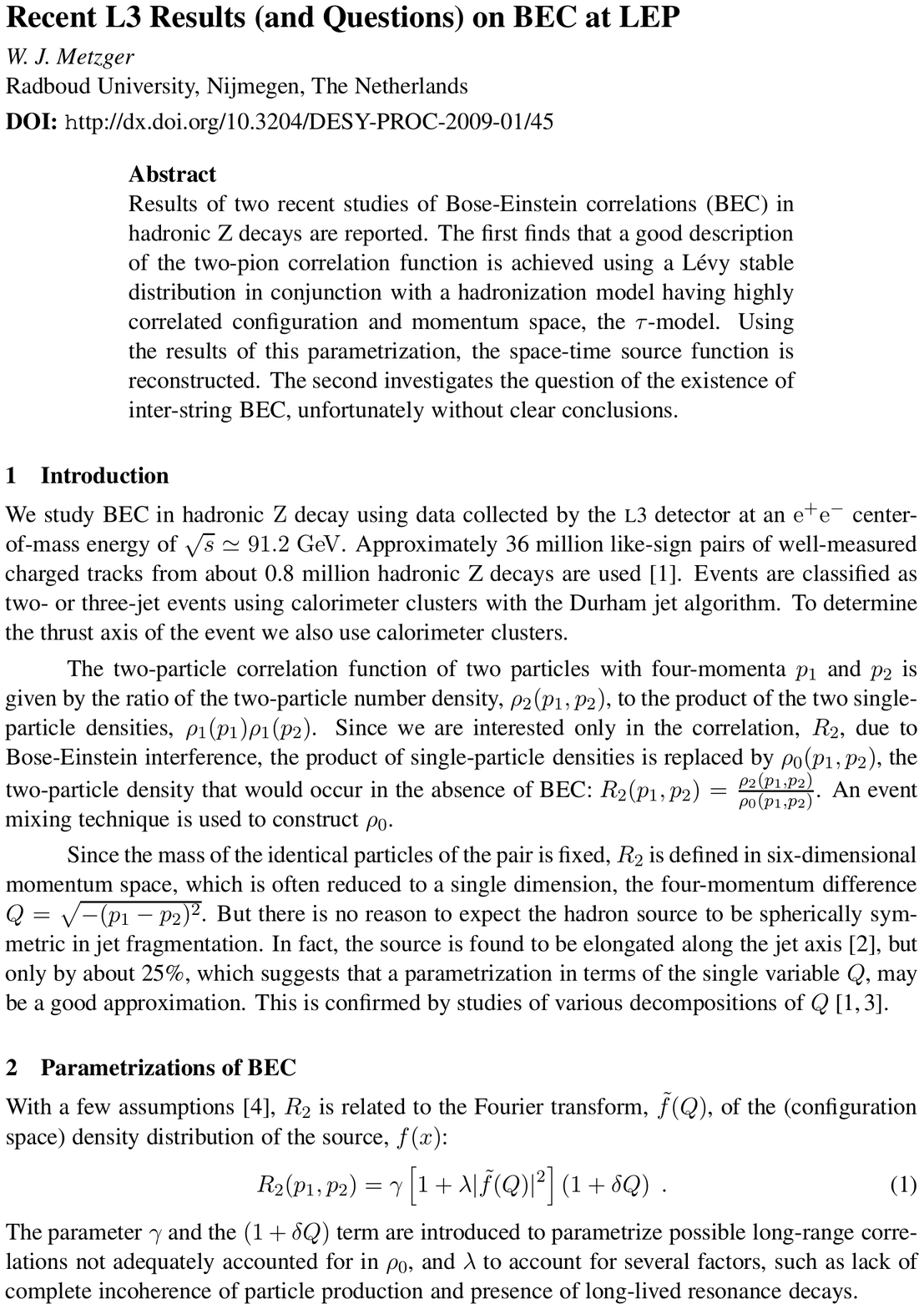} 

\coltoctitle{Bose-Einstein or HBT correlations in high energy reactions}
\coltocauthor{T. Cs\"org\H{o}}
\Includeart{\CAUT}{\CTIT}{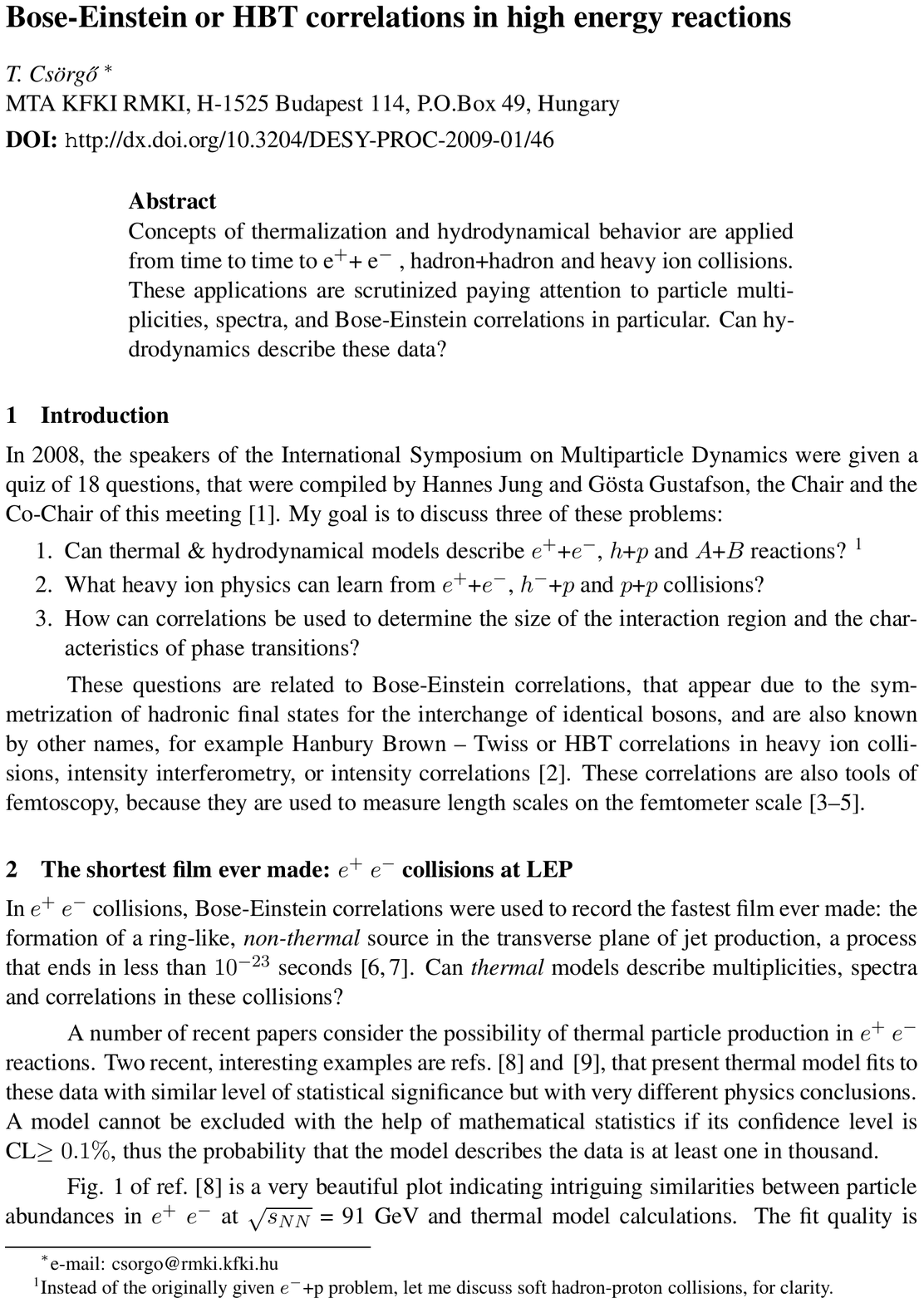} 

\coltoctitle{QCD and Monte Carlo generators}
\coltocauthor{Z. Nagy}
\Includeart{\CAUT}{\CTIT}{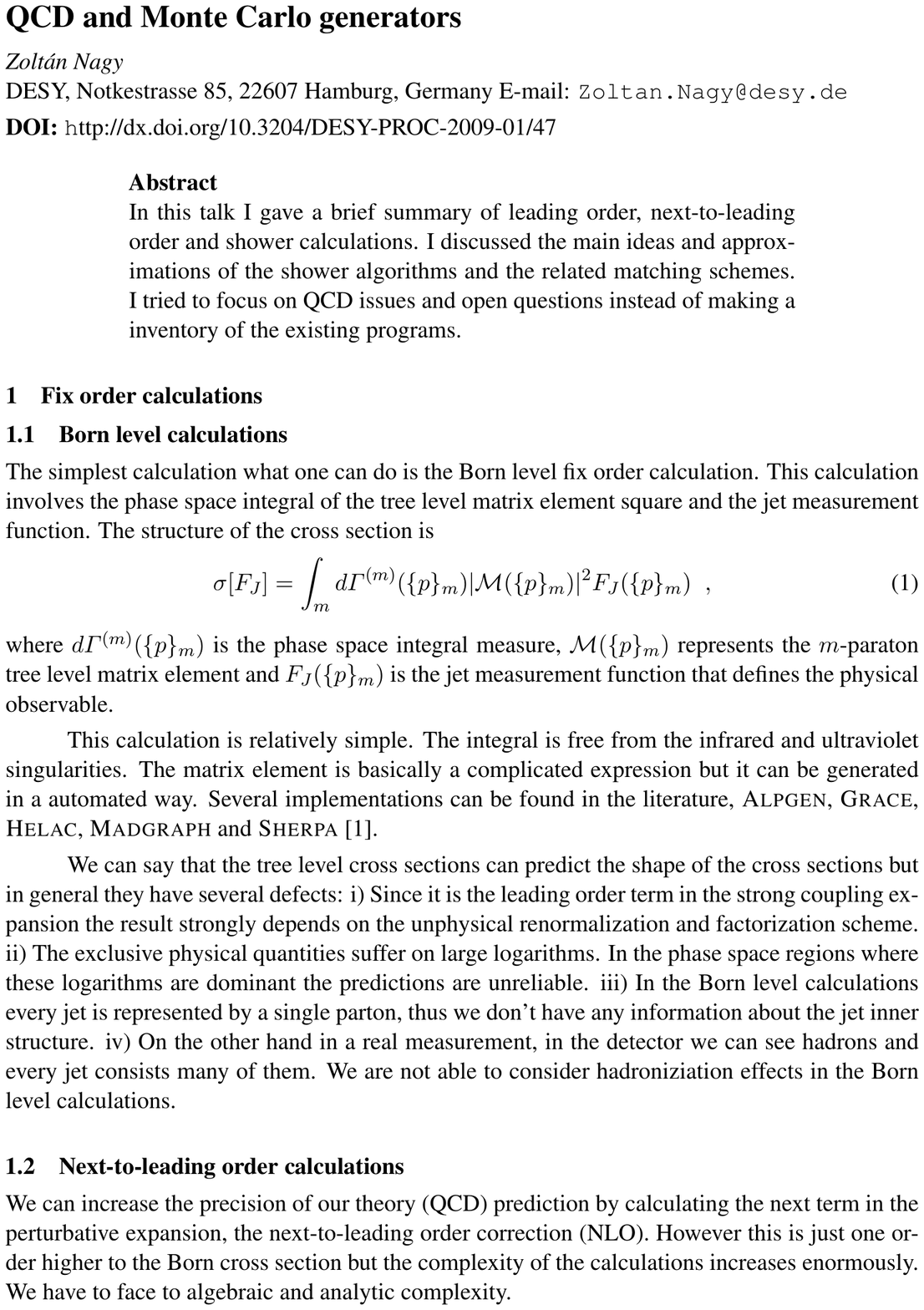} 

\coltoctitle{On factorization scheme suitable for NLO Monte Carlo event generators}
\coltocauthor{K. Kolar}
\Includeart{\CAUT}{\CTIT}{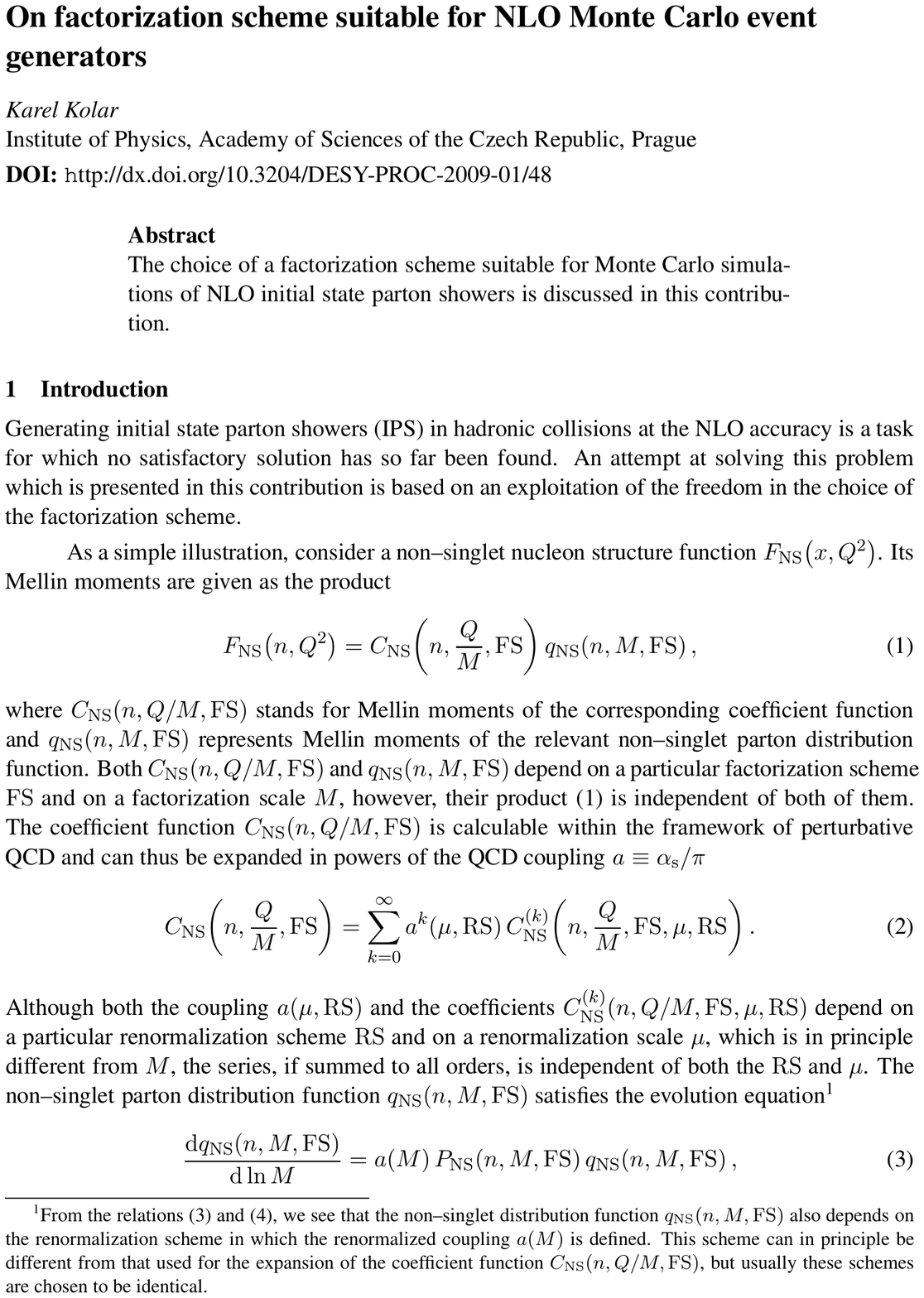} 

\coltoctitle{Review on recent developements in jet finding}
\coltocauthor{J. Rojo}
\Includeart{\CAUT}{\CTIT}{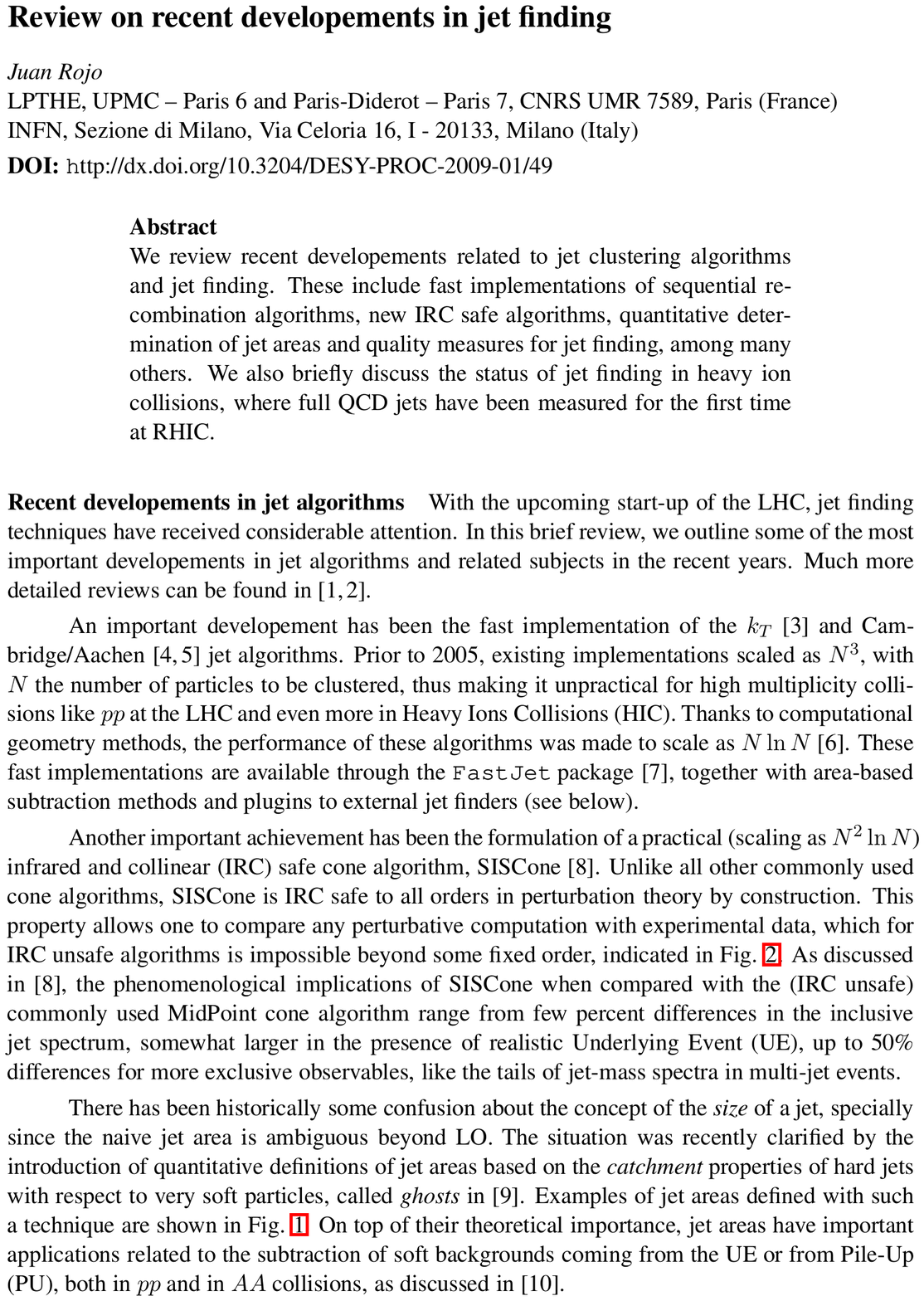} 

\coltoctitle{Multi-particle production and TMD distributions}
\coltocauthor{F. Hautmann}
\Includeart{\CAUT}{\CTIT}{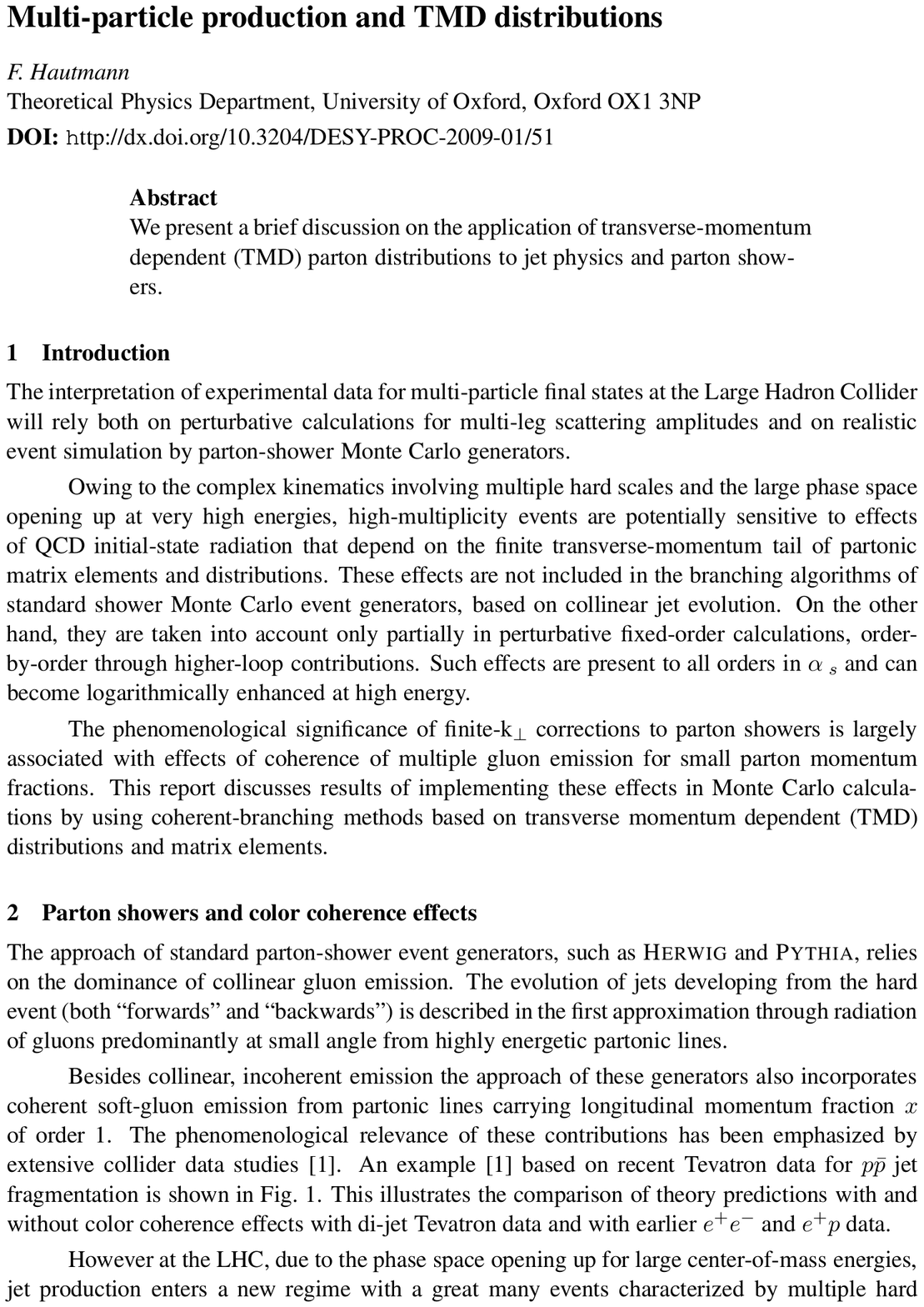} 

\coltoctitle{Bose-Einstein study of position-momentum correlations
of charged pions in hadronic Z$^0$ decays}
\coltocauthor{C. Ciocca}
\Includeart{\CAUT}{\CTIT}{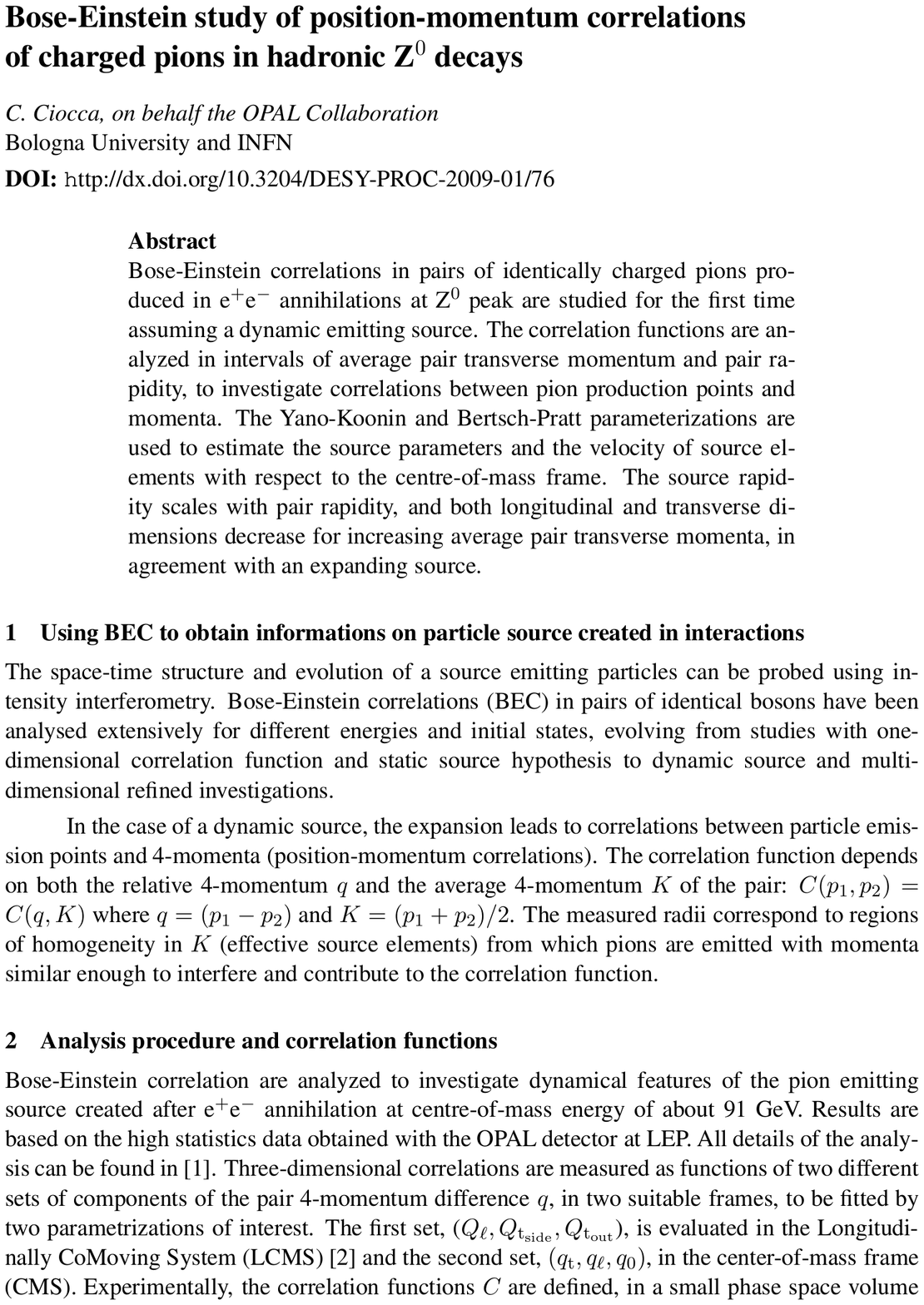} 

\coltoctitle{Squeezed correlations among particle-antiparticle pairs}
\coltocauthor{S.S. Padula}
\Includeart{\CAUT}{\CTIT}{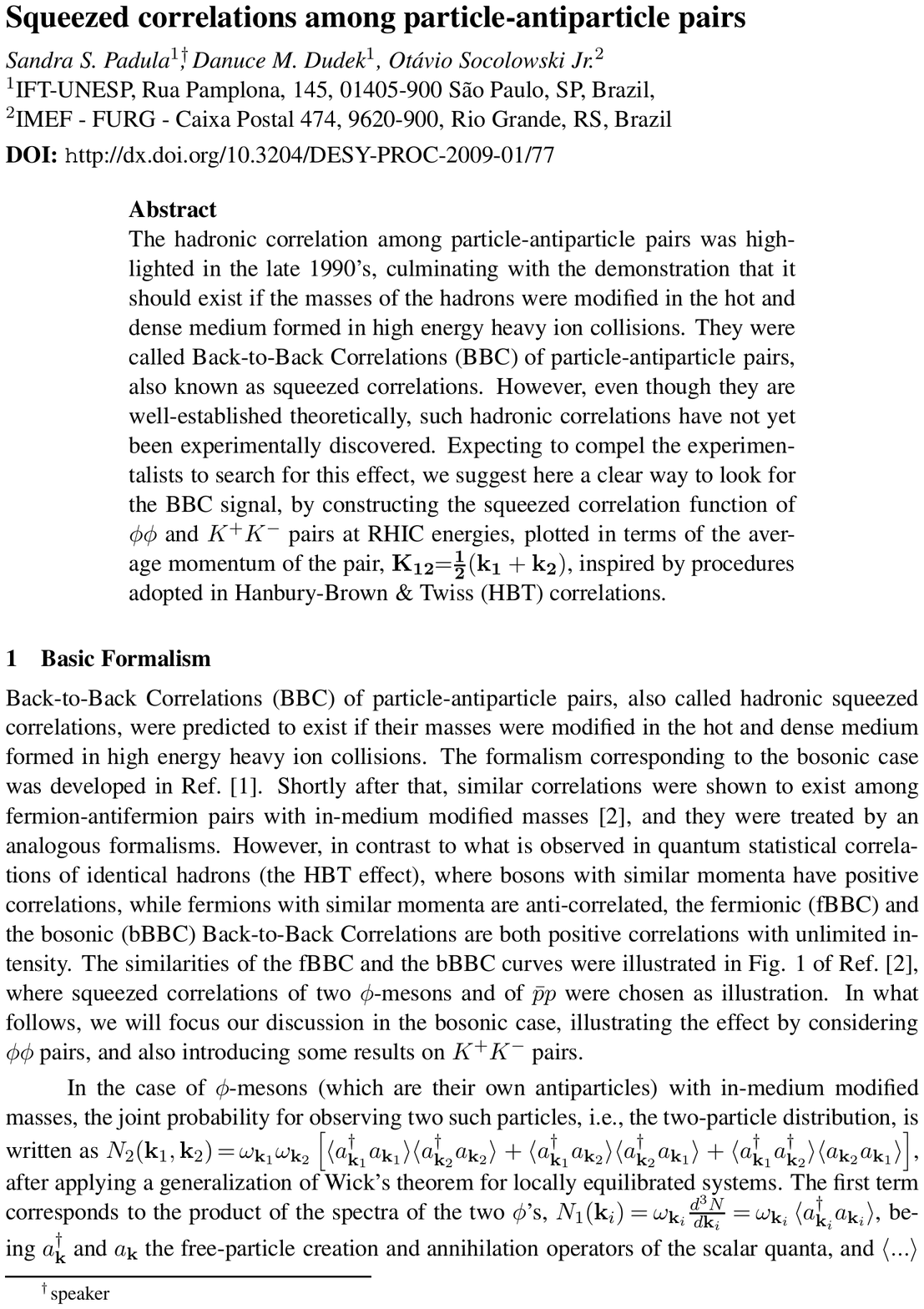} 

\coltoctitle{An approach to QCD phase transitions via multiplicity fluctuations and correlations}
\coltocauthor{K. Homma}
\Includeart{\CAUT}{\CTIT}{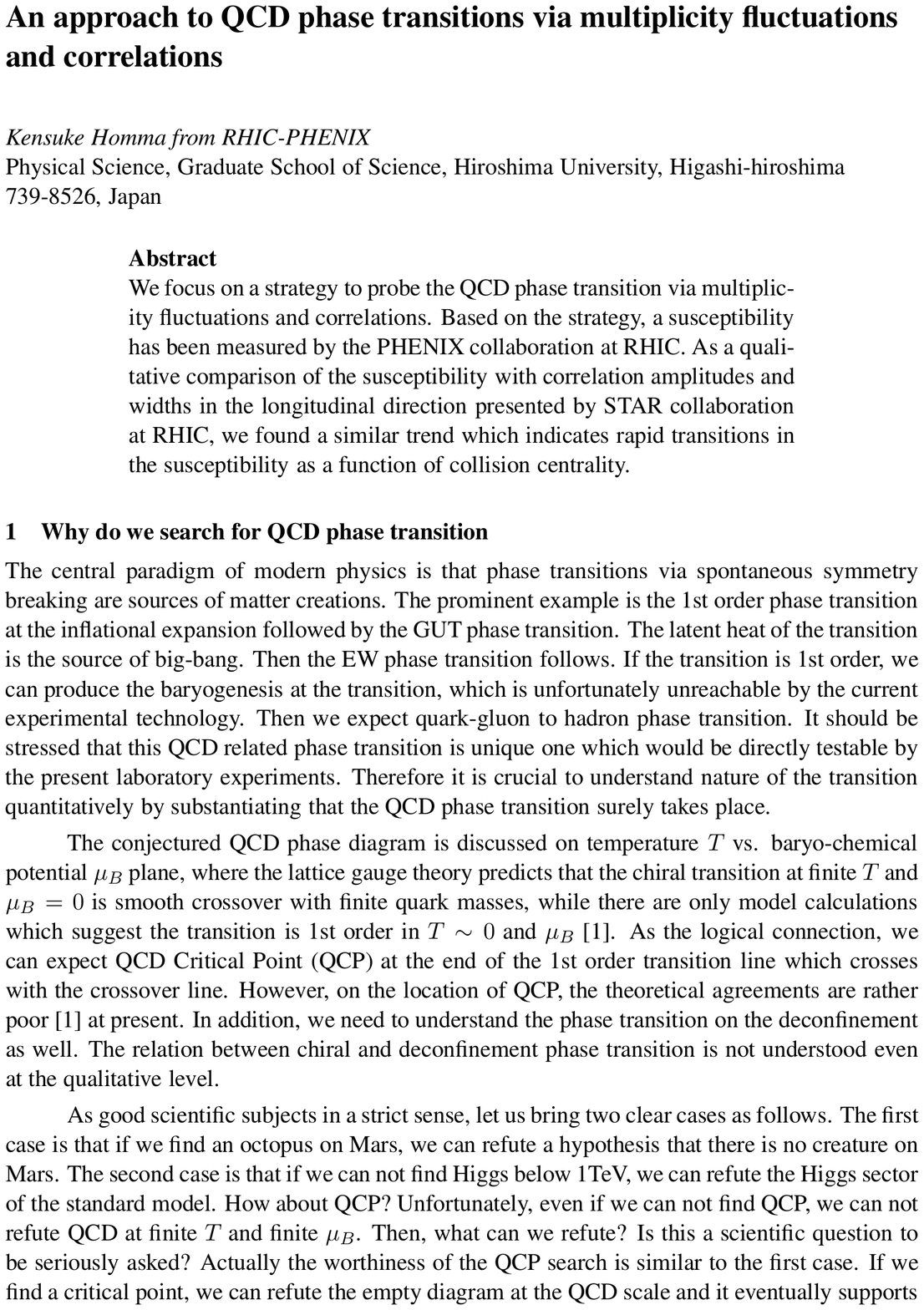} 

\coltoctitle{Antibaryon to Baryon Production Ratios 
 in Pb-Pb and p-p collision at LHC energies 
 of the DPMJET-III Monte Carlo}
\coltocauthor{F.W.Bopp, J.Ranft, R.Engel, S.Roesler}
\Includeart{\CAUT}{\CTIT}{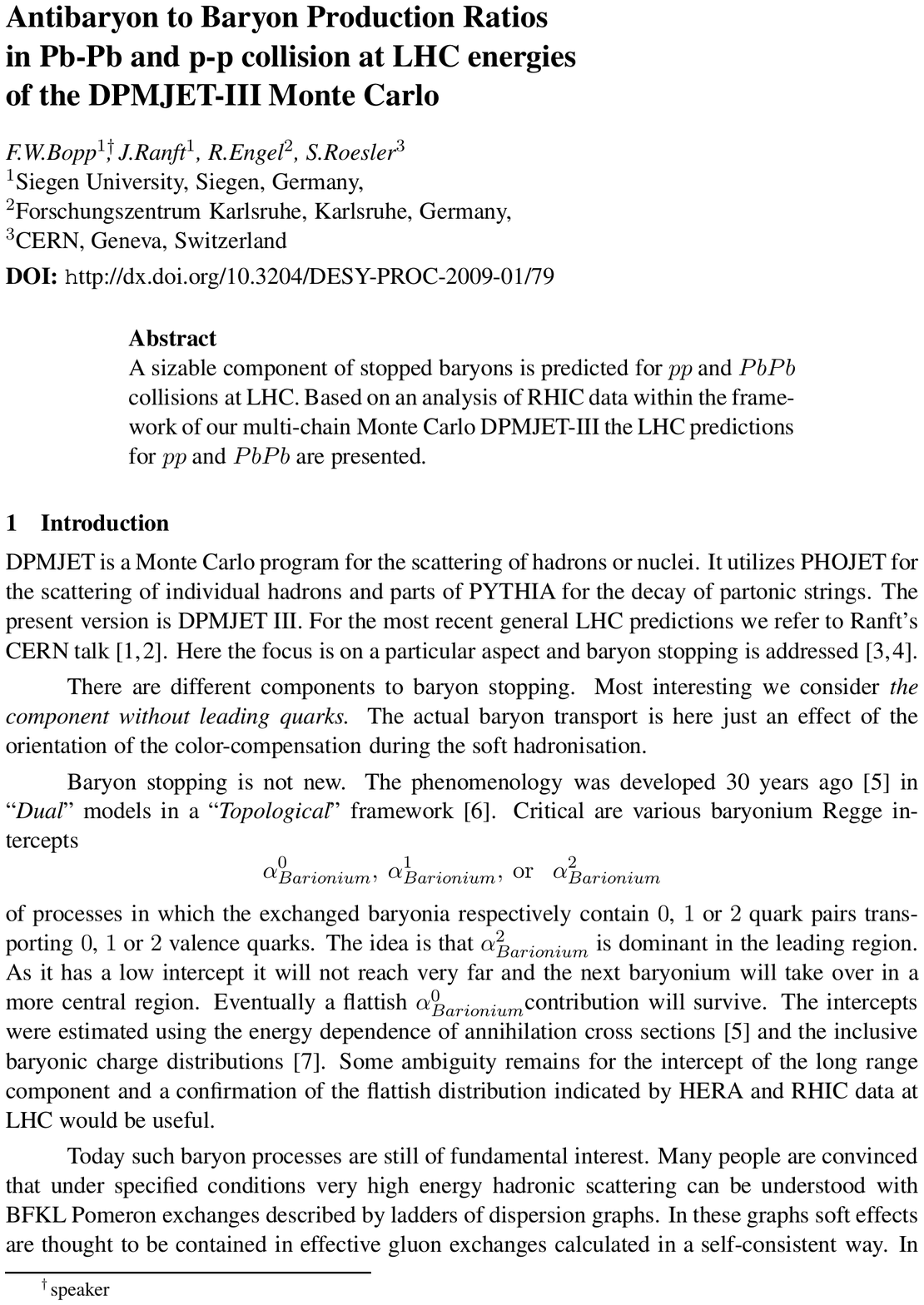} 

% \coltoctitle{Second Talk}
% \coltocauthor{David Summerset}
% \Includeart{\CAUT}{\CTIT}{summerset_david}

\end{papers} % end of individual articles/papers

%% file: include-new-physics.tex
\chapter[\large  WG: New Physics]{ Working Group \\ New Physics}
{\Large\bfseries Convenors:}
\vspace{10mm}

{\Large\itshape 
G. Weiglein (Durham)\\
A. DeRoeck (CERN)}

\CLDP
\begin{papers} % start of individual articles/papers

%\coltoctitle{Higgs production at the LHC: progress on higher-order contributions for signal and backgrounds}
%\coltocauthor{Ch. Anastasiou}
%\Includeart{\CAUT}{\CTIT}{../} 

\coltoctitle{New signatures and challenges for the LHC}
\coltocauthor{M. Strassler}
\Includeart{\CAUT}{\CTIT}{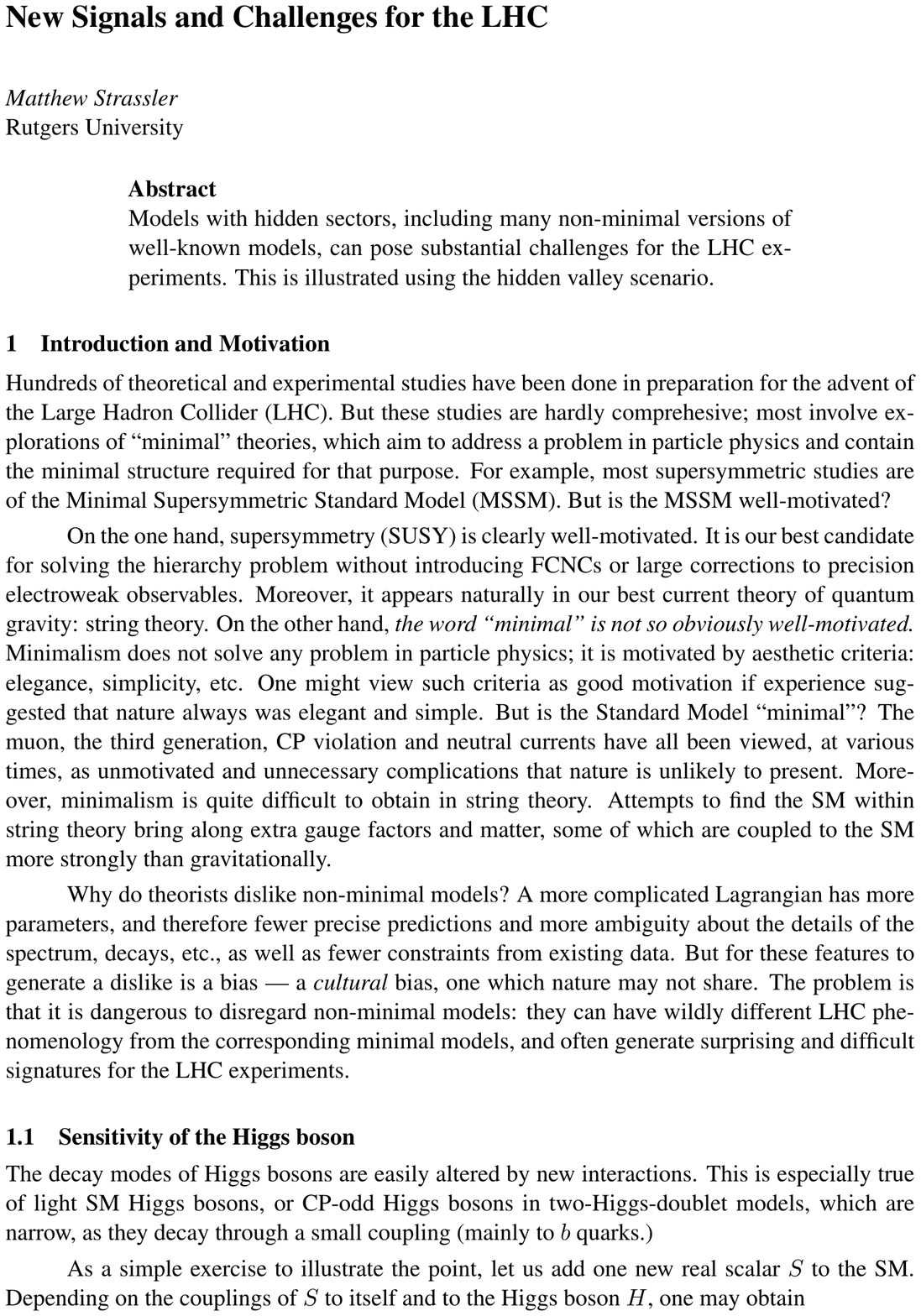} 

\coltoctitle{New physics search in the LHCb era}
\coltocauthor{T. Hurth}
\Includeart{\CAUT}{\CTIT}{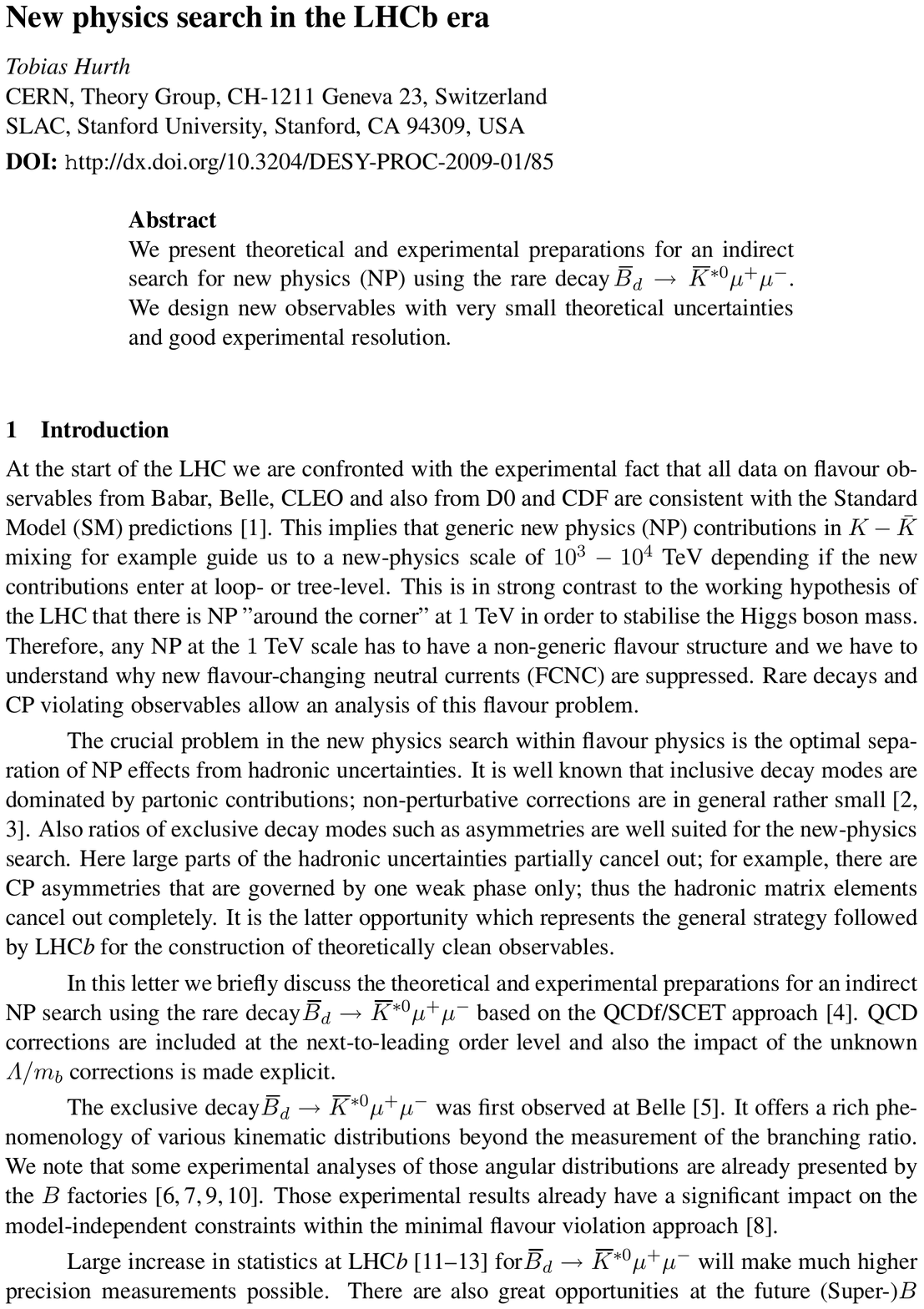} 

\coltoctitle{Searches for Physics beyond the Standard Model}
\coltocauthor{A. Meyer}
\Includeart{\CAUT}{\CTIT}{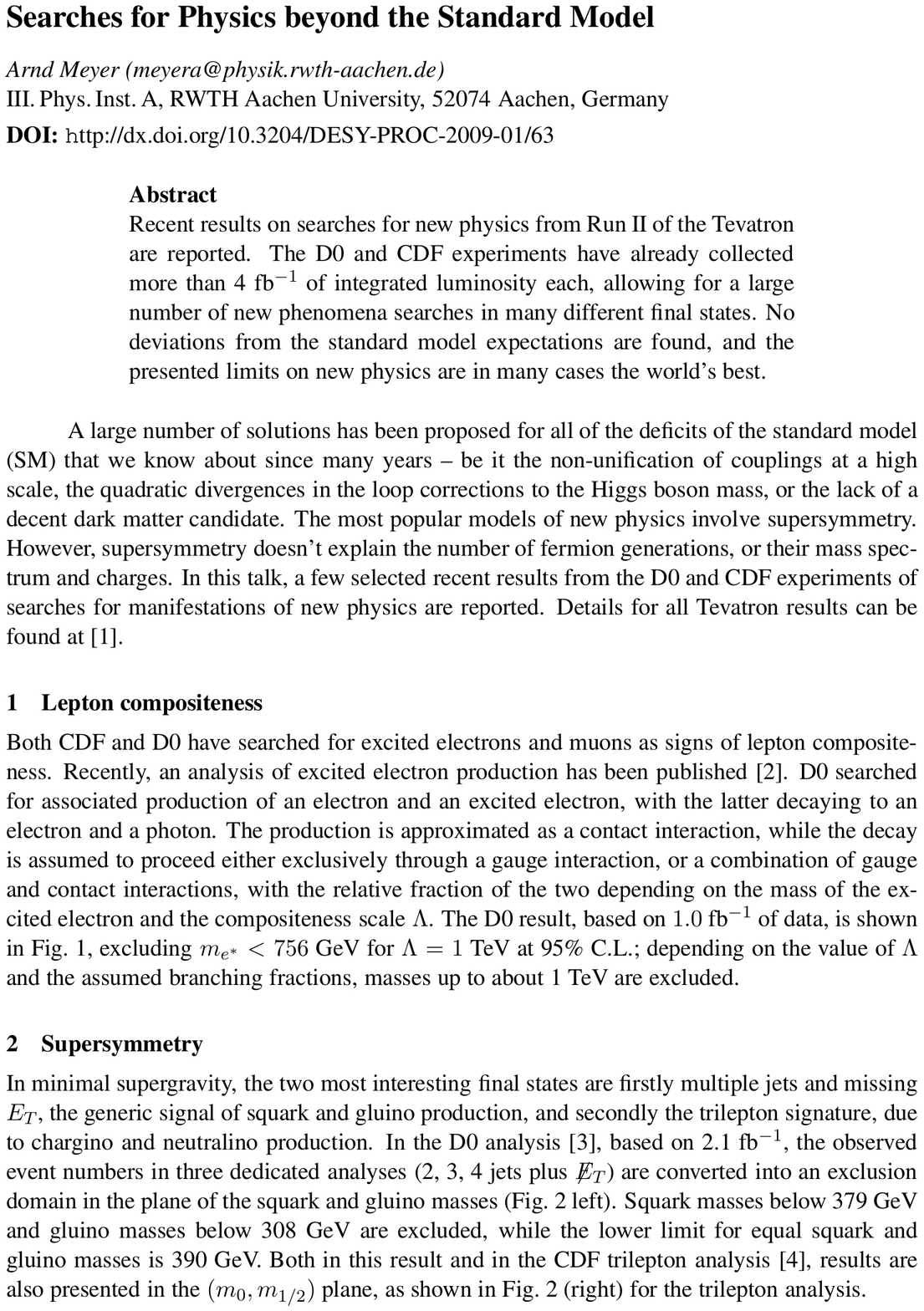} 

\coltoctitle{Discovery potential at the LHC: channels relevant for SM Higgs}
\coltocauthor{I. Tsukerman}
\Includeart{\CAUT}{\CTIT}{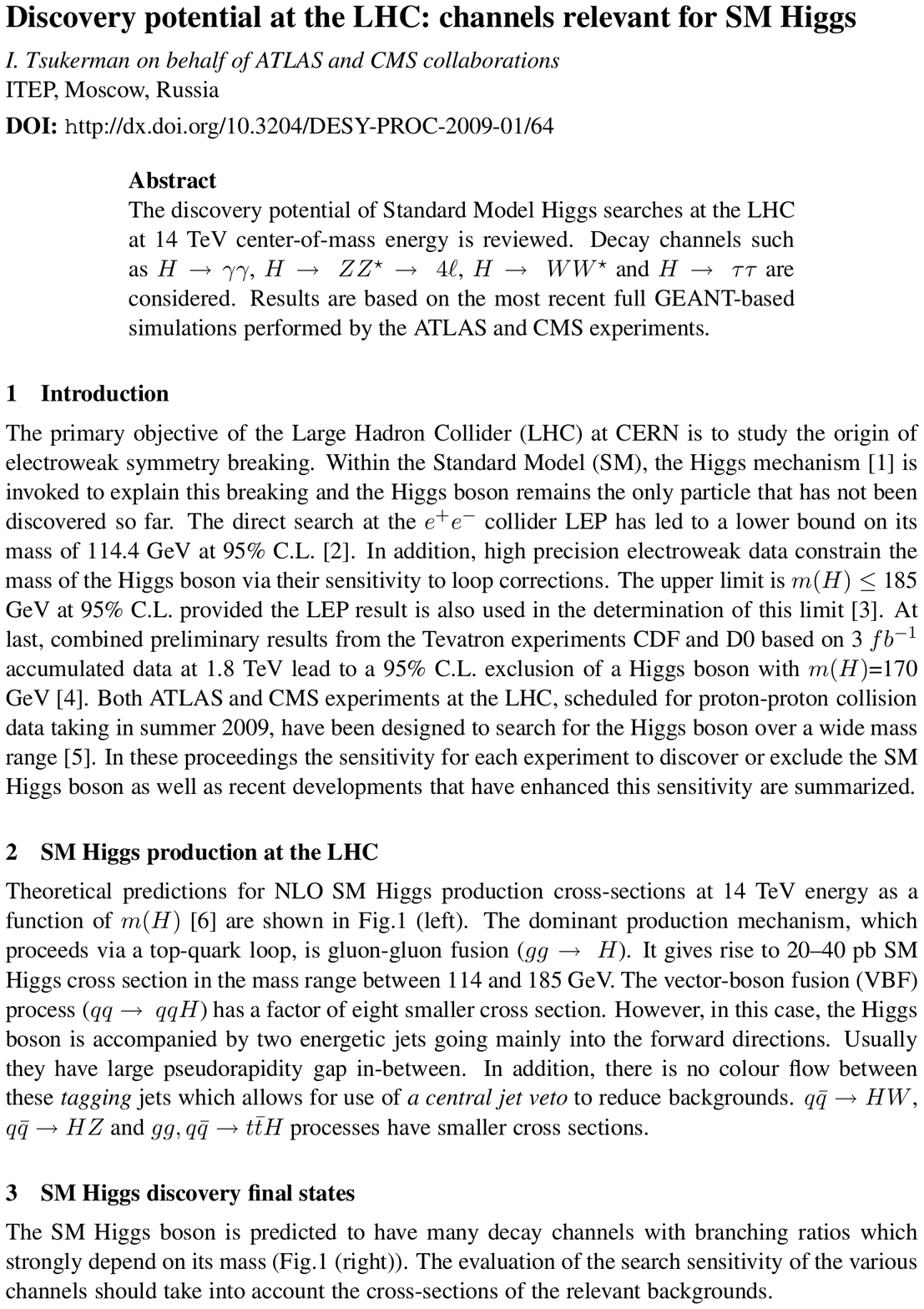} 

\coltoctitle{Central Exclusive Production of BSM Higgs bosons at the LHC}
\coltocauthor{S. Heinemeyer, V.A. Khoze,
M.G. Ryskin, M. Ta\v{s}evsk\'{y} G. Weiglein}
\Includeart{\CAUT}{\CTIT}{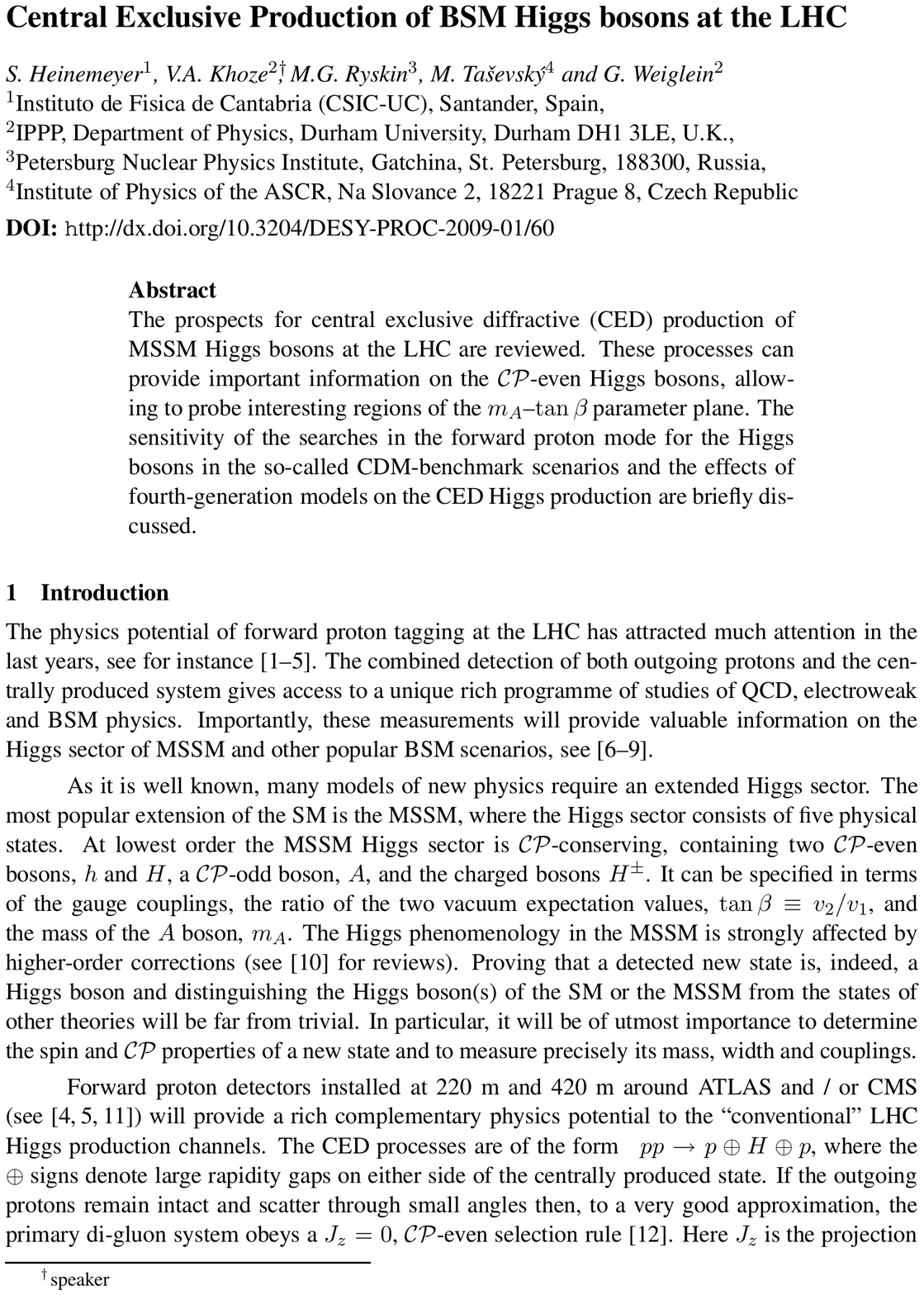} 

\coltoctitle{A bottom-up strategy for reconstructing the 
underlying MSSM parameters at the LHC}
\coltocauthor{J.-L. Kneur}
\Includeart{\CAUT}{\CTIT}{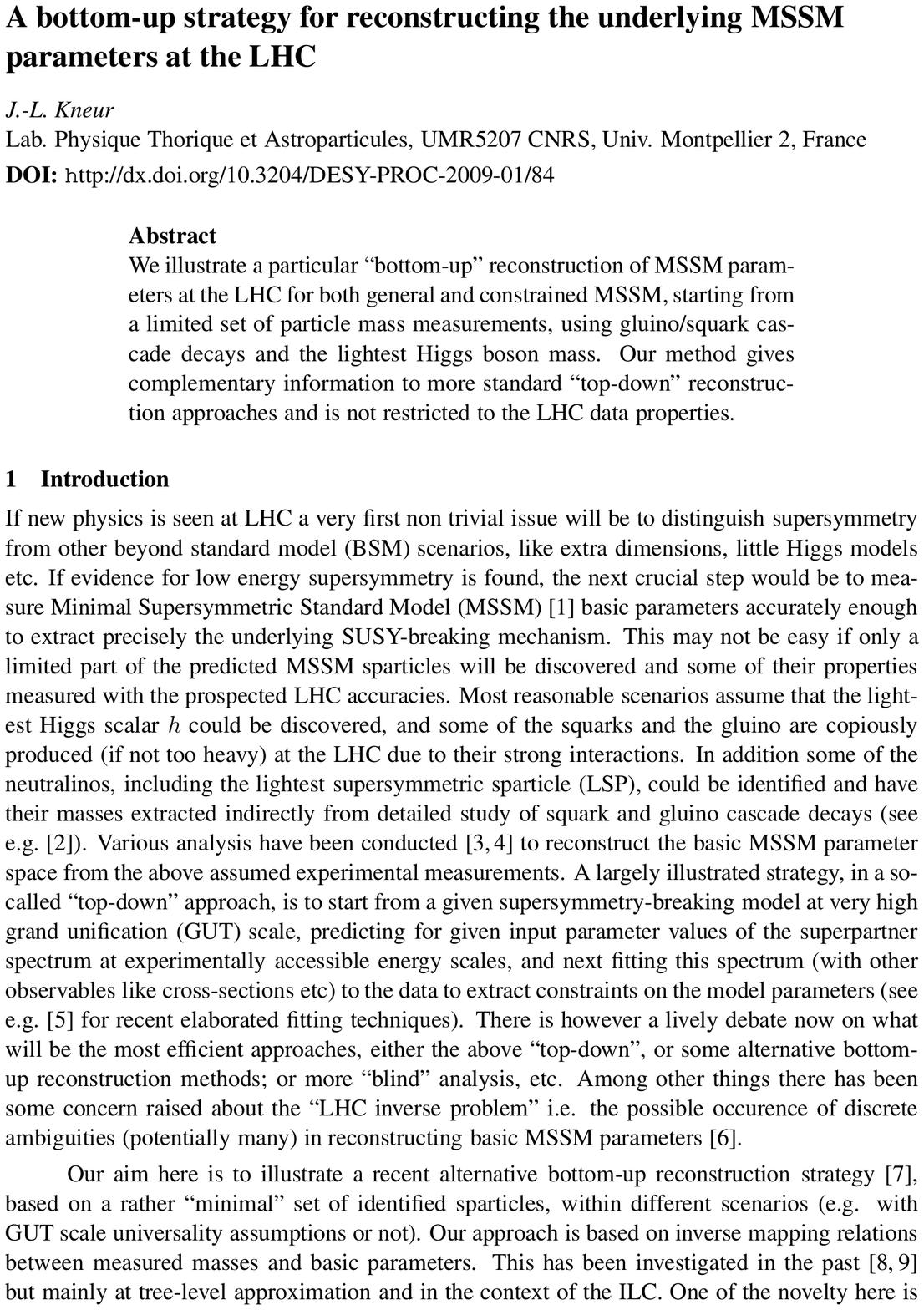} 

\coltoctitle{Supersymmetry and other beyond the Standard Model physics:
  Prospects for determining mass, spin and CP properties}
\coltocauthor{W. Ehrenfeld}
\Includeart{\CAUT}{\CTIT}{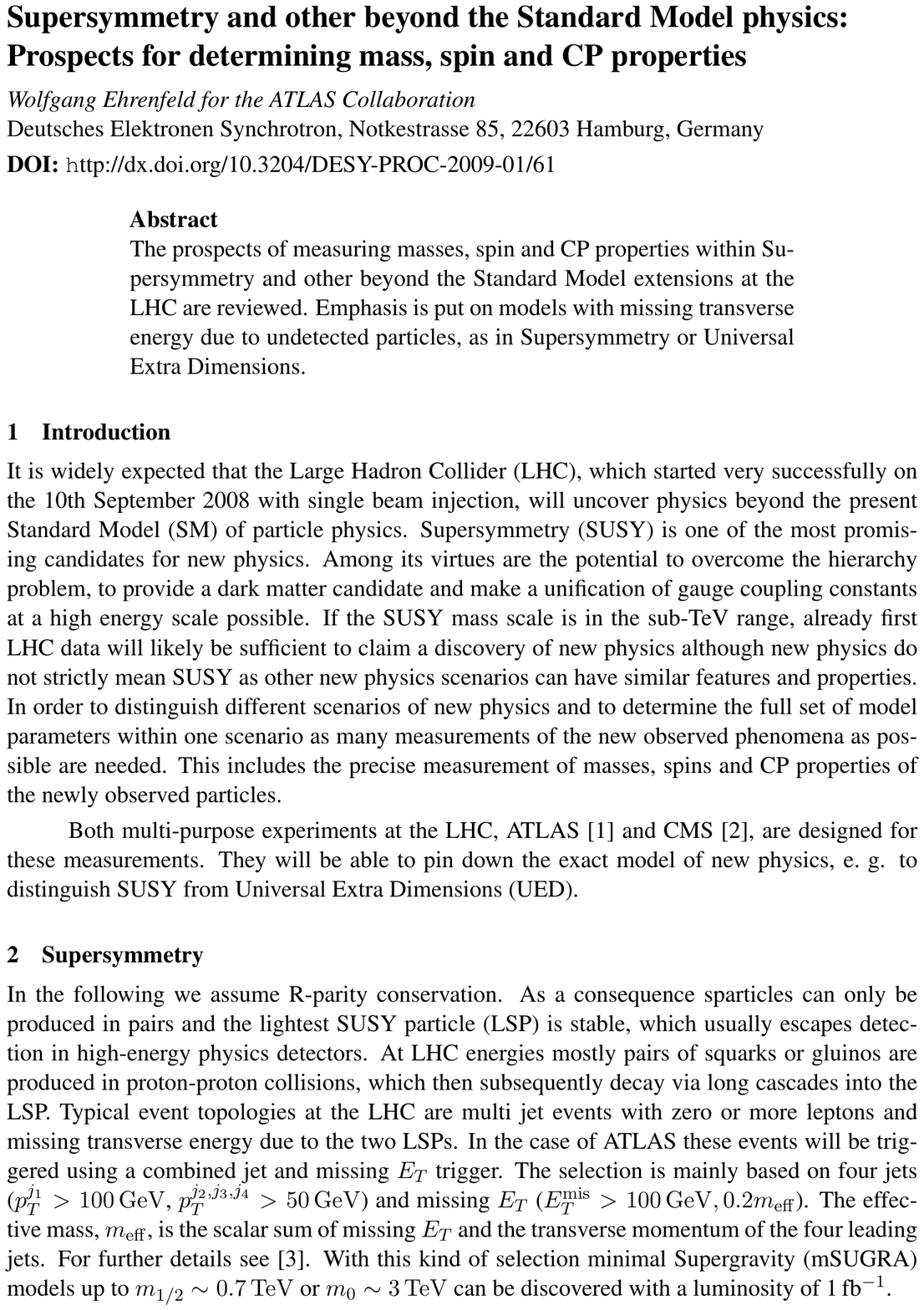}

% \coltoctitle{Second Talk}
% \coltocauthor{David Summerset}
% \Includeart{\CAUT}{\CTIT}{summerset_david}

\end{papers} % end of individual articles/papers

%% file: include-recent-developments.tex
\chapter[\large  WG: Recent Developments]{ Working Group \\ Recent Developments}
{\Large\bfseries %Convenors:}
\vspace{10mm}

{\Large\itshape 
%Mandy Cooper-Sarkar (Oxford), \\
%Anna Kulesza (DESY), \\
%Ken Hatakeyama (Rockefeller)}

\CLDP
\begin{papers} % start of individual articles/papers

\coltoctitle{Crossover
between hadronic and partonic phases and liquid property of
sQGP}
\coltocauthor{M. Xu , M. Yu Meiling, L .Liu}
\Includeart{\CAUT}{\CTIT}{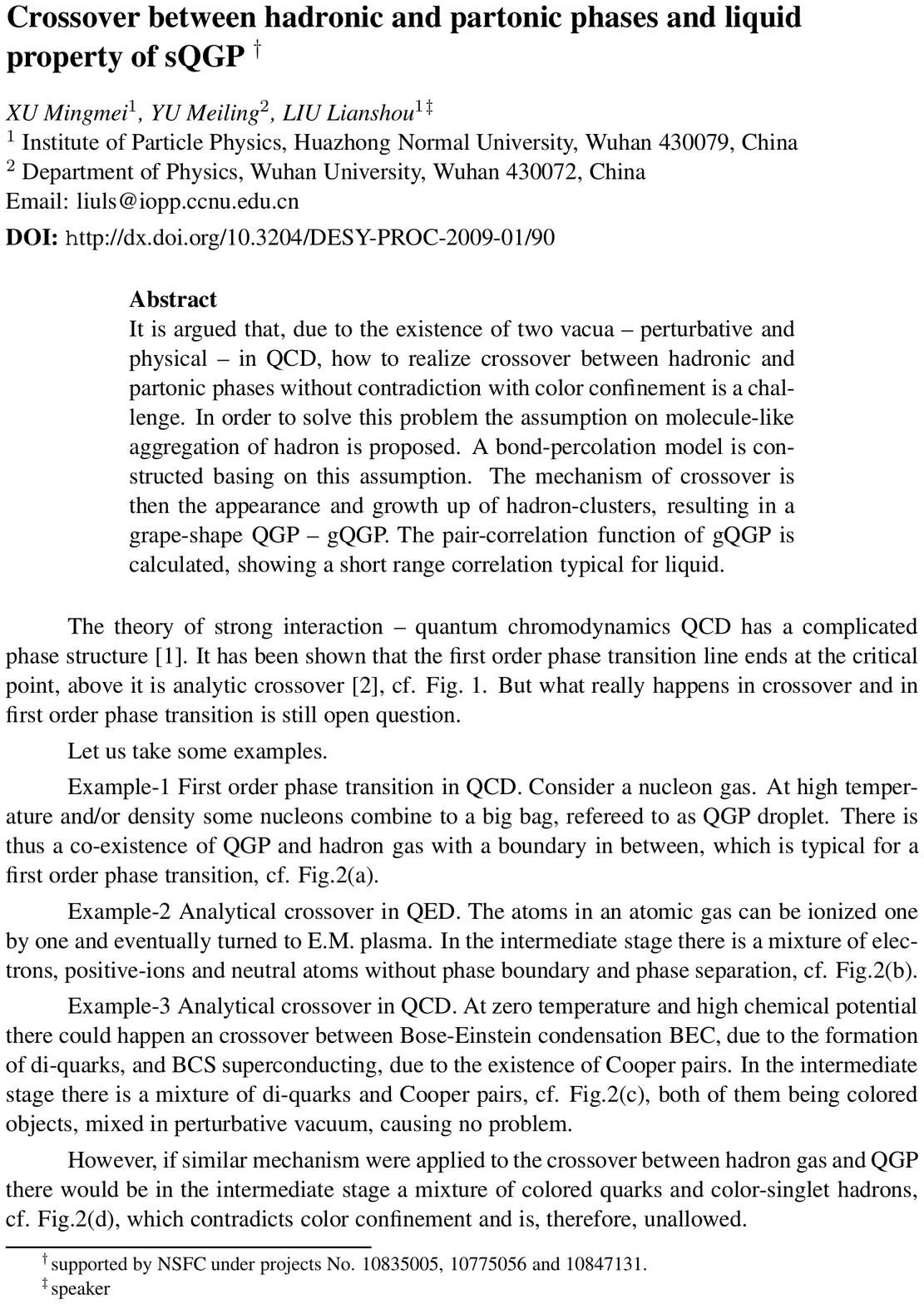} 

\coltoctitle{Nuclear Effects in High-$p_T$
Hadron Production at Large $x$}
\coltocauthor{J.~Nemchik,  M.~\v Sumbera}
\Includeart{\CAUT}{\CTIT}{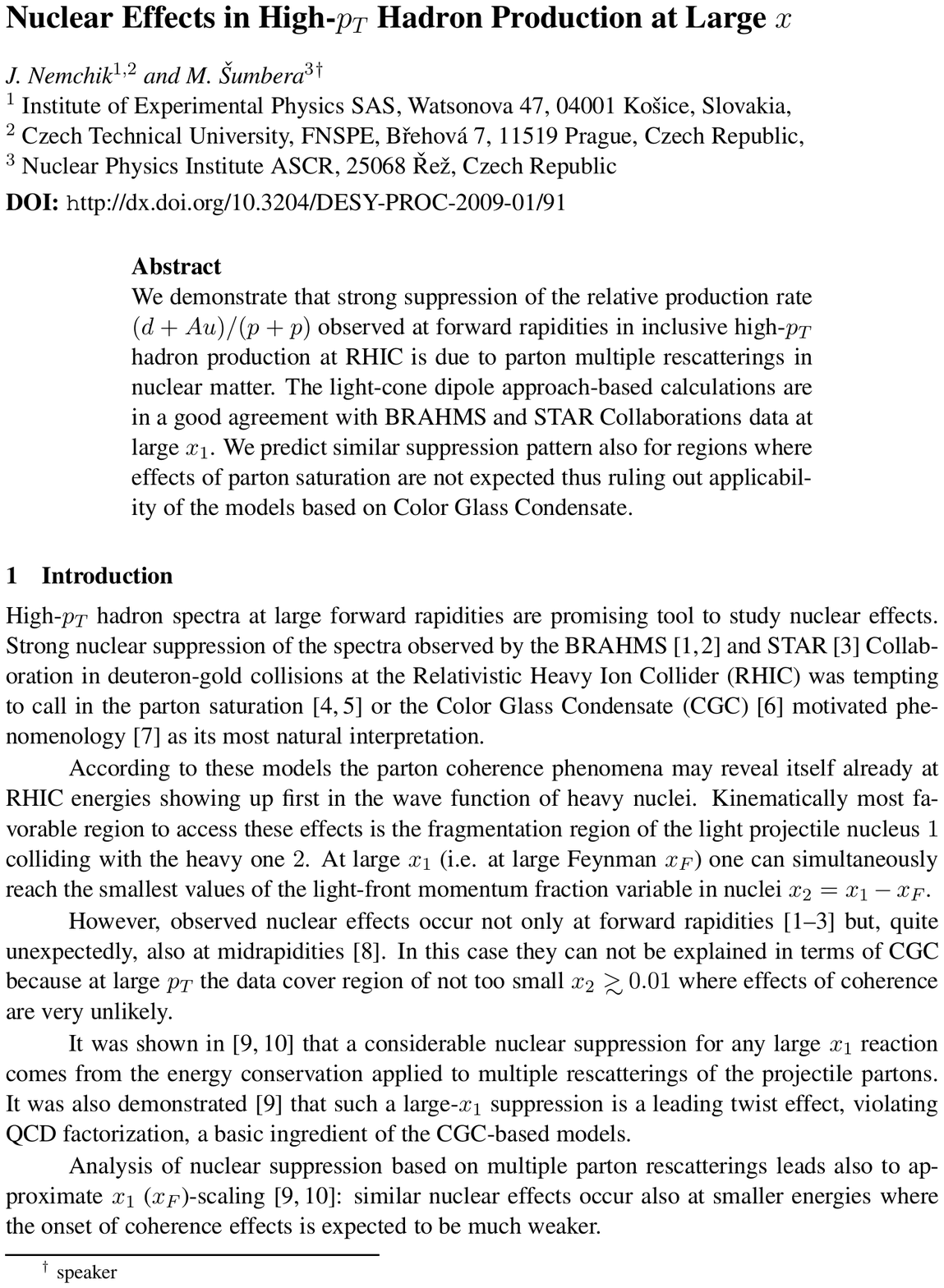} 

\coltoctitle{The nimbus of away-side jets}
\coltocauthor{I.M. Dremin}
\Includeart{\CAUT}{\CTIT}{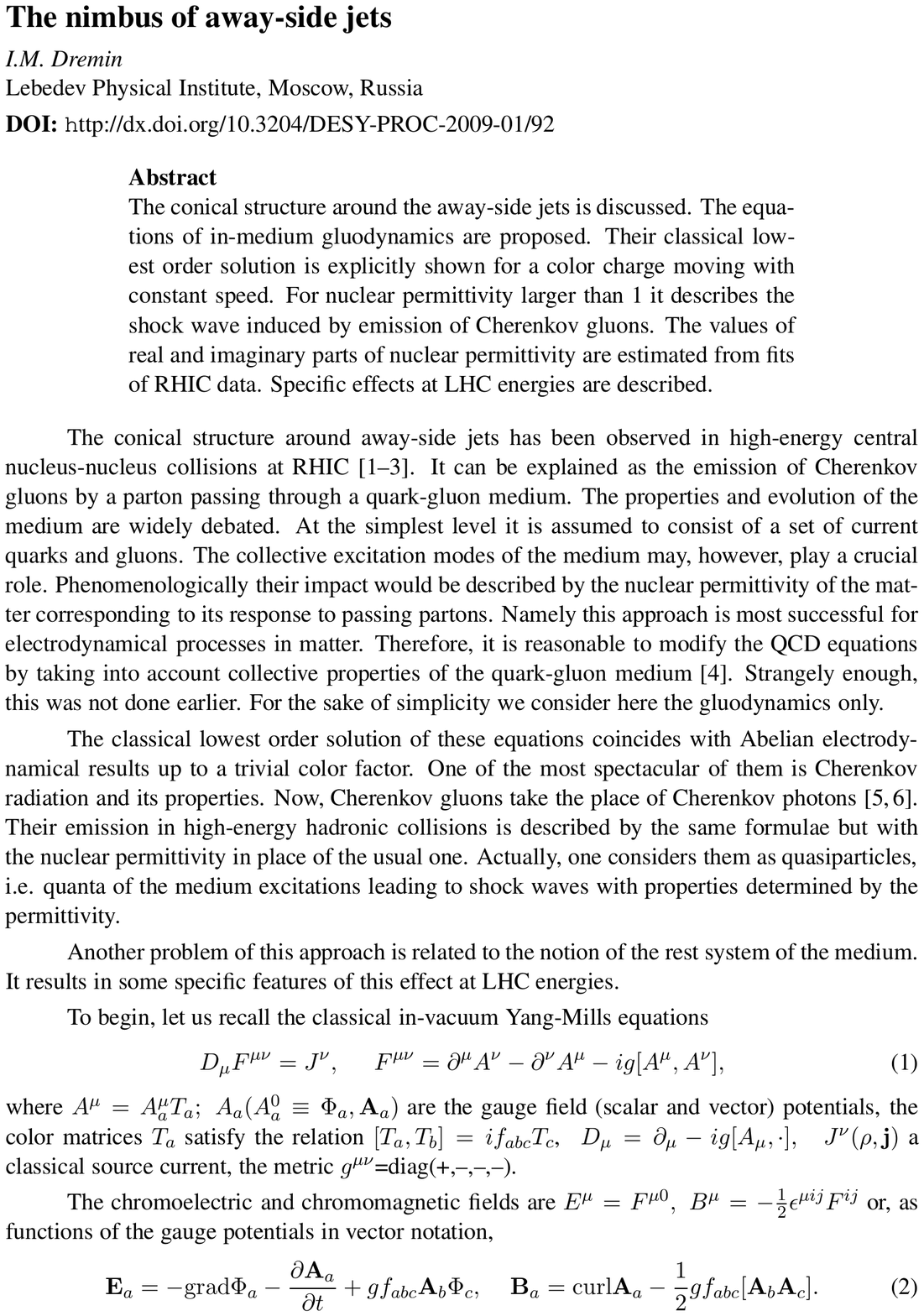} 

\coltoctitle{Baryon stopping as a test of geometric scaling}
\coltocauthor{Y. Mehtar-Tani,  G.  Wolschin}
\Includeart{\CAUT}{\CTIT}{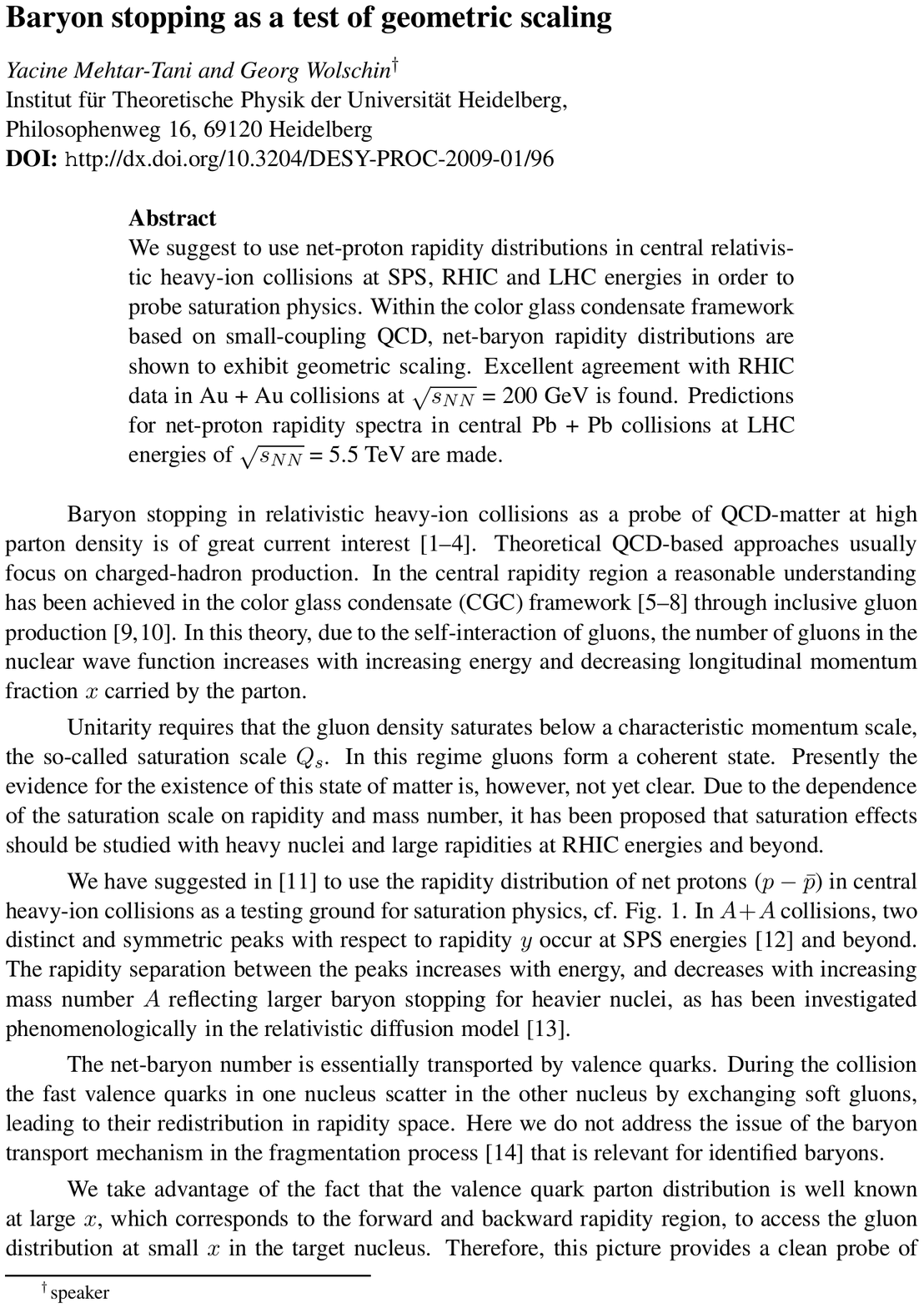} 

\coltoctitle{Transport Coefficients for Non-Newtonian Fluids and Causal
Dissipative Hydrodynamics}
\coltocauthor{T. Kodama, T. Koide}
\Includeart{\CAUT}{\CTIT}{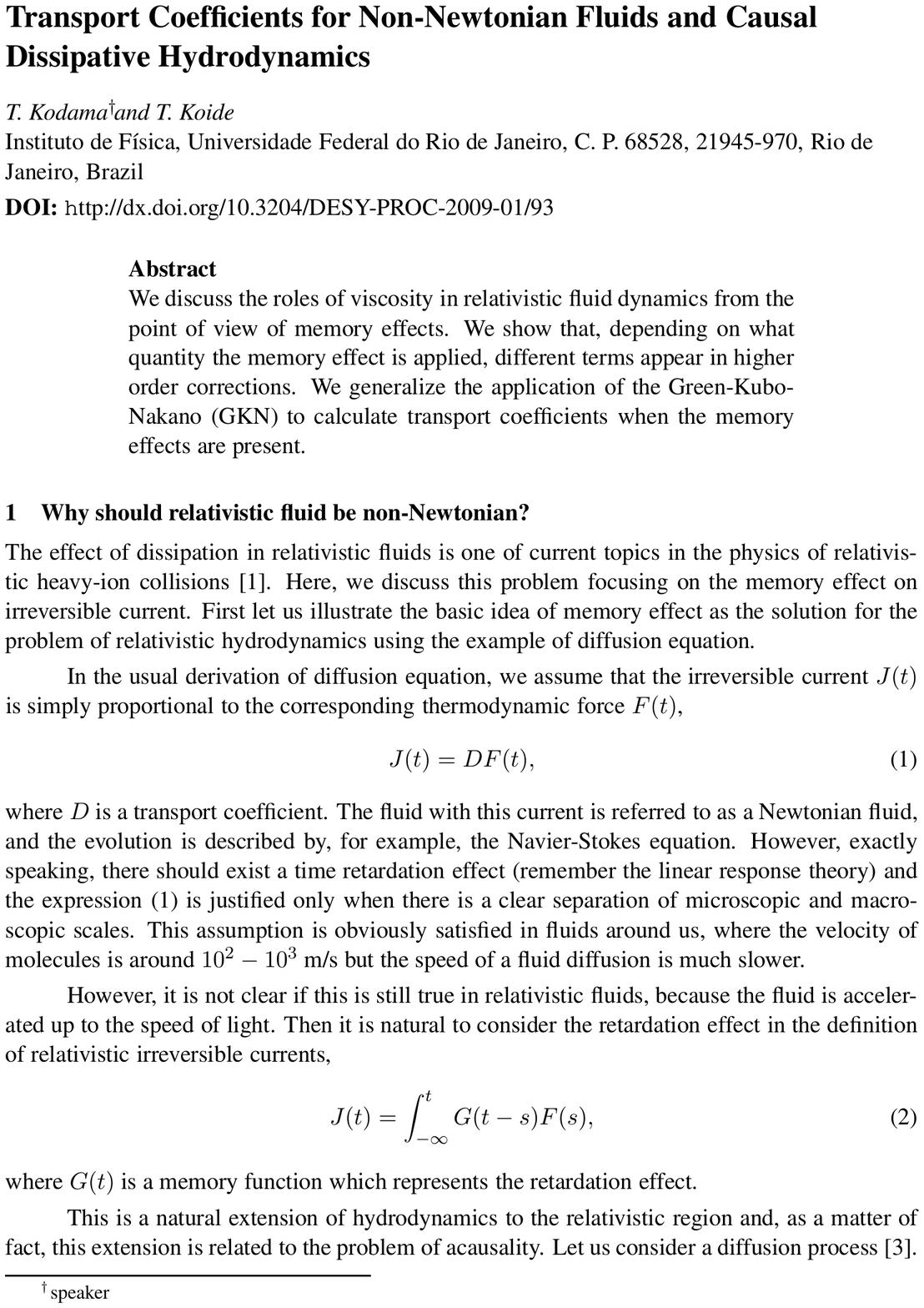} 

\coltoctitle{Study of extremely high multiplicity events in the SVD-2 experiment}
\coltocauthor{E.~Kokoulina,
 A.~Kutov, V.~Ryadovikov}
\Includeart{\CAUT}{\CTIT}{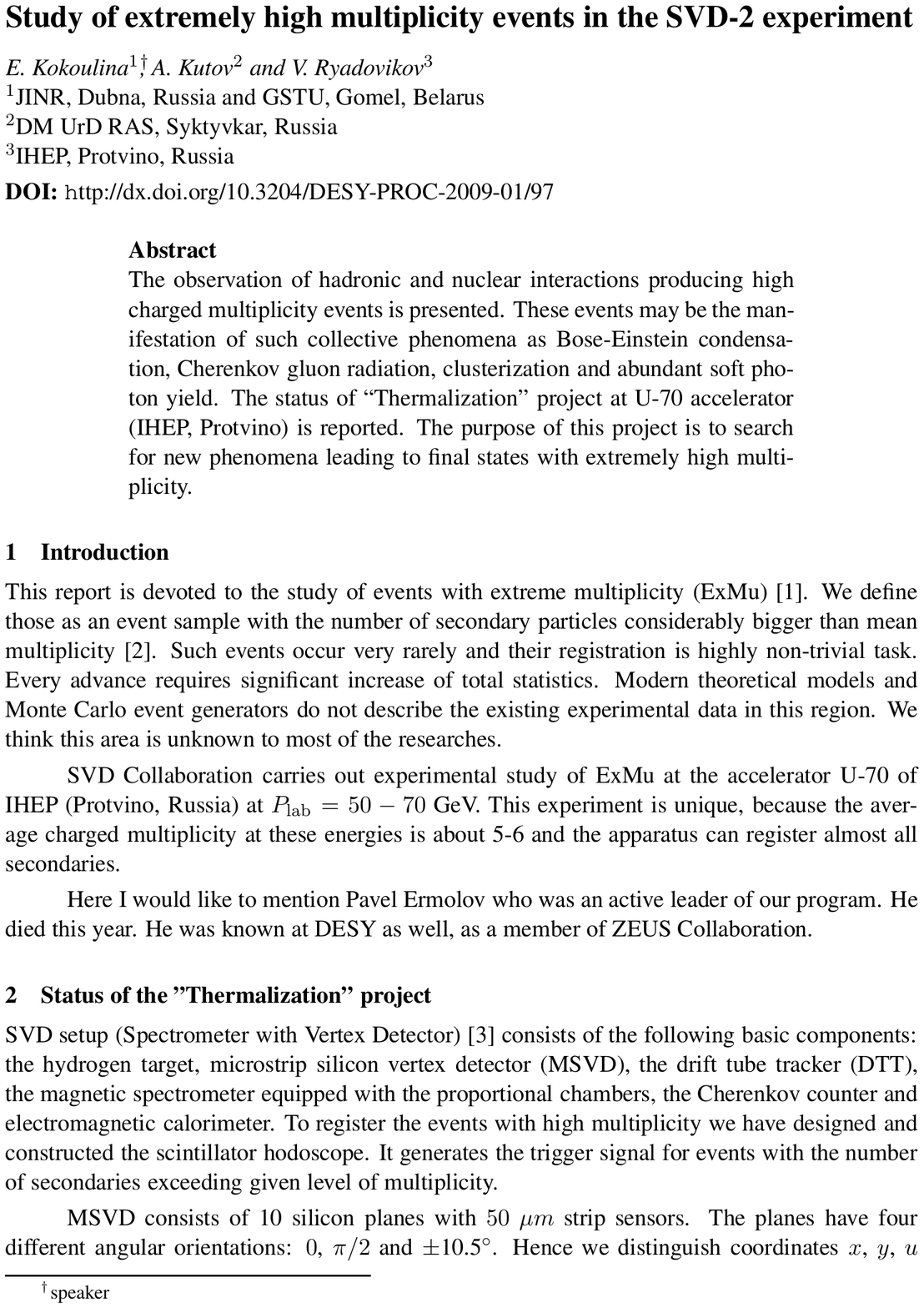} 

\coltoctitle{Soft photon production in matter in two particle green's function consideration}
\coltocauthor{A.V. Koshelkin}
\Includeart{\CAUT}{\CTIT}{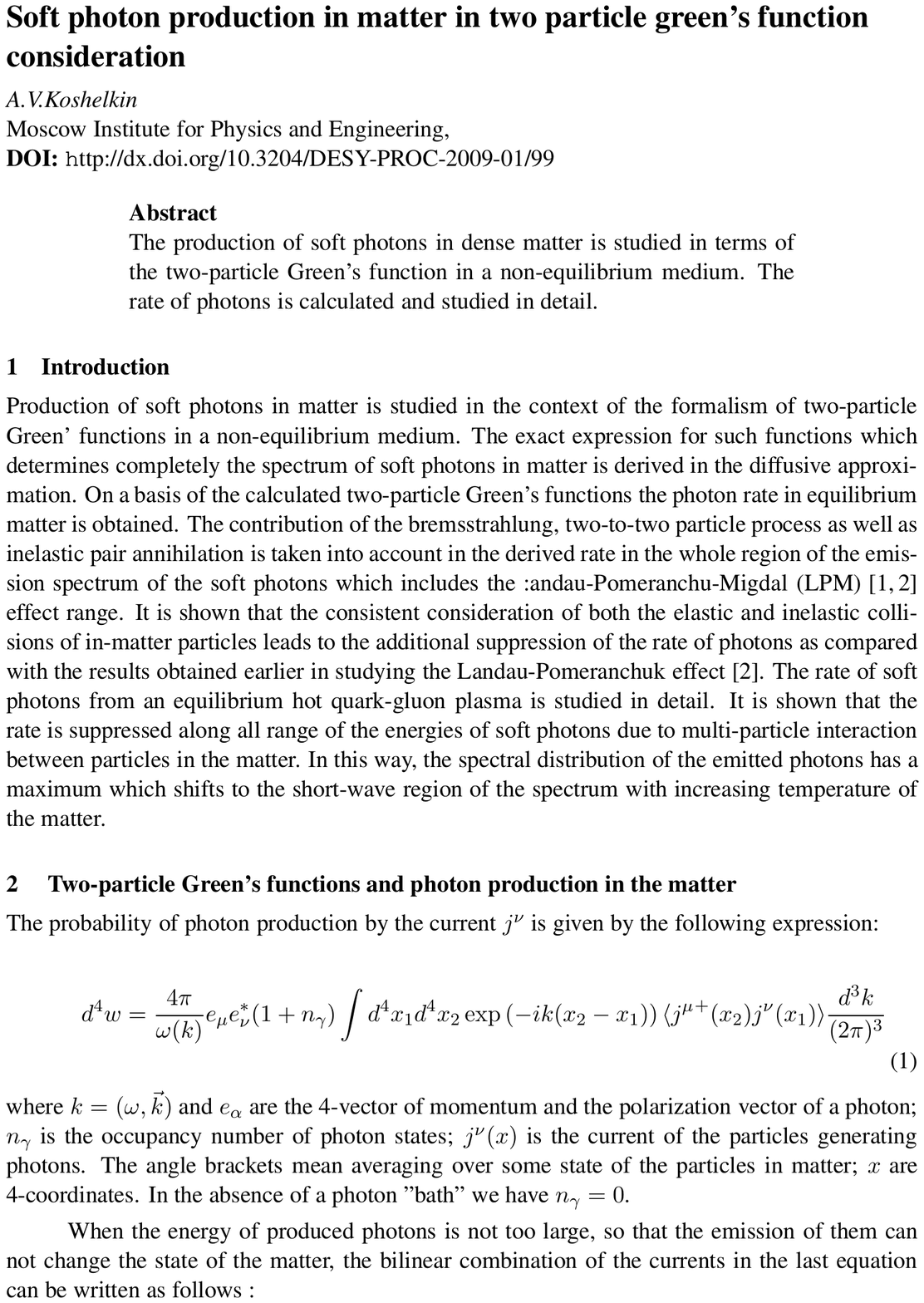} 

\coltoctitle{New Scaling Behavior of low-$p_T$ Hadron Production in proton-(anti)proton collisions at RHIC and Tevatron}
\coltocauthor{M. Tokarev, I. Zborovsk}
\Includeart{\CAUT}{\CTIT}{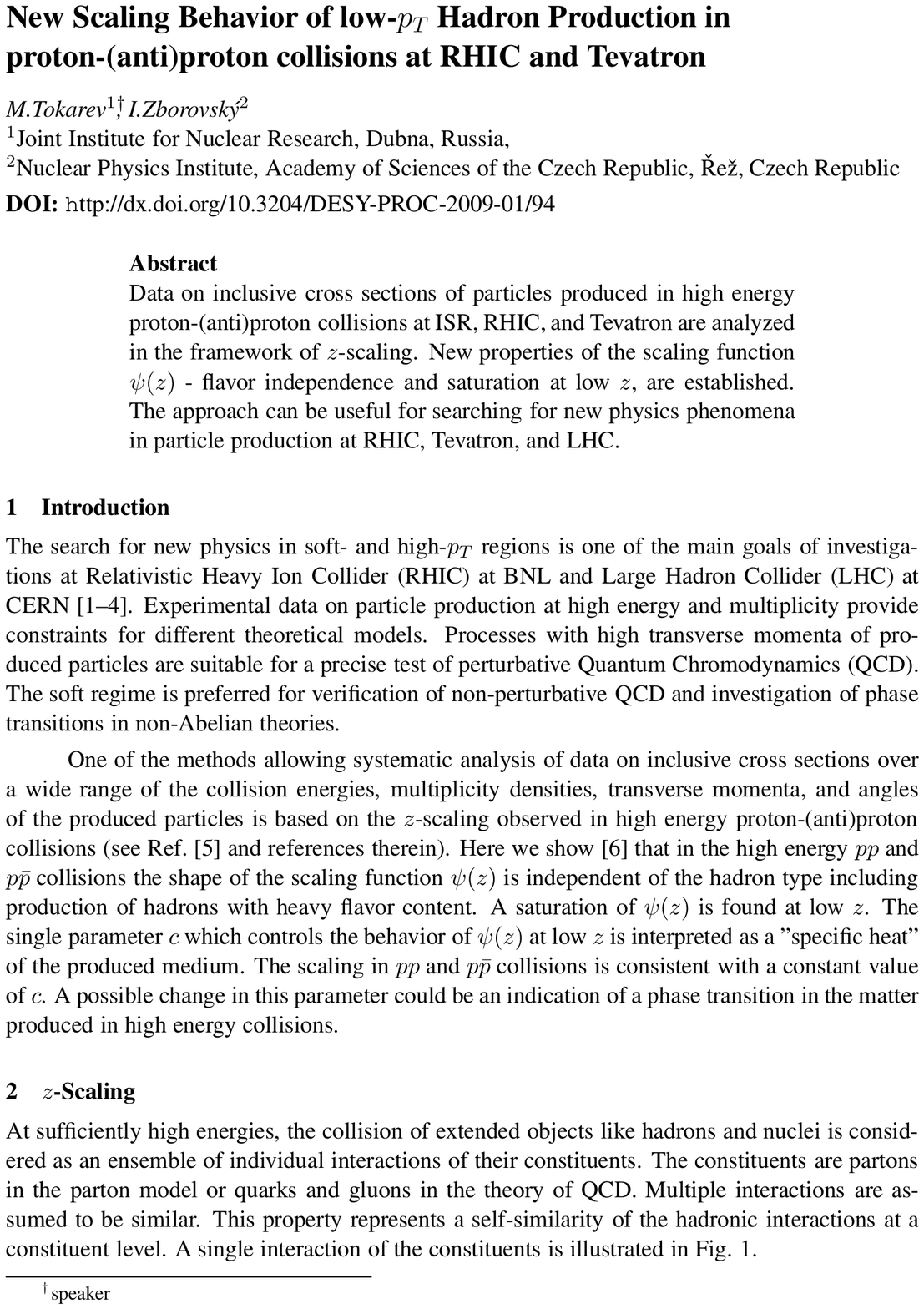} 

\coltoctitle{Strongly Interacting Massive Particles at LHC}
\coltocauthor{O.I.~Piskounova,
 A.B.~Kaidalov}
\Includeart{\CAUT}{\CTIT}{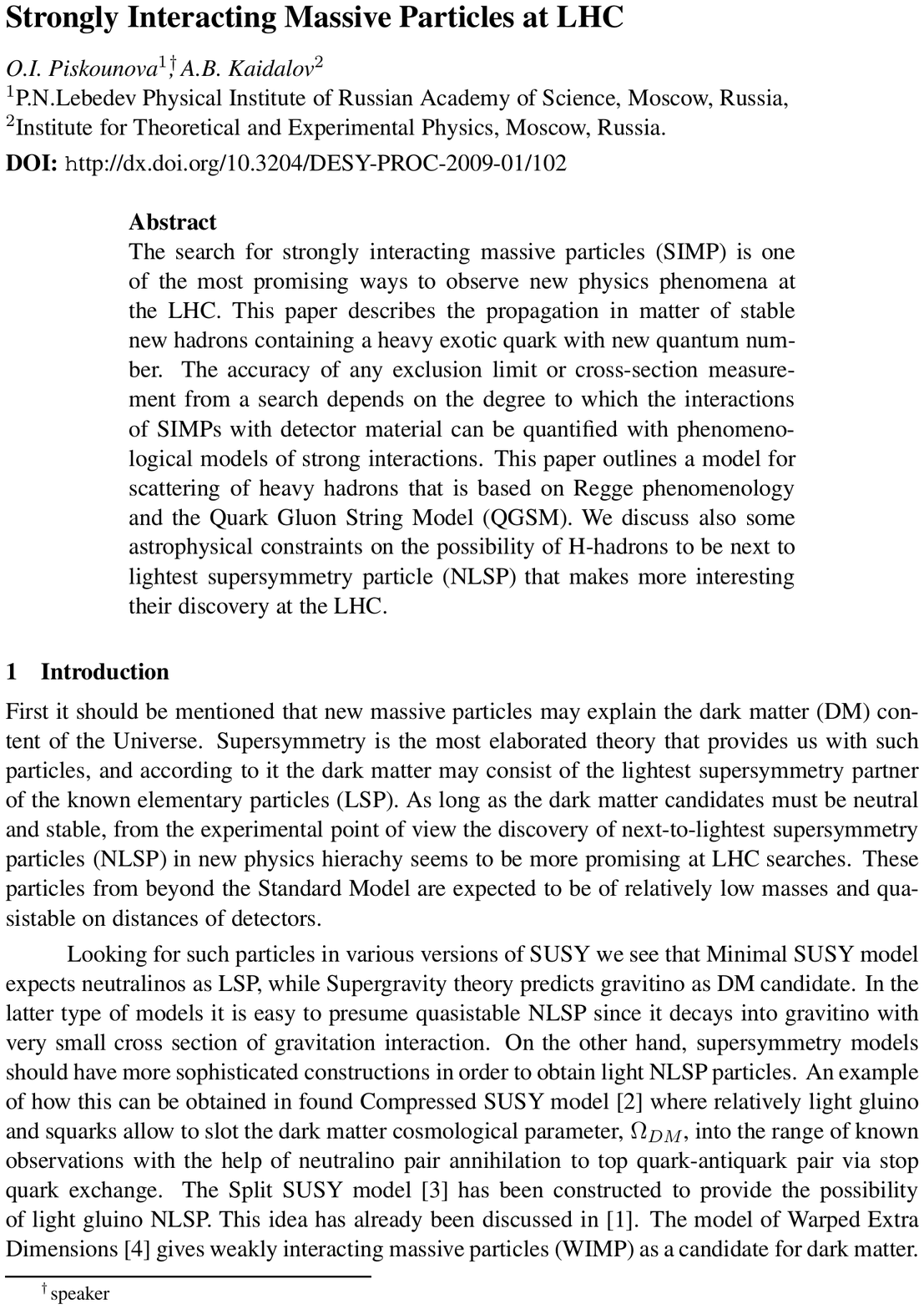} 

\coltoctitle{Saturation in lepton- and hadron induced reactions}
\coltocauthor{L.L.~Jenkovszky}
\Includeart{\CAUT}{\CTIT}{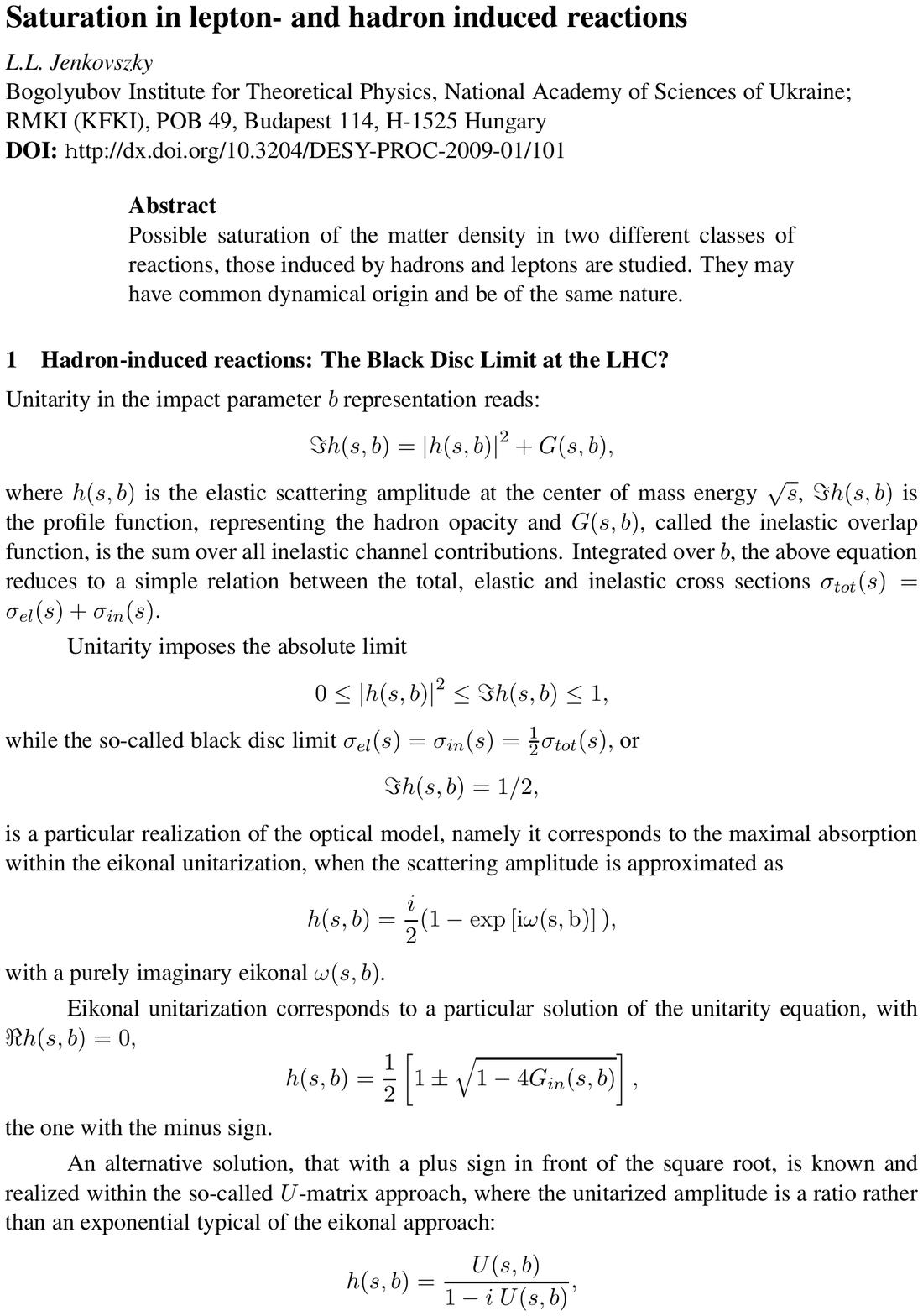} 

\coltoctitle{Multi-parton interactions and underlying events from Tevatron to LHC}
\coltocauthor{P. Bartalini,
F. Ambroglini, L. Fan\`o, R. Field, L. Garbini, D. Treleani}
\Includeart{\CAUT}{\CTIT}{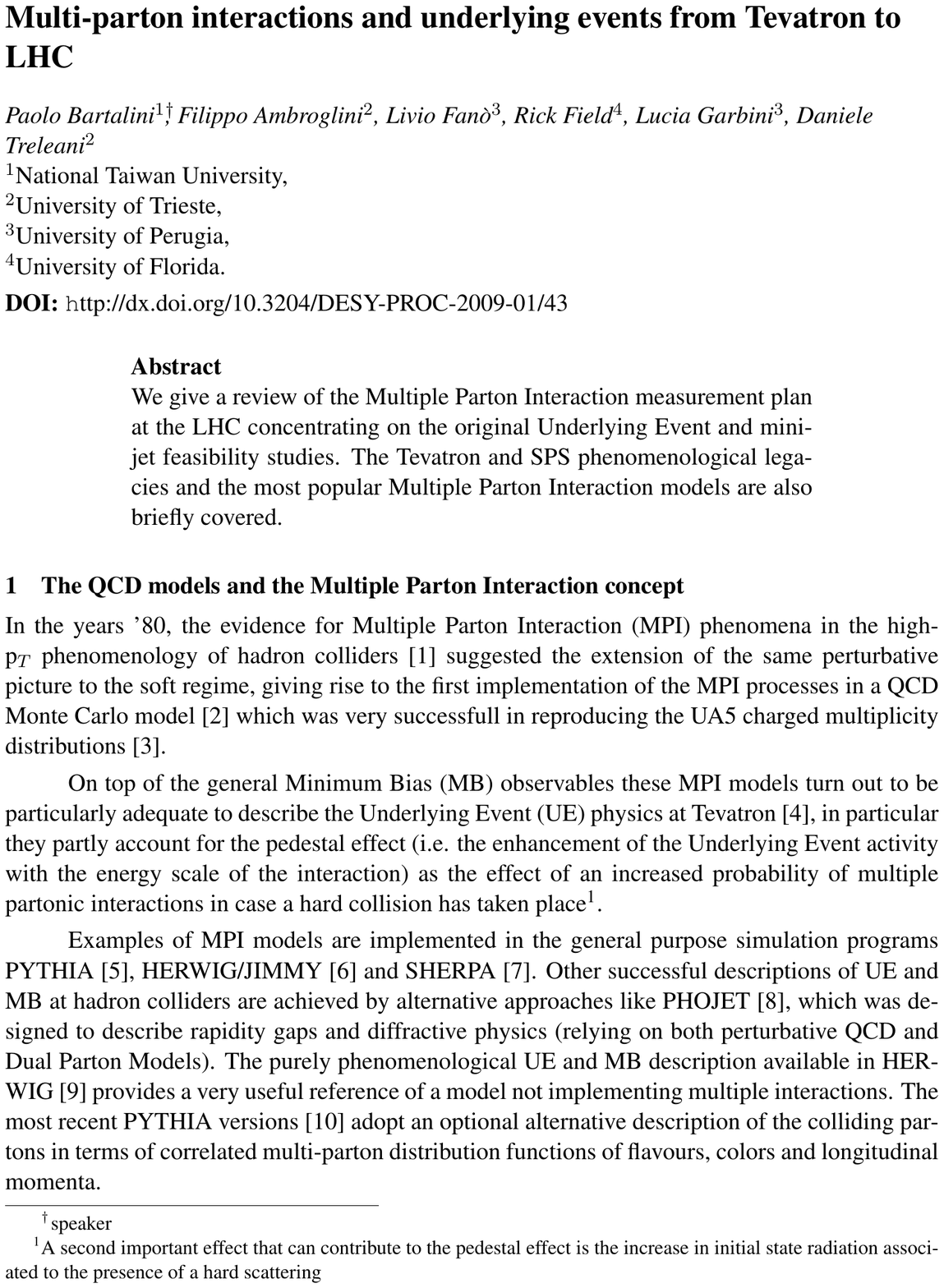} 

\coltoctitle{Multiplicities and the Underlying Event}
\coltocauthor{D. Kar}
\Includeart{\CAUT}{\CTIT}{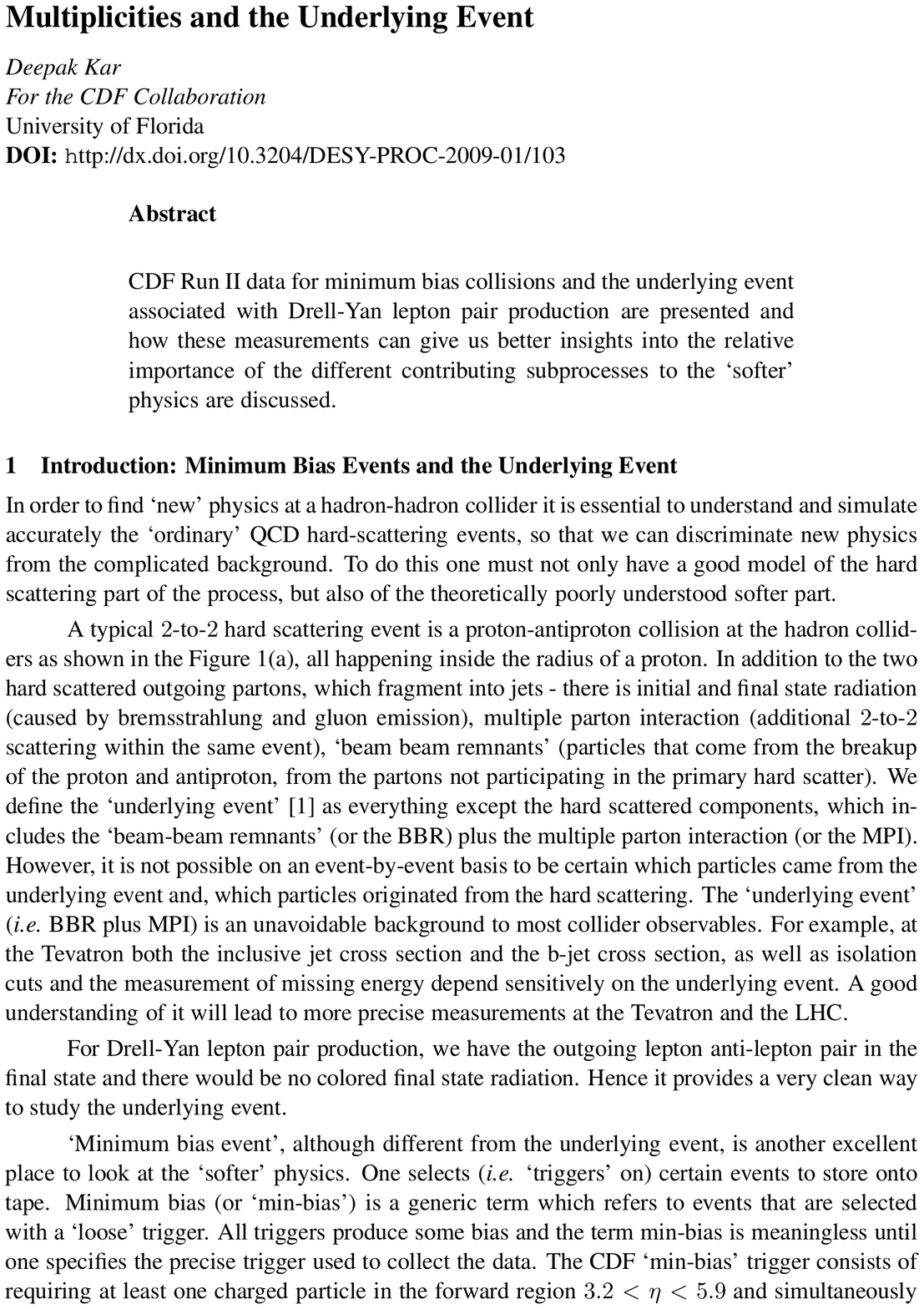} 

\coltoctitle{Saturation effects in final states due to CCFM with absorptive boundary}
\coltocauthor{K. Kutak, 
 H. Jung}
\Includeart{\CAUT}{\CTIT}{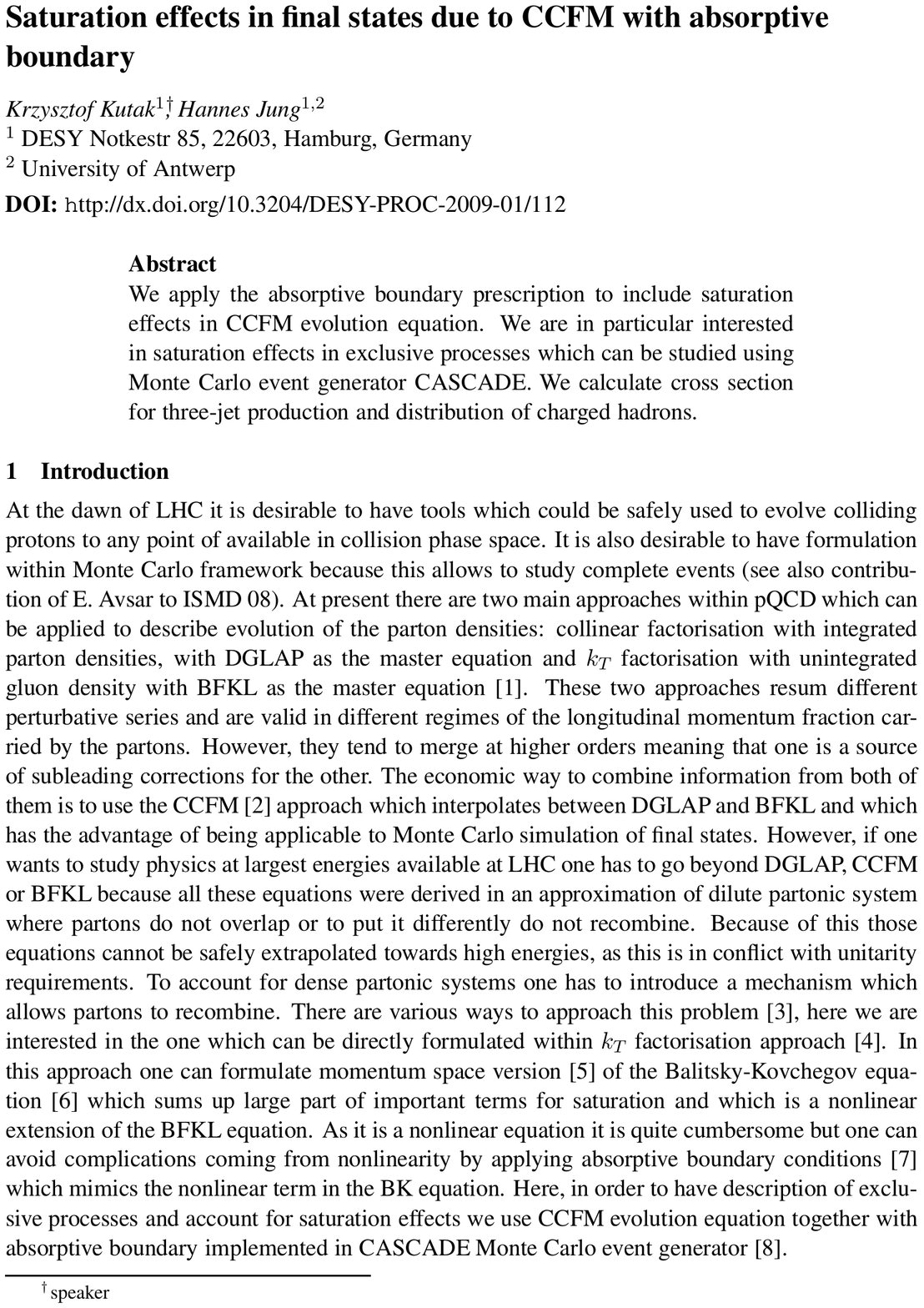} 

\coltoctitle{Photoproduction total cross-sections
at very high energies and the Froissart bound}
\coltocauthor{Y. N. Srivastava, 
A. Achilli, R. Godbole, A. Grau, G. Pancheri}
\Includeart{\CAUT}{\CTIT}{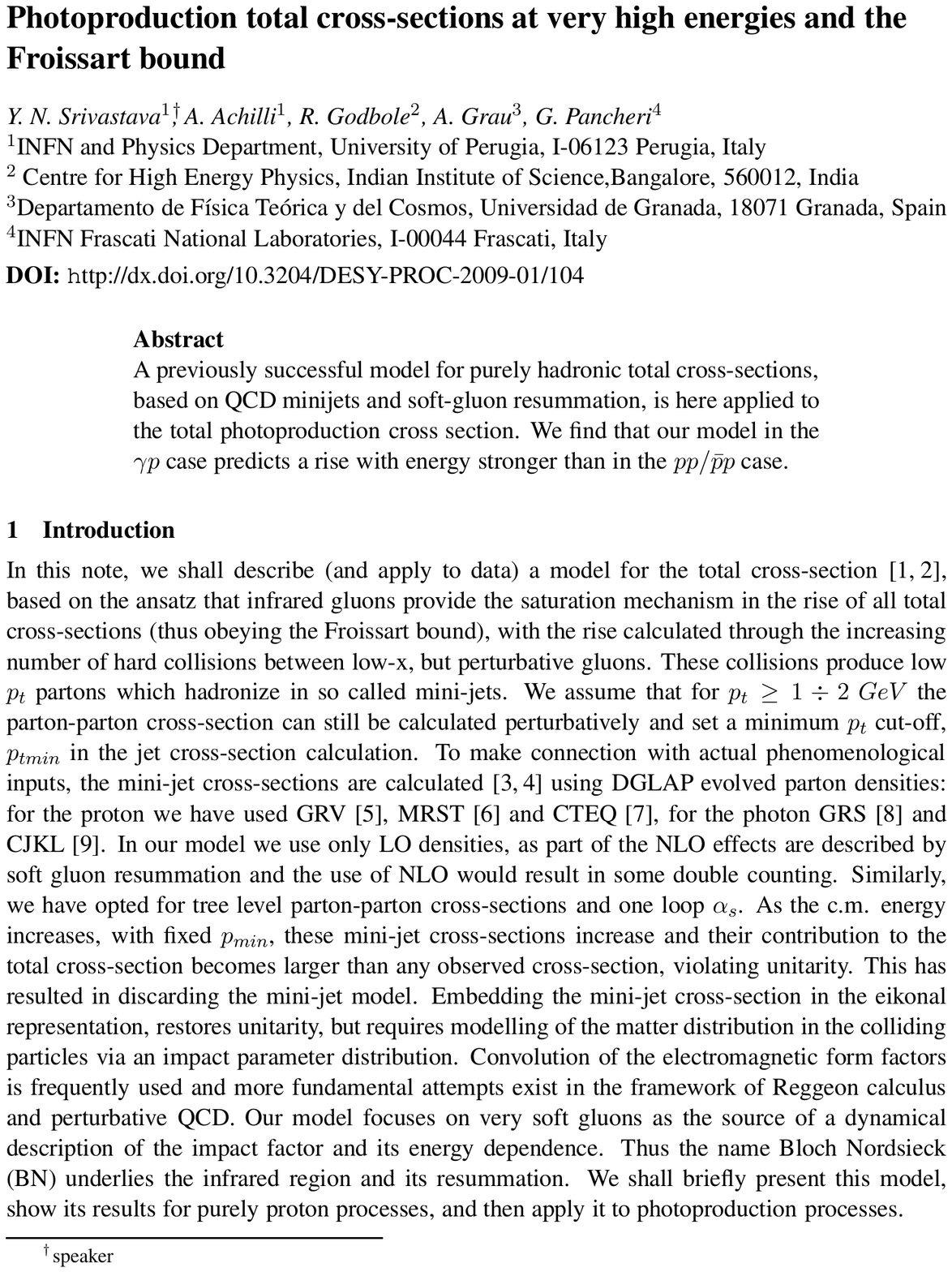} 

\coltoctitle{Monte Carlo and large angle gluon radiation}
\coltocauthor{G. Marchesini}
\Includeart{\CAUT}{\CTIT}{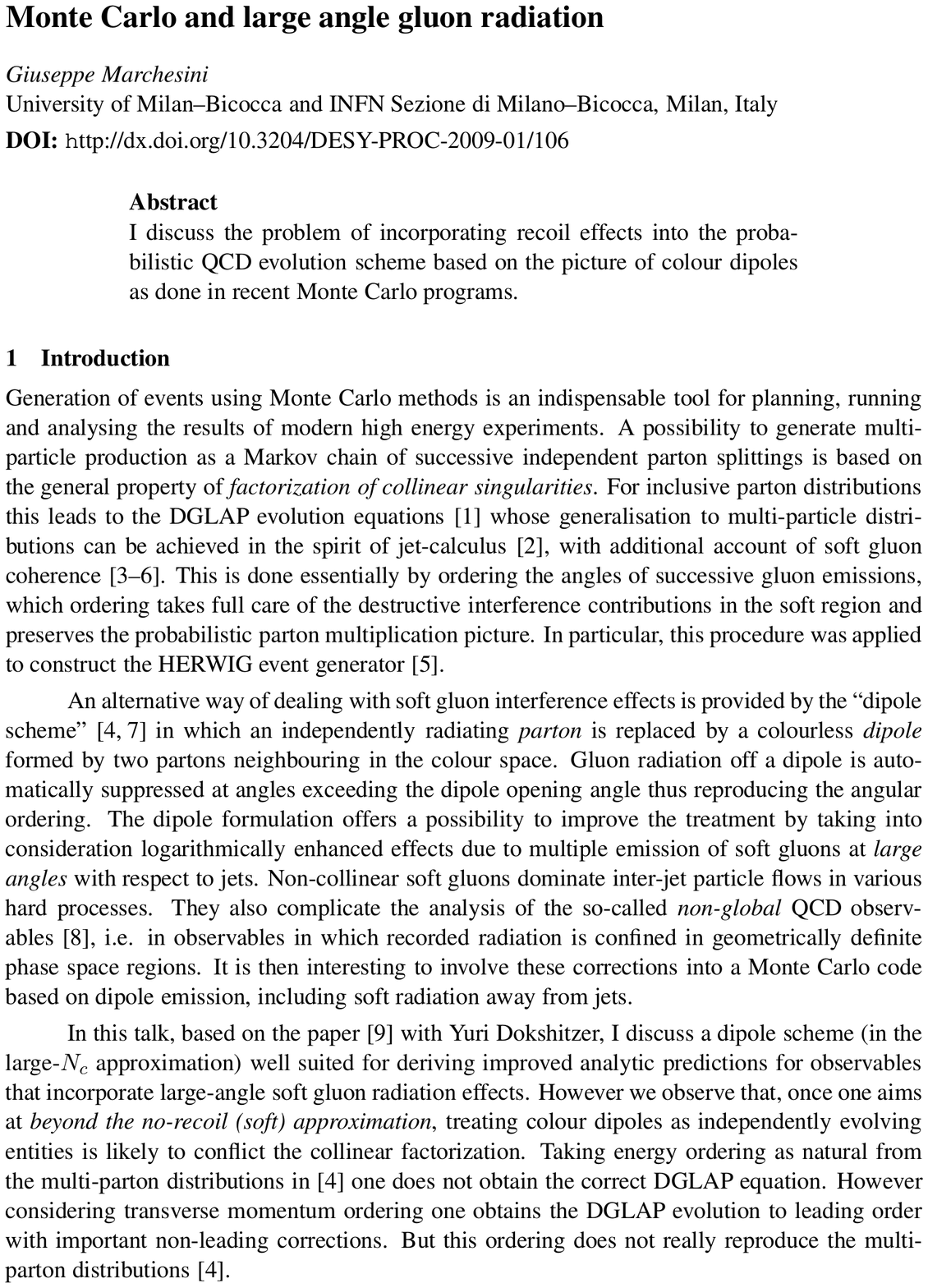} 

\coltoctitle{Production amplitudes in $N=4$ SUSY and Mandelstam cuts}
\coltocauthor{J. Bartels, L.N. Lipatov}
\Includeart{\CAUT}{\CTIT}{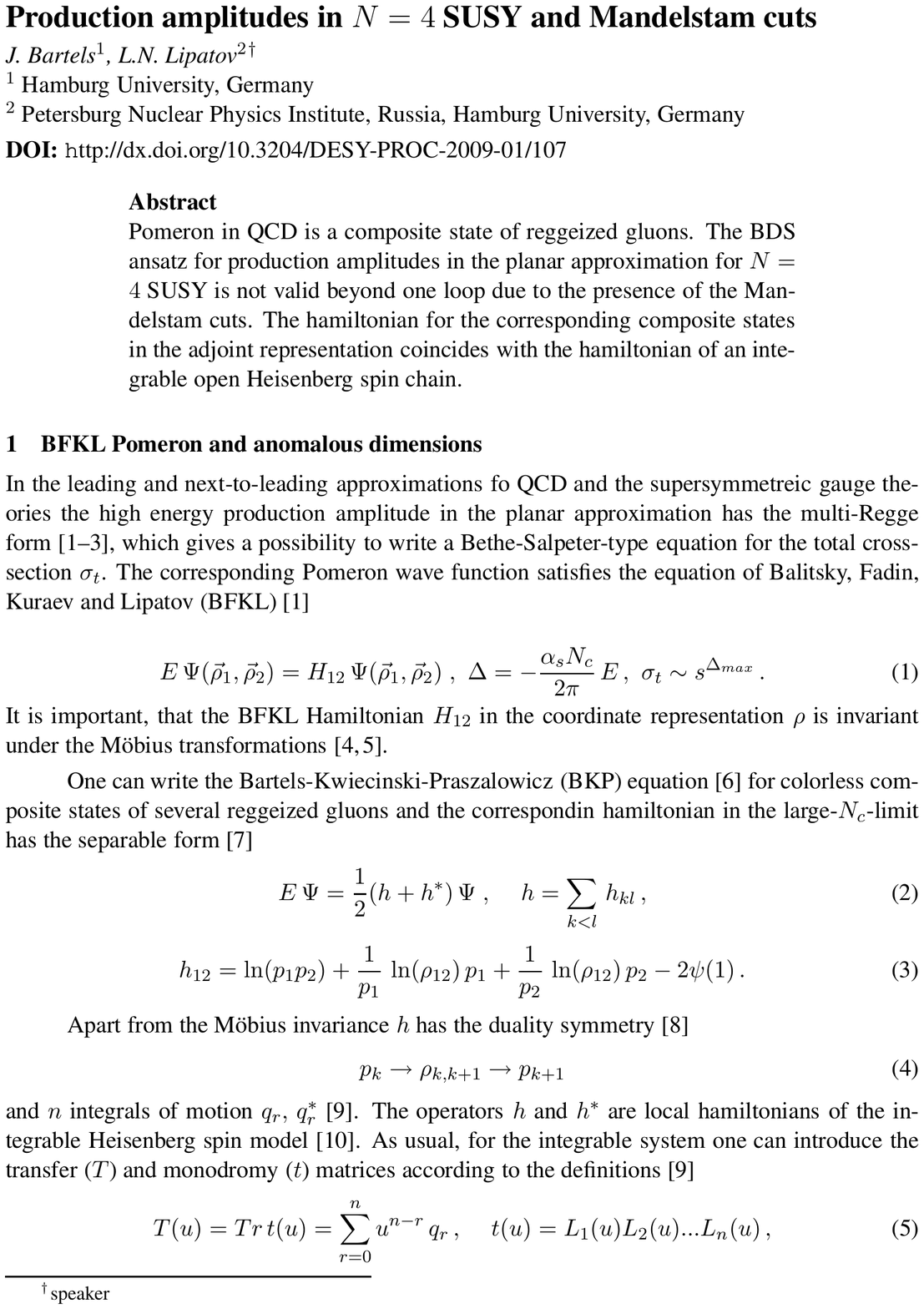} 

\coltoctitle{High energy scattering in QCD vs. tiny black holes}
\coltocauthor{L. \'Alvarez-Gaum\'e, C. G\'omez, 
A. Sabio Vera, A. Tavanfar, M. A. V\'azquez-Mozo}
\Includeart{\CAUT}{\CTIT}{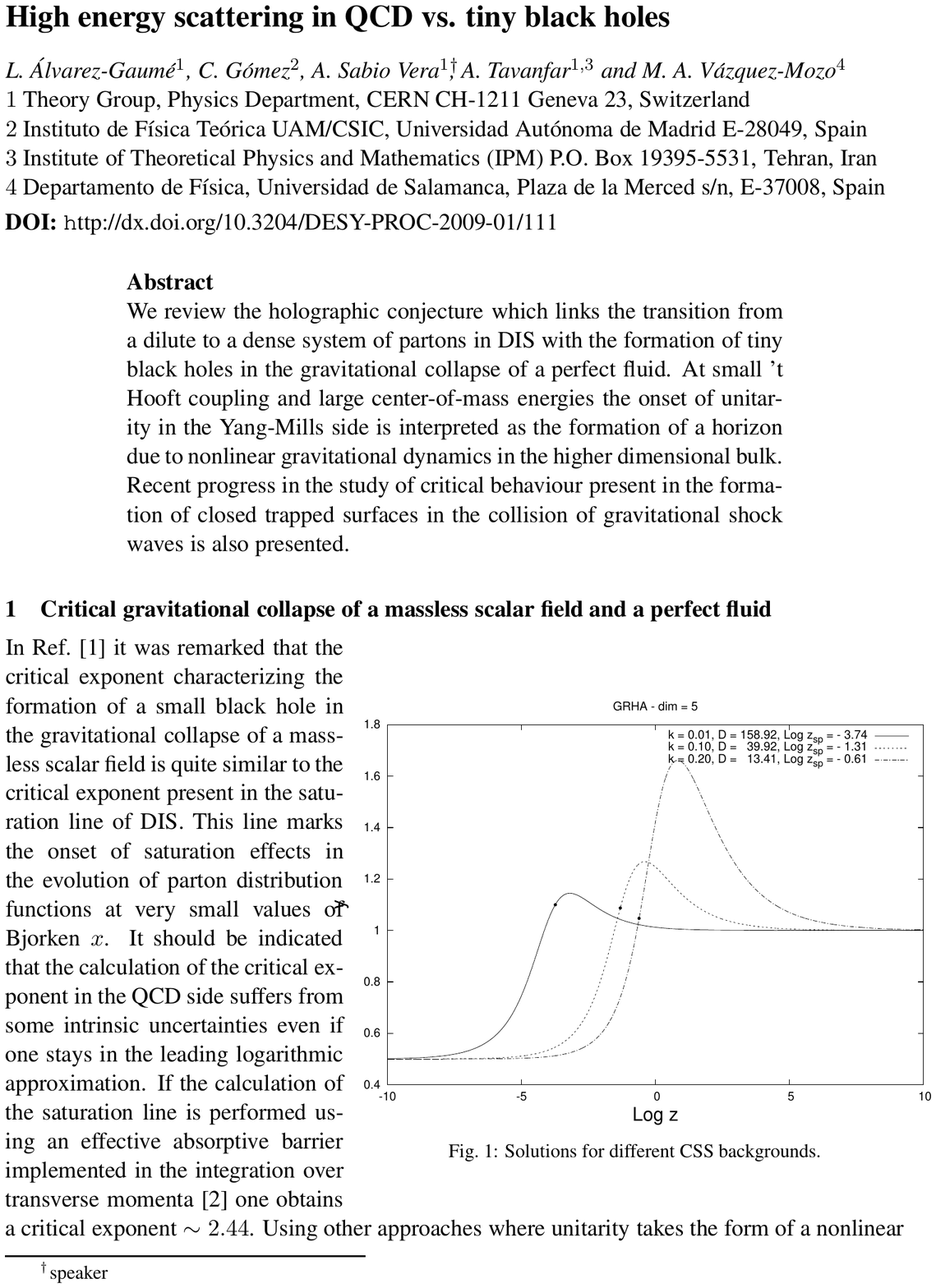} 

% \coltoctitle{Second Talk}
% \coltocauthor{David Summerset}
% \Includeart{\CAUT}{\CTIT}{summerset_david}

\end{papers} % end of individual articles/papers

%% file: include-discussion.tex
\chapter[\large  Discussion Session]{ Discussion Session}
%{\Large\bfseries Convenors:}
%\vspace{10mm}

%{\Large\itshape 
%Mandy Cooper-Sarkar (Oxford), \\
%Anna Kulesza (DESY), \\
%Ken Hatakeyama (Rockefeller)}

\CLDP
\begin{papers} % start of individual articles/papers

\coltoctitle{Discussion Session}
\coltocauthor{M. Diehl,  K. Golec-Biernat, A. Cooper-Sakar, A. De Roeck}
\Includeart{\CAUT}{\CTIT}{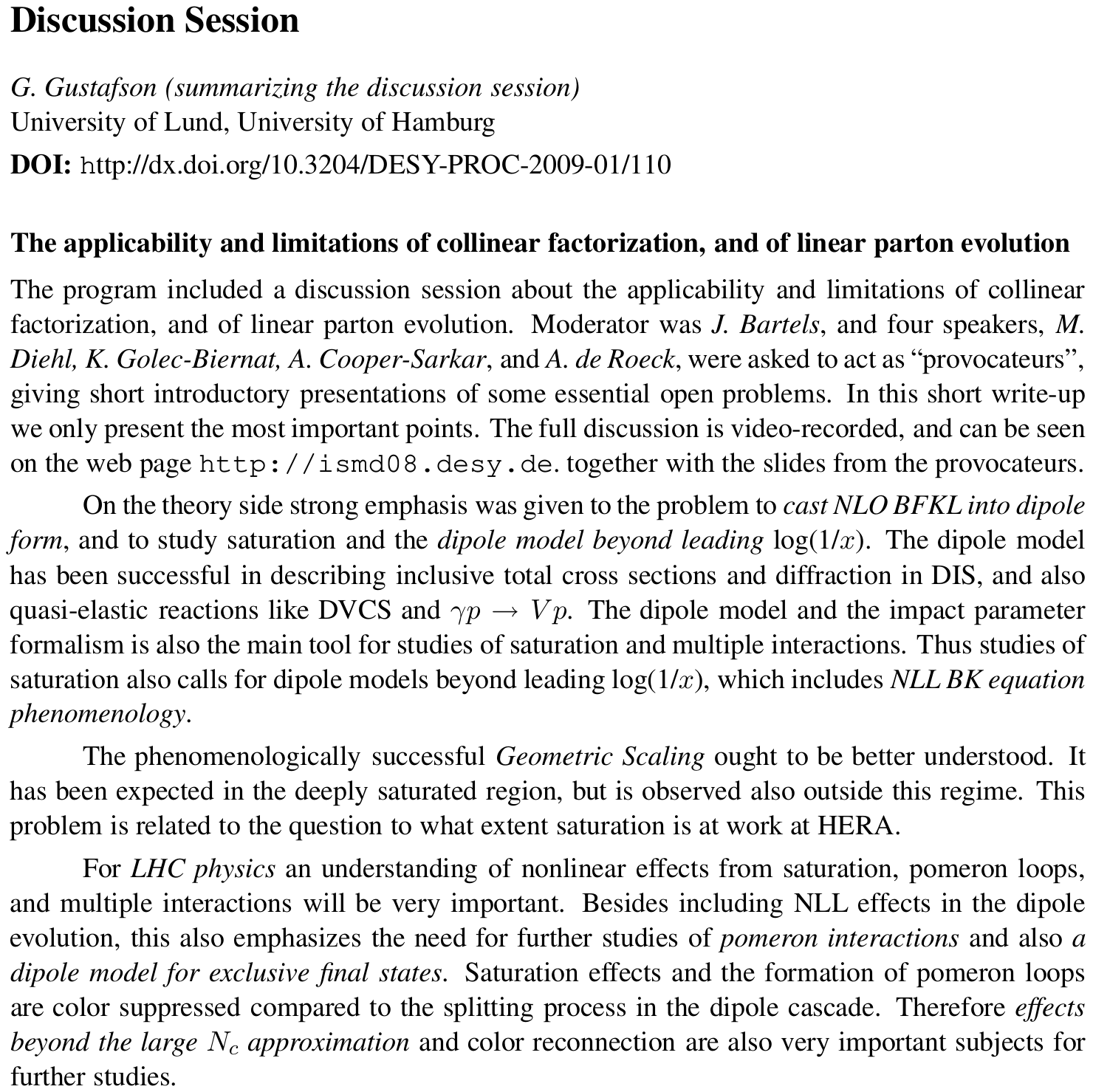} 

\coltoctitle{Have we seen anything beyond (N)NLO DGLAP at HERA? }
\coltocauthor{A. Cooper-Sarkar}
\Includeart{\CAUT}{\CTIT}{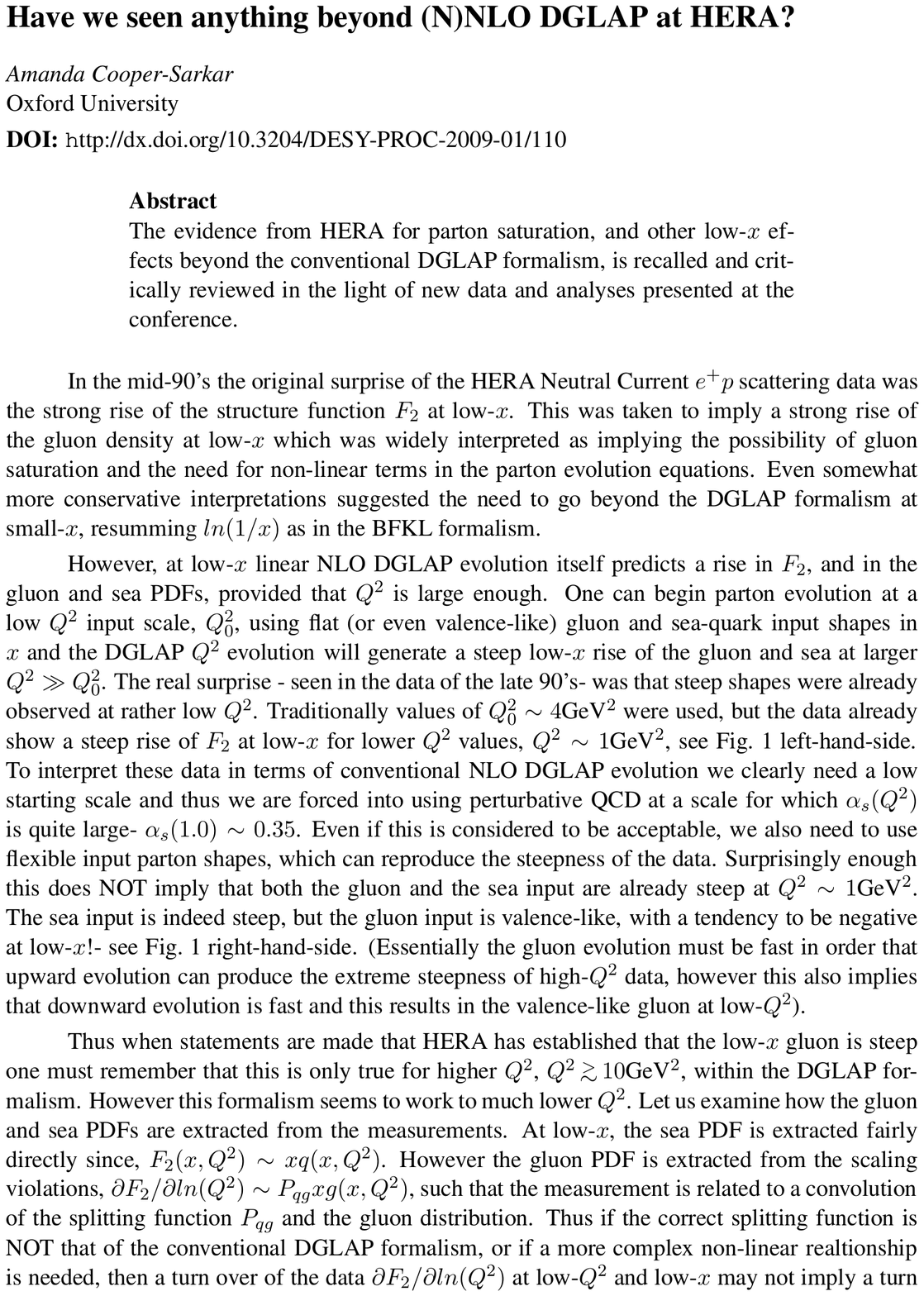}

\coltoctitle{Saturation: what do we need}
\coltocauthor{A. De Roeck, H. Jung}
\Includeart{\CAUT}{\CTIT}{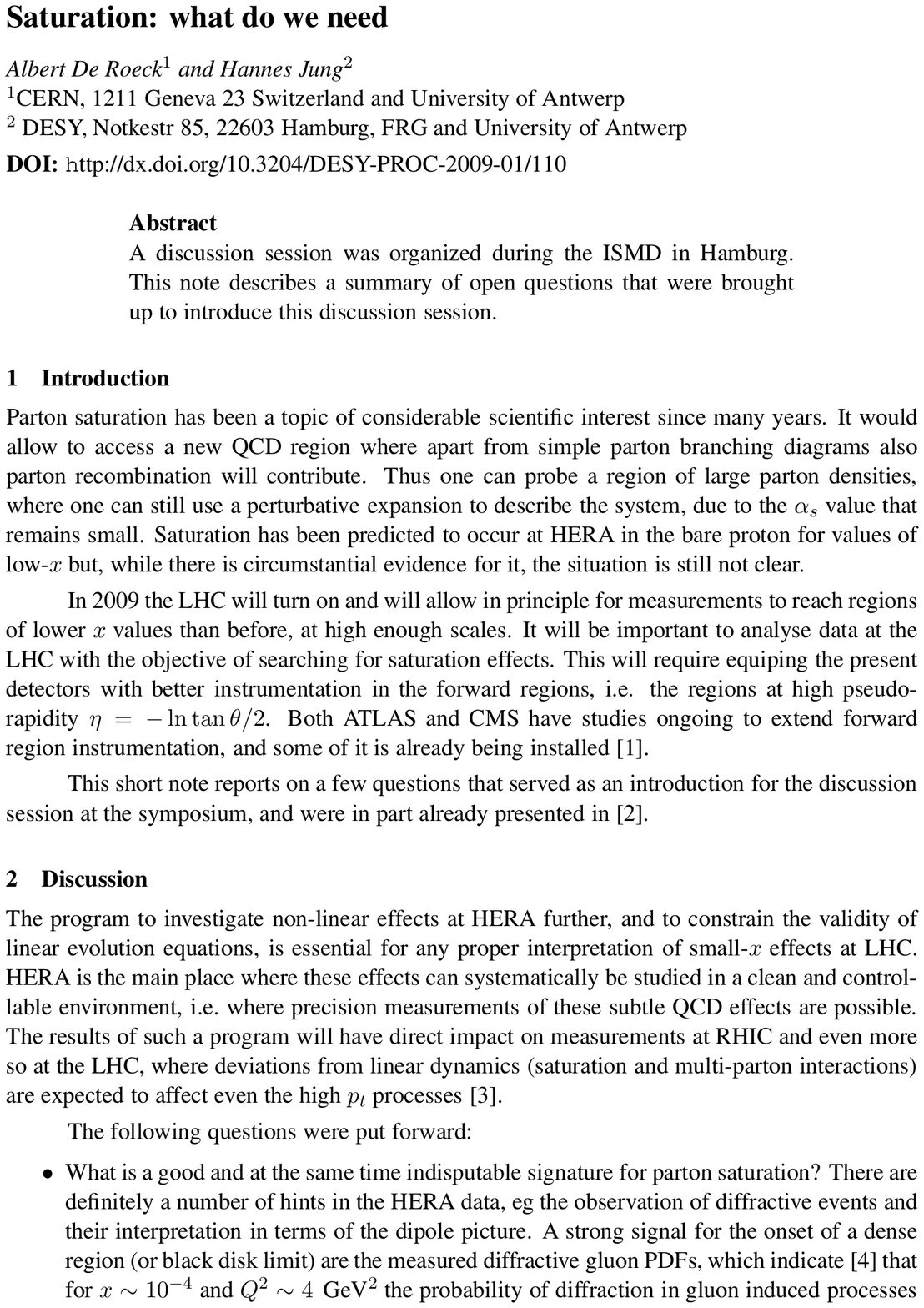} 

% \coltoctitle{Second Talk}
% \coltocauthor{David Summerset}
% \Includeart{\CAUT}{\CTIT}{summerset_david}

\end{papers} % end of individual articles/papers

%% file: include-summary.tex
\chapter[\large  Summaries]{Summaries}
\begin{papers} % start of individual articles/papers

\coltoctitle{Experimental Summary}
\coltocauthor{ P. van Mechelen}
\Includeart{\CAUT}{\CTIT}{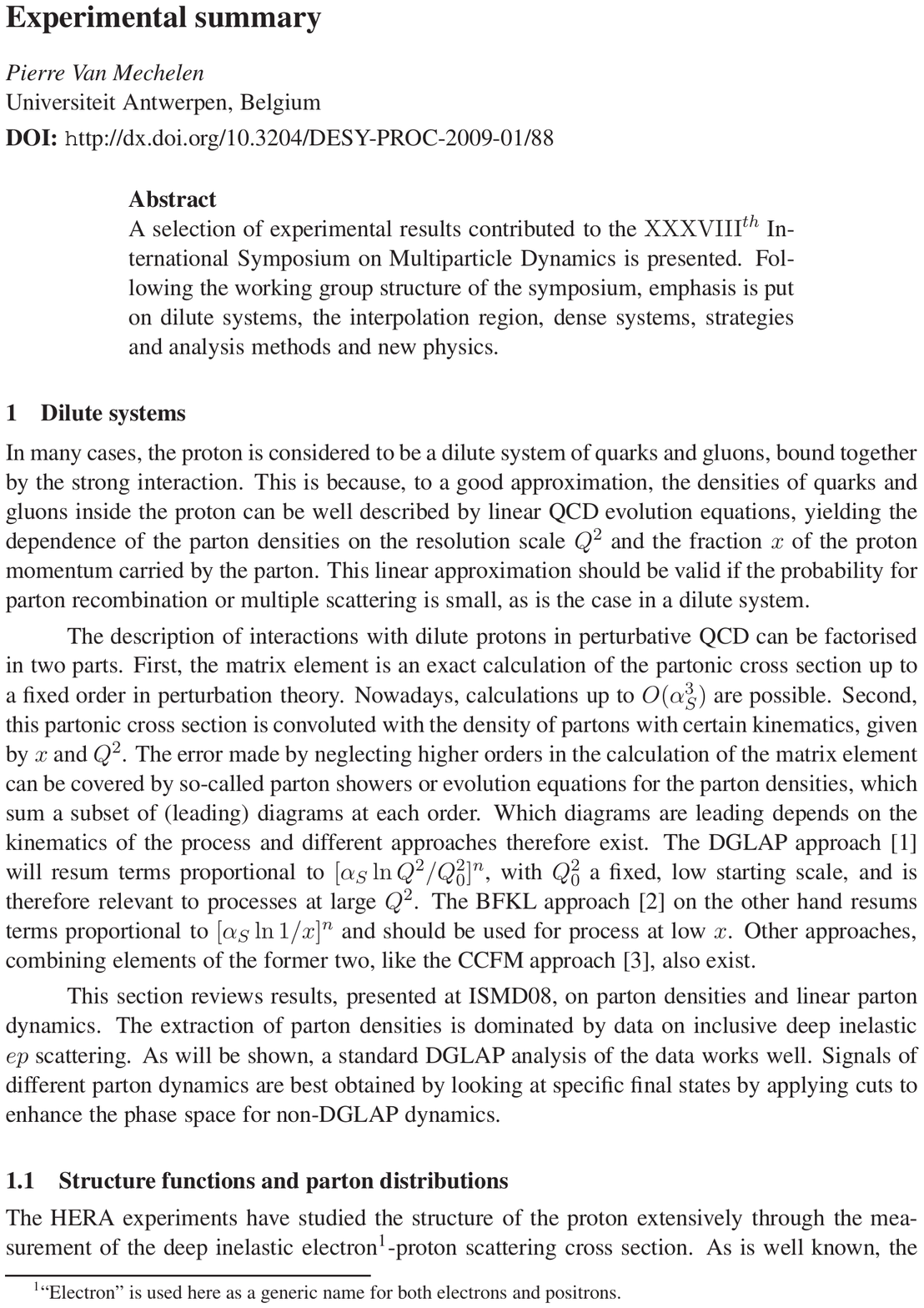} 

\coltoctitle{Theory Summary}
\coltocauthor{Y. V.\ Kovchegov}
\Includeart{\CAUT}{\CTIT}{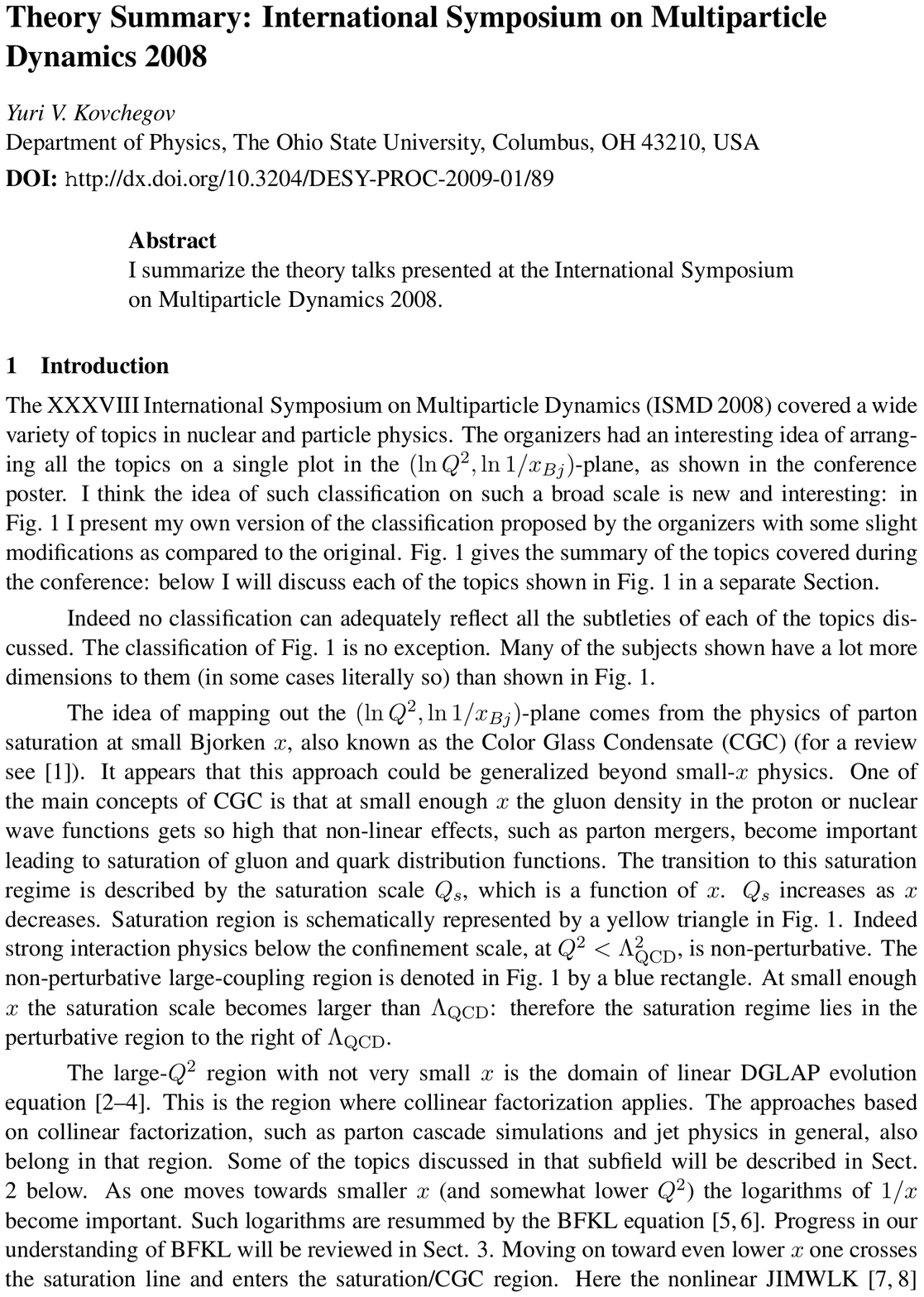} 

% \coltoctitle{Second Talk}
% \coltocauthor{David Summerset}
% \Includeart{\CAUT}{\CTIT}{summerset_david}

\end{papers} % end of individual articles/papers